%% file: cryptoknight.tex
\documentclass[10pt,journal,transmag]{IEEEtran}

\usepackage[printonlyused,nolist]{acronym}
\usepackage{tikz}
\usepackage{graphicx}
\usepackage{tikzscale}
\usepackage{hhline}
\usepackage{multicol}
\usepackage{colortbl}
\usepackage{caption}
\usepackage{pgfplotstable}
\usepackage{pgfplots}
\usepackage{wrapfig}
\usepackage{booktabs}
\usepackage{amsmath,amsfonts,amssymb,amsthm,bm}
\usepackage{algorithm}
\usepackage{algorithmic}

\definecolor{blue}{RGB}{38,139,210}
\definecolor{cyan}{RGB}{42,161,152}
\definecolor{green}{RGB}{0,102,0}
\definecolor{violet}{RGB}{108,113,196}
\definecolor{white}{RGB}{255,255,255}
\definecolor{yellow}{RGB}{244,229,66}
\definecolor{red}{RGB}{220,50,47}
\definecolor{base01}{RGB}{88,110,117}
\definecolor{base02}{RGB}{7,54,66}
\definecolor{base03}{RGB}{0,43,54}

\newcommand{\R}{\mathbb{R}}

\ifCLASSOPTIONcompsoc
  \usepackage[nocompress]{cite}
\else
  \usepackage{cite}
\fi

\begin{document}

\title{Deep Learning Based Cryptographic Primitive Classification}

\author{\IEEEauthorblockN{Gregory D. Hill}
\IEEEauthorblockA{Division of Computing and Mathematics\\
Abertay University\\
Dundee, Scotland\\
Email: gregorydhill@outlook.com}
\and
\IEEEauthorblockN{Xavier J. A. Bellekens}
\IEEEauthorblockA{Division of Computing and Mathematics\\
Abertay University\\
Dundee, Scotland\\
Email: x.bellekens@abertay.ac.uk}}

\markboth{}%
{Shell \MakeLowercase{\textit{et al.}}: Bare Demo of IEEEtran.cls for Computer Society Journals}

\maketitle

\begin{abstract}
Cryptovirological augmentations present an immediate, incomparable threat. Over the last decade, the substantial proliferation of crypto-ransomware has had widespread consequences for consumers and organisations alike. Established preventive measures perform well, however, the problem has not ceased. Reverse engineering potentially malicious software is a cumbersome task due to platform eccentricities and obfuscated transmutation mechanisms, hence requiring smarter, more efficient detection strategies. The following manuscript presents a novel approach for the classification of cryptographic primitives in compiled binary executables using deep learning. The model blueprint, a \ac{DCNN}, is fittingly configured to learn from variable-length control flow diagnostics output from a dynamic trace. To rival the size and variability of contemporary data compendiums, hence feeding the model cognition, a methodology for the procedural generation of synthetic cryptographic binaries is defined, utilising core primitives from \emph{OpenSSL} with multivariate obfuscation, to draw a vastly scalable distribution. The library, CryptoKnight, rendered an algorithmic pool of AES, RC4, Blowfish, MD5 and RSA to synthesis combinable variants which are automatically fed in its core model. Converging at 91\% accuracy, CryptoKnight is successfully able to classify the sample algorithms with minimal loss.
\end{abstract}

\begin{IEEEkeywords}
Deep Learning, Convolutional Neural Network, Cryptovirology, Ransomware, Binary Analysis
\end{IEEEkeywords}

\section{Introduction}\label{sec:introduction}


\IEEEPARstart{T}{he} idea of cryptovirology was first introduced by \cite{502676} to describe the offensive nature of cryptography for extortion-based security threats. It comprises a set of revolutionary attacks that combine strong (symmetric and asymmetric) cryptographic techniques with unique viral technology. The fierce proliferation of `crypto-ransomware' is rather troubling for a number of reasons. Designed to infect, encrypt and lock-down available hosts, this category of malware has had disastrous consequences for many~\cite{snow_2016}~\cite{chiu2017}. For those who can afford to reclaim their private data, the financial loss is typically quite substantial, despite the fact that there is no guaranteed recovery. Ultimately, without a backup, there is little that can be done. Preventative frameworks have been proven to effectively halt unusual activity~\cite{kharraz2015cutting}~\cite{scaife2016cryptolock} by closely monitoring the file system's input / output, but administrators are not always likely to follow best practices~\cite{birmingham2017}. In any case, unknown weaknesses can still be exploited to further an attacker's goal.

The cryptovirological landscape has evolved in recent years. A distinct growth has been noted in the overall number of targeted attacks and variants~\cite{intel_security_2016}. A long-term study of 1,359 ransomware samples~\cite{kharraz2015cutting} observed between 2006 and 2014 found a distinct number of variants with cryptographic capabilities. During analysis, these instances were found to utilise both standard and customised cryptography with generational enhancements, specifically in terms of key generation and management. Tailored cryptosystems particularly limit the scale of effective analysis~\cite{lutz2008}. Infamous variants are known to have employed well-established documented algorithms~\cite{lutz2008}~\cite{Gröbert2011}, but a number of deviations are significant. In some instances, malware variants have employed custom `cryptography', lesser-known algorithms (such as the Soviet/Russian symmetric block cipher GOST~\cite{GOSTIBM}), common substitution-ciphers, or exclusively, encoding mechanisms.

The field of malware analysis seeks to determine the potential impact of malicious software by examining it in a controlled environment. Investigators find flaws otherwise unknown to current identification technologies - sourcing keys and blocking further infection~\cite{snow_2016}. When reverse engineering a potentially malicious executable, several issues should be addressed. Possible problems include the accuracy of analysis, quality of the application's obfuscation, and the lifetime of findings. The analysis of a binary can typically be considered from two viewpoints: static or dynamic. Static analysis is performed in a non-runtime environment, therefore examination is relatively safe; however potential morphism restricts the accuracy of results~\cite{Lestringant:2015:AIC:2714576.2714639}~\cite{moser2007limits}. Alternatively, dynamic analysis~\cite{luk2005pin} sequentially assesses a binary throughout its execution, which can provide significantly more accurate results and contend with obfuscatory measures~\cite{XuCrypto2017}~\cite{li2014cipherxray}, but if not properly handled, samples could prove somewhat hazardous. This manuscript focuses on the latter methodology. 

Cryptographic algorithm identification facilitates malware analysis in a number of ways, but in this case, when assessing ransomware strains, it yields a starting point for investigation. This is essential when analytical time is restricted. With the uncertainty surrounding an application's custom, undocumented or established cryptosystem, analysts struggle to maintain complete awareness of the field - which makes this task ideal for automation. To effectively model cryptographic execution, previous research has relied on a number of assumptions and observations. These features do not necessarily always depict cryptographic code, but provide a baseline for analysis. For instance, cryptographic algorithms naturally involve the use of bitwise integer arithmetic and logical operations. These activities frequently reside in loops, for example, block ciphers typically loop over an input buffer to decrypt it block-by-block. \cite{lutz2008} also postulated that any encrypted data is likely to have a higher information entropy than decrypted data.


Deep learning studies intricate \acp{ANN} with multiple hidden computational layers ~\cite{lecun2015deep} that effectively model representations of data with multiple layers of abstraction, in which the high-level representations can amplify aspects of the input that are important for discrimination. These techniques have been used amongst others to identify network threats~\cite{7746067} or encrypted traffic on a network~\cite{Hodo:2017:MLA:3098954.3106068}~\cite{fruehwirt2014using}. A \ac{CNN}, is a specialised architecture of \ac{ANN} that employs a convolution operation in at least one of its layers~\cite{lecun1989generalization}~\cite{DBLP:journals/corr/HodoBHTA17}. A variety of substantiated \ac{CNN} architectures have been used to great effect in computer vision~\cite{lecun2010convolutional} and even \ac{NLP}, with empirically distinguished superiority in semantic matching~\cite{hu2014convolutional}, compared to other models. 


\emph{CryptoKnight} is developed in coordination with this methodology. We introduce a scalable learning system that can easily incorporate new samples through the scalable synthesis of customisable cryptographic algorithms. Its entirely automated core architecture is aimed to minimise human interaction, thus allowing the composition of an effective model. We tested the framework on a number of externally sourced applications utilising non-library linked functionality. Our experimental analysis indicates that CryptoKnight is a flexible solution that can quickly learn from new cryptographic execution patterns to classify unknown software. This manuscript presents the following contributions:

\begin{itemize}
	\item Our unique convolutional neural network architecture fits variable-length data to map an application's time-invariant cryptographic execution.
	\item Complimented by procedural synthesis, we address the issue of this task's disproportionate latent feature space.
	\item The realised framework, CryptoKnight, has demonstrably faster results compared to that of previous methodologies, and is extensively re-trainable.
\end{itemize}

\begin{figure*}
	\includegraphics[width=\textwidth]{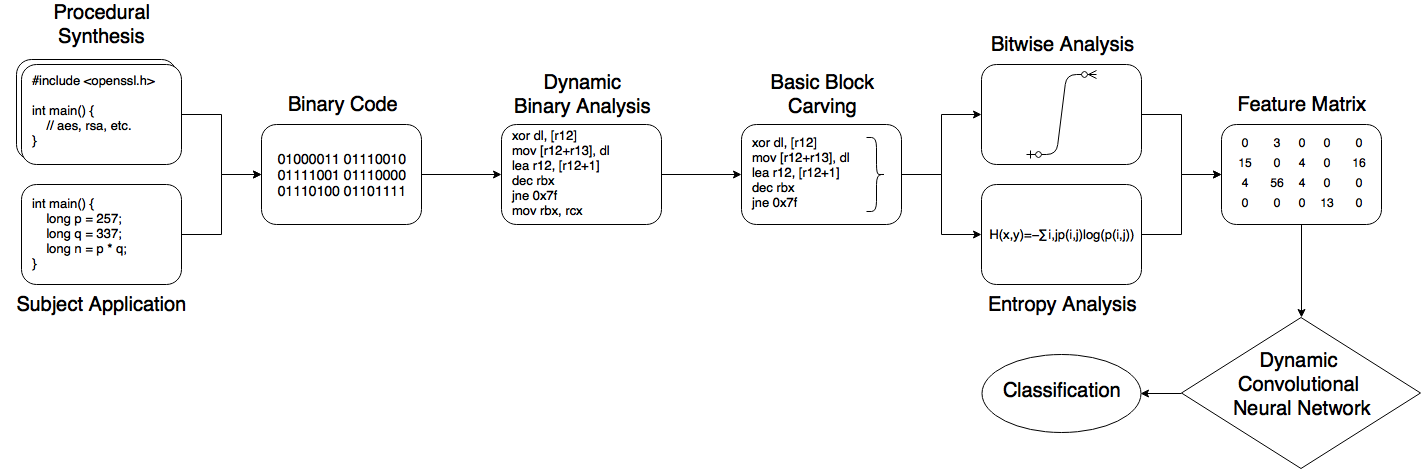}
	\caption{Framework Architecture}
	\label{fig:design}
\end{figure*}

\section{Related Work}

The cryptovirological threat model has rapidly evolved over the last decade. A number of notable individuals and research groups have attempted to address the problem of cryptographic primitive identification. We will discuss the consequences of their findings here and address intrinsic problems.

\subsection{Heuristics}

Heuristical methods~\cite{pearl1984heuristics} are often utilised to locate an optimal strategy for capturing the most appropriate solution. These measures have previously shown great success in cryptographic primitive identification. A joint project from ETH Z\"{u}rich and Google, Inc.~\cite{lutz2008} detailed the automated decryption of encrypted network communication in memory, to identify the location and time a subject binary interacted with decrypted input. From an execution trace which dynamically extracted memory access patterns and control flow data~\cite{lutz2008}, was able to identify the necessary factors required to retrieve the relevant data in a new process. His implementation was successfully able to identify the location of several decrypted webpages in memory fetched using \emph{cURL} and \emph{OpenSSL}, and successfully extracted the decrypted output from a \emph{Kraken} malware binary. The entropy metric was found to negatively affect the recognition of simple substitution ciphers as they do not typically effect the information entropy, which was also found to affect the analysis of \emph{GnuPG}.

Gr\"{o}bert at al.~\cite{felixccc}~\cite{Gröbert2011} utilised fine-grained dynamic binary analysis to generate a high-level control flow graph for evaluation using three heuristics. The chains heuristic measured the ordered concatenation of all mnemonics in a basic block, comparing known signatures; the mnemonic-const heuristic extent the former method by assessing the combination of instructions and constants; and the verifier heuristic confirmed a relationship between the input and output (I/O) of a permutation block. Trialled on \emph{cURL}, the tool detected both \ac{RSA} and \ac{AES} in a traced \ac{SSL} session. When tested on a real-world malware sample,\emph{GpCode}, the operation successfully detected the cryptosystem and extracted its keys. However, the trace took fourteen hours with an extra eight hours for analysis.

The \emph{Crypto Intelligence System}~\cite{matenaarcis} conglomerates a number of these heuristical measures to counter the problem of situational dependencies. As cryptography can be reduced to Rice's Theorem, it is suggested that different heuristics can provide more accurate readings. Evaluating the detection methods used by \cite{lutz2008} and \cite{felixccc}, Matenaar et al. found that \cite{caballero2009dispatcher} presented the least false positives in their tests. However, in presenting these comparisons, the problem with heuristics is shown as in that they do not always generalise suitably.

\subsection{Machine Learning}

Attempting to address a difficulty with past methodologies, one thesis~\cite{DBLP:journals/corr/Hosfelt15} studied the suitability of machine learning. While automated and highly efficient, thresholds~\cite{Gröbert2011}~\cite{Lestringant:2015:AIC:2714576.2714639}~\cite{XuCrypto2017} often require manual adjustment to manage the identification of new algorithmic samples. Hosfelt sought to emphasise the ease of model retraining by analysing the performance of: Support Vector Machine, Kernels, Naive Bayes, Decision Tree and K-means Clustering. The study was met with varying success, but ultimately suffered from a limited sampling of the latent feature space, preventing adequate scaling for more complex data - i.e. multi-purpose applications that may use cryptography in addition to other functions that may unintentionally obfuscate the control flow. \emph{EldeRan}~\cite{sgandurra2016automated} similarly assessed the suitability of automated dynamic analysis for ransomware. A regularized logistic regression classifier was utilised in conjuncture with a dataset of 582 ransomware and 942 `good' applications - thus producing an \ac{AUC} of 0.995. While the results in this case were positive, the methodology required a large number of malicious samples to have been run to generate a sufficient distribution.

\subsection{Obfuscation}

A tool by the name of \emph{CryptoHunt}~\cite{XuCrypto2017} was recently developed to identify cryptographic implementations in binary code despite advanced obfuscation. The implementation tracked the dynamic execution of a reference binary at instruction level, to further identify and transform loop bodies into boolean formulas. Each formula was designed to successfully abstract the particular primitive, but remain compact to describe the most emblematic features. Unlike sole I/O verification, performed by~\cite{Gröbert2011}, this semantic depth more prominently revealed distinguishable features regardless of obfuscation. \emph{Aligot}~\cite{calvet2012aligot} was also designed for obfuscatory resilience, but instead chose to focus on I/O. Both tools performed well on a variety of samples, but required pre-existing reference implementations and manual integration.

\subsection{Data Flow Analysis}

Two papers studied representational patterns of cryptographic data through dynamic analysis~\cite{6878965}~\cite{Zhao2011}. By closely monitoring an application's I/O each methodology aimed to pinpoint a cryptographic algorithm that matched a similar pattern. \cite{li2014cipherxray} alternatively assessed the `avalanche effect' as a unique discriminatory feature, where a small change in the input would dramatically alter the output. Although effective, none of the methods would likely adapt to unique obfuscations. 

A unique and fairly effective approach for the identification of symmetric algorithms in binary code was based on subgraph isomorphism and static analysis \cite{Lestringant:2015:AIC:2714576.2714639}. Lestringant et al. resolved each cryptographic algorithm to a \ac{DFG}, normalising their structure without breaking semantics. The proposed subgraph isomorphism step then assessed which signatures were contained within the normalised \ac{DFG}. For the targeted sample pool (XTEA, \ac{MD5}, \ac{AES}), their implementation reached 100\% accuracy. Unfortunately, the formula relied on the manual selection of appropriate signatures which distinguish the applicable algorithms. For these three instances, generation was elementary, but would not realistically scale to the dimensions sought in this paper.

\section{Overview}

With any given subject application, this methodology aims to automatically verify the existence of pre-determined cryptographic signatures in unknown binary code. The intention is to provide a solution that can easily generalise when presented with new conditions, without manually adjusting a number of thresholds. The full system, as in Figure~\ref{fig:design}, is comprised of three higher-level stages.

\begin{enumerate}
	\item Procedural generation guides the synthesis of unique cryptographic binaries with variable obfuscation and alternate compilation.
	\item Assumptions of cryptographic code aid the discrimination of diagnostics from the dynamic analysis of synthetic or reference binaries, to build an `image' of execution.
	\item A \ac{DCNN} fits variable-length matrices for ease of training and the immediate classification of new samples.
\end{enumerate}

\section{Synthesis}
To construct a reasonably sized dataset with enough variation to satisfy the abstraction of cryptographic primitives, it is not enough to simply hand-write a small number of applications with little diversity in terms of operational outliers. For example, extracting features from the execution of a manually implemented single-purpose binary may give an appropriate feature vector, but re-running the extraction process will not provide any variation for repeating labels, outside of environmental setup. This methodology leverages procedural generation to include elements that provide some obfuscation without directly altering the intended control flow. For the three main algorithmic categories - symmetric, asymmetric and hashing - interpretation should correlate the related components to dynamically construct a unique executable.

\subsection{Artefacts}

\emph{OpenSSL} is an open source cryptographic library that provides an \ac{API} for accessing its algorithmic definitions. Review of its documentation revealed a number of similarities in the intended implementation of the each function. These specificities are either: \emph{variable}, differ for each primitive; or \emph{constant}, true for each category. This approach exclusively examined \emph{C} in experimentation, but \emph{C++} can also be integrated.

Via appropriate headers, an application first imports the libraries which provide the expectant functionality when later compiled. In this case, from either \emph{OpenSSL} or the \emph{C} Standard Library. Each primitive naturally requires contrastive functionality, so this is variable. Within the scope of an application's main body however, each symmetric algorithm requires the specification of a key and~\ac{IV}, asymmetric algorithms require a certificate declaration whereas hashing definitions do not expect either. These rules are categorically constant, therefore, their definitions can be specified by type. Next, a plaintext sequence is loaded into memory - directly or from a file - and ciphertext memory is allocated, also constant. Each sample will then employ its unique algorithm through differing declarations, reading in the plaintext, key or~\ac{IV}.

\subsection{Obfuscation}

Two primary transformation mechanisms were highlighted by \cite{XuCrypto2017}. The first technique discusses the abstraction of relevant data groups to decrease their perceptible mapping. For example, a multidimensional array may be concatenated into a single column and either expanded or accessed as necessary. The second technique concerns itself with the splitting of variables \cite{drape2009intellectual} to disguise their representation. Colloquially known as `data aggregation' and `data splitting', these methods partially obfuscate the data flow without subtracting from a application's distinct activity.

Outside of such distinct obfuscation, the inclusion of structured loops, arithmetical or bitwise operations create discriminatory irregularity in an otherwise translucent process. Analogously, many `training' images for computer vision will contain noise that suitably detract from the trivial classification of its subject. This process aims to replicate such uncertainty.

\subsection{Interpretation}

Formatting each respective variable artefact to allow ease of parsing in a similarly structured markup tree will allow the interpretation of unique cryptographic applications with alternate obfuscation. Stochastically generated keys, \acp{IV} and plaintext will add additional variation into each image. Algorithm~\ref{alg:generate} outlines the pseudo-code for this procedure.

\begin{algorithm}
	\caption{Cryptographic Synthesis}
	\label{alg:generate}
	\begin{algorithmic}[1]
		\renewcommand{\algorithmicrequire}{\textbf{Input:}}
		\renewcommand{\algorithmicensure}{\textbf{Output:}}
		\REQUIRE cryptographic constants \& variables
		\ENSURE  application code
		\\ \textit{New Sample} :
		\STATE select obfuscation - (aggregation, split, normal)
		\STATE write to file:
		\\ \textit{import statements}
		\\ \textit{abstracted keys}
		\\ \textit{encryption routines}
		\STATE inject randomised arithmetic
		\RETURN relative location 
	\end{algorithmic} 
\end{algorithm}

When compiling the resultant collection of cryptographic applications, data variability can be further increased. With alternate compilers or optimisation options, the resultant object code will dramatically fluctuate. Real-world instances are rarely compiled identically so multivariate output can provide further assurance for generalisation.

\section{Feature Extraction}

Cryptographic execution is time-invariant. Therefore, a reference binary may employ its associated functions at any point within a trace. Unintentional obfuscation of the control flow will negatively affect discriminatory performance, so granularity needs to be high. An underlying problem of previous work~\cite{DBLP:journals/corr/Hosfelt15}. This following approach opts to draw appropriate features from a reference binary using dynamic instrumentation via Intel's Pin \ac{API}. Through the disassembly of run-time instruction data, this section's outlined measurements principally assess the activity's importance in relation to the assumptions of cryptographic code.

\subsection{Basic Blocks \& Loops}

A \ac{BBL} is a sequential series of instructions executed in a particular order, exclusively defined when there is one branch in (entry) and one branch out (exit). A \ac{BBL} ends with one of the following conditions.

\begin{itemize}
	\item unconditional or conditional branch - direct / indirect.
	\item return to caller.
\end{itemize}

Each instruction is evaluated in a linear trace, if any criteria are met, it is marked as a \emph{tail}. The following instruction is delimited as the \emph{head} of the subsequent sequence, but can similarly be identified as a \emph{tail}. The `stack' of execution stores relevant data from each instruction, and two boolean expressions indicate the predetermined blocks. As each \ac{BBL} is dynamically revealed, non-executed instructions will unfortunately not be observed~\cite{Gröbert2011}, but we can monitor indirect branches.

Many high level languages share a distinctly strict definition of a loop, contrarily, common interpretations of amorphous code are loose. Extending previous definitions~\cite{650542}~\cite{moseley2006loopprof} we hence delineate a loop upon the immediate re-iteration of any \ac{BBL}, as output from Algorithm~\ref{alg:trace}.

\begin{algorithm}
	\caption{Instruction Sequencing, BBL Detection}
	\label{alg:trace}
	\begin{algorithmic}[1]
		\renewcommand{\algorithmicrequire}{\textbf{Input:}}
		\renewcommand{\algorithmicensure}{\textbf{Output:}}
		\REQUIRE run-time hooks
		\ENSURE  path of execution
		\\ \textit{Callback} :
		\STATE head = false
		\IF {(last instruction is tail)}
		\STATE head = true
		\ENDIF
		\IF {(instruction is branch, call or return)}
		\STATE tail = true
		\ELSE
		\STATE tail = false
		\ENDIF
		\IF {(write)}
		\STATE get entropy (memory write)
		\ENDIF
		\RETURN stack 
	\end{algorithmic} 
\end{algorithm}

\subsection{Instructions}

Conventional architectures use a common instruction format, interpretable as: \textit{opcode operand (destination/source)}. In x86, there are zero to three operands (separated by commas), two of which specify the destination and source. For example, when AES performs a single round of an encryption flow it calls the instruction \textit{66 0f 38 dc d1} which can be disassembled as \textit{aesenc xmm2, xmm1}. Directly operating on the first operand, in this case \textit{xmm2}, it performs a round of AES encryption using the 128-bit round key from \textit{xmm1}. Although this is a interesting example, the \ac{AES-NI} architecture presents a problem for later generalisation as cryptographic acceleration prevents detailed analysis.

Alternate object code is typically quite distinctive, especially in cryptographic code. A primitive may employ a number of operations, in any order, and it is important not to dwell on specificities - i.e. exact semantics. For each instruction in the `carved' linear trace we weight its ratio of bitwise operations upon the cross-correlation of prominent operators from a pool of cryptographic routines for discriminatory emphasis.

\subsection{Entropy}

As characterized by~\cite{renyi1961measures}, the associated uncertainty of a finite discrete probability distribution $p = (p_1, p_2, \ldots, p_n)$ can be measured using Shannon's Entropy. Suppose $p_k > 0(k=1, 2, \ldots, n)$ and $\sum_{k=1}^n p_k = 1$, distribution $p$ is measured by quantity $H[p] = H(p_1, p_2, \ldots, p_n)$ hence defined as:

\begin{equation}\label{eqn:shannon}
H(p_1, p_2, \ldots, p_n)=\sum_{k=1}^{n} p_k log_2 \frac{1}{p_k}
\end{equation}

\begin{figure}
	\begin{center}
		\input{ent-blow.tex}
		\qquad
		\input{ent-rc4.tex}
	\end{center}
	
	\begin{center}
		\input{ent-md5.tex}
		\qquad
		\input{ent-rsa.tex}
	\end{center}
	
	\begin{center}
		\input{ent-aes.tex}
	\end{center}
	\caption{Entropy Scoring}
	\label{fig:ent-scoring}
    \medskip
    \small
	\centering
	The relative entropy scoring (\emph{y}) of each \ac{BBL} (\emph{x}) within a trace from a sample set of algorithms.
\end{figure}
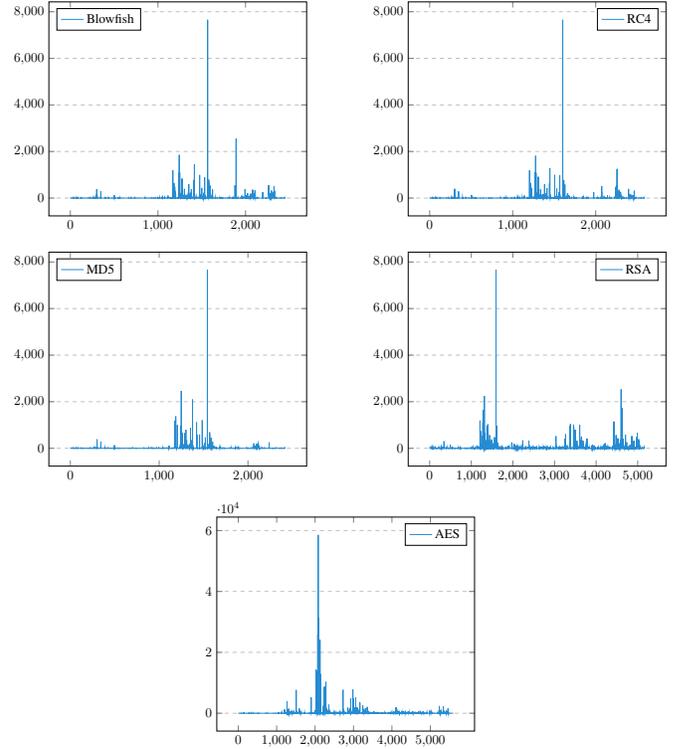

Upon detecting a memory write, the respective location's contents can be replicated. Casting its value to distribution $p$ will allow the immediate calculation of $H[p]$. Related memory can then be deleted to prevent unnecessary exhaustion. Each \acp{BBL} absolute entropy increase / decrease can then be scored by its relation to prior activations over opposing registers and then summated as in Figure~\ref{fig:ent-scoring} where each \ac{BBL} is $\in \mathbb{W}$.

\section{Model}
Founded in earlier research~\cite{DBLP:journals/corr/KalchbrennerGB14}, this proposed definition of \ac{DCNN} treats the input in a manner similar to that of a sentence. For each word, embeddings are defined as $\mathbf{d}$, where $\mathbf{d}_i$ corresponds to the total weight of a particular operation, multiplied by its entropic score. Feature vector $\mathbf{w}_{i} \in \mathbb{W}^{d}$ is a therefore a column in sentence matrix $\mathbf{s}$ such that $\mathbf{s} \in \mathbb{W}^{d \times s}$. Let's say $\mathbf{w}=128$, exclusively assessing pure arithmetic impact would make $\mathbf{d}=12$, so $\mathbf{s}$ would equal $12 \times 128$. The model itself combines a number of one wide convolutions with (dynamic) k-max pooling and folding to map variable-length input. Topologically presented in Figure~\ref{fig:model}, the model combines a number of one wide convolutions with (dynamic) k-max pooling and folding to map variable-length input. 

Mathematically, convolution is an operation on two functions of a real-valued argument, in this case a vector of weights $\mathbf{m} \in \mathbb{W}^{m}$ and a vector of inputs $\mathbf{s} \in \mathbb{W}^{s}$, to yield a new sequence $\mathbf{c}$. With \textit{kernel} $\mathbf{m}$, the convolved sequence in a one-wide convolution takes the form $\mathbf{c} \in \R^{s+m-1}$, thus preserving the number of defined embeddings.

Equation~\ref{eqn:pooling} describes the selection of $k$, a distinct subset of $\mathbf{s}$ that most relevantly depicts an $l$-th order’s progression. Based on the total number of convolutional layers $L$, the current layer $l$, the projected sentence length $s$ and the predefined final pooling parameter $k_{top}$, any particular layer's selection is delimited as:

\begin{equation}\label{eqn:pooling}
k_l=max(k_{top},[\frac{L-l}{L}s])
\end{equation}

With value $k$ and sequence $\mathbf{p} \in \R^{p}$ of length $p \geq k$ the base \textit{k-max pooling} operation selects the subsequence $\mathbf{p}_{max}^k$ of the $k$ most active features. Completely unreactive to positional variation, it not only preserves the original perspective, but can also distinguish repetitious features. Since identifiable cryptographic routines may execute at any point in a sequential trace, this operation fits perfectly.

Another significant phase in this procedure simplifies the way the model perceives complex dependencies over rows. Veritable feature independence is removed through component-wise summation of every two rows, shrinking $d/2$. Nested between a convolutional layer and (dynamic) k-max pooling, a folding layer halves the respective matrix.

While the model itself can automatically scale with new variants, a number of hyper-parameters may need to be adjusted to better suit the sample pool. Manual tuning is an inefficient, costly process that frequently offers no advantage. \cite{bergstra2012random} empirically show that randomly chosen trials are more efficient for hyper-parameter optimization than manual or grid search over the same domain - calculating the most viable constants in a fraction of the time. Taking the shuffled Cartesian product of all hyper-parameter subsets allows the ambiguated selection of distinct constants for trial on a small number of epochs.

\input{cnn_arch}

\section{Experimentation}

Table~\ref{fig:algorithms} presents a range of popular algorithms which were selected for experimental analysis. These primitives are widely utilised for both legitimate and fraudulent purposes.

CryptoKnight was configured to build two feature maps from its first convolution, with a dropout probability of $0.6$, through a round of \textit{k}-max pooling and non-linear activation. Eleven hidden one-dimensional wide convolutions interspersed with more tanh non-linear activations and \textit{k}-max pooling further reduced the feature space. Due to the number of embeddings the model utilised two folding operations simultaneously prior to the penultimate pooling layer, after its convolution. The final convolution built two additional feature maps and then pooled on the stipulated topmost magnitude of~56. A linear transformation, of output size~8, was then applied with softmax to fully connect the model. Filter widths were specified for the first and last convolutional layers of~20~and~12 respectfully, the remaining hidden layers shared a width of~10. In total, the model employed 14 convolutions.

\begin{center}
	\begin{tabular}{ r l }
		Category 	& Algorithm \\ \toprule
		Symmetric 	& AES \\  
		& RC4 \\  
		& Blowfish \\  
		Asymmetric 	& RSA  \\
		Hashing 	& MD5 \\
	\end{tabular}
	\captionof{table}{Cryptographic Algorithms}
	\label{fig:algorithms}
\end{center}

Based on a frequency analysis of simplified opcodes within the sample pool, the mnemonics \emph{add}, \emph{sub}, \emph{inc}, \emph{dec}, \emph{shr}, \emph{shl}, \emph{and}, \emph{or}, \emph{xor}, \emph{pxor}, \emph{test} \& \emph{lea} were selected for weighting. Therefore the final design matrix, of embedding~$12$, contained a variable number of vectors corresponding to the numeration and associated weightings of each basic block in a subject binary. A distribution of size~$n = 750$ was drawn in which~75\% was used for training and~25\% remained for testing. After 200 epochs, the model successfully converged at~91.3\% accuracy with a minimal loss of $0.29$. Figure~\ref{fig:accuracy} shows the test accuracy over 200 epochs, and Figure~\ref{fig:loss} displays its simultaneous loss. An additional collection was then drawn for validation. Table~\ref{tbl:confusion} diagnoses the pre-trained model's associated confusion on an additional $n = 300$ samples.

\input{accuracy.tex}
\input{loss.tex}
\input{accuracy-noentropy.tex}

We collected five open source implementations, presented in Table~\ref{tbl:github}, from GitHub~\footnote{https://github.com} for~\ac{RC4} and Blowfish. The resultant collection of binaries leveraged pure (non-library linked) cryptographic functionality to assess our method. 

\newpage
\input{confusion-validation.tex}

No representational~\ac{RSA} or~\ac{MD5} instances were identified at time of testing. \ac{AES} was overlooked due to the model having been trained on cryptographically accelerated binaries. The classification rate varied with hyper-parameter optimisations and distribution sizes, but was able to correctly classify~$4/5$~RC4 and~$4/5$~Blowfish samples in our tests.

\begin{center}
	\begin{tabular}{ r l }
		Algorithm & Source \\ \toprule
		RC4 		& github.com/ogay/rc4 \\  
					& github.com/tomaslu/rc4 \\  
					& github.com/Maihj/RC4 \\ 
					& github.com/anthonywei/rc4/tree/master/c \\  
					& github.com/shiffthq/rc4 \\ 
		Blowfish 	& github.com/jdiez17/blowfish \\  
					& github.com/Rupan/blowfish \\  
					& github.com/rkhullar/cipher-c \\ 
					& github.com/genterist/simpleBlowFishX \\  
					& github.com/JaseP88/Blowfish-Algorithm \\ 
	\end{tabular}
	\captionof{table}{Open Source Implementations}
	\label{tbl:github}
\end{center}

Analysing \textit{GnuPG 1.4.20}, the software was directed to encrypt an empty text document using 256-bit~AES. With a trace time of~1 minute~40~seconds, CryptoKnight predicted the utilisation of \ac{RSA} with \ac{AES}.

\section{Discussion}\label{discussion}

Traditional cryptographic identification techniques are inherently expensive~\cite{lutz2008}~\cite{Gröbert2011}~\cite{Lestringant:2015:AIC:2714576.2714639}~\cite{XuCrypto2017} and heavily rely on human intuition. CryptoKnight was built to reduce this associated error-prone interaction. With refined sampling of the latent feature space, a procedurally synthesised distribution allowed our \ac{DCNN} to map proportional linear sequences with a finer granularity than that of conventional architectures without overfitting, CryptoKnight converged at~91\% accuracy without extensive hyper-parameter optimisation. The model ultimately fit the synthetic distribution with veritable ease, with performance on par to that of \cite{DBLP:journals/corr/KalchbrennerGB14}.

The impediment of dynamic binary instrumentation was made clear by \cite{Gröbert2011} who highlighted an extensive twenty two hour trace and analysis time. CryptoKnight's analysis time also varied, but not quite to this extent. Hence, for the sample binaries, analysis took up to a maximum of around one minute. Consequently, large collections saw exponential draw times of indefinable length. Since manual analysis often takes invariably longer than an adequate draw time, this footprint is arguably marginal. Once trained, the proposed framework would be most beneficial as part of an analyst's toolkit - to quickly verify any cryptographic instances.

The \ac{DCNN} was intended to map time-invariant cryptographic execution despite control or data flow obfuscation. An intrinsic problem of initial work and the approach by used \cite{DBLP:journals/corr/Hosfelt15}. While successful, this formulation still has a few preliminary issues to address. The predefined operator embeddings explicitly define the entire feature set, therefore new samples which perhaps deviate from traditional operation may be unidentifiable. Should the framework not immediately generalise to an additional algorithm, correlating its most predominant operators by interchanging embeddings or enlarging the scope should enhance cognition.

A fundamental part of CryptoKnight's supervised design is that it can only classify known samples. New cryptographic algorithms must be added to the generation pool - a process this framework has strived to simplify, but an unavoidable limitation of the proposed architecture. This also makes custom cryptography more difficult to classify, as no high-level reference implementations would feasibly exist to import. Integrating an unsupervised component into the core model could facilitate the detection of non-cryptographic signatures. This, or alternatively advanced synthesis could negate the need for procedural generation entirely, to further reduce the presently expensive time requirement and/or aid in the classification of customised cryptography. A multi-class element that learns application invariant primitives would also be prove beneficial as CryptoKnight was exclusively trained on pre-combined functions. A similar model could be re-purposed to decompile binaries with more accuracy than traditional methods - which typically only manage simplistic control flow. 

The entropy metric assumes that a cryptographic function's associated uncertainty is higher than that of conventional interaction. In the case of \cite{lutz2008}, this negatively affected the recognition of simple substitution ciphers, however it is unlikely to affect CryptoKnight in the same way due to its scoring mechanics and demonstrably high accuracy in subjective tests without the metric. Figure~\ref{fig:accuracy-ent} proves that the entropy metric does not impact classification rate, converging at 83.3\% - \~ 8\% difference.

Problems with cryptographic acceleration played an important role in the detection of native \ac{AES} implementations. Intel's \ac{AES-NI} extension was proposed in 2008 for boosting the relative speed of encryption and decryption on microprocessors from Intel and AMD. \cite{akdemir2010breakthrough} describe the instruction set with regard to its breakthrough performance increase. The six instructions, prefixed by \ac{AES}, directly perform each of the cipher's operations on the \ac{SSE} XMM registers, however, the natural progression of the cipher could not be fully observed.

\section{Conclusion}
Despite advanced countermeasures, the cryptovirological threat has significantly increased over the last decade, incentivising the aforementioned research. Our research demonstrated that cryptographic primitive classification in compiled binary executables could be achieved successfully using a dynamic convolutional neural network. We also demonstrated our implementation and achieved ~91\% accuracy without extensive hyper-parameter optimisation. Moreover, our implementation is fundamentally more flexible than that of previous work, marginalising the error prone human element. The framework successfully detected every implementation of RC4 and Blowfish (including externally sourced, native (non-library) written compositions) and maintained a distinctively high accuracy on synthetic implementations. Future work includes the detection of cryptographic functions in parallel with the execution of a subject binary.


%



\ifCLASSOPTIONcaptionsoff
  \newpage
\fi

\bibliographystyle{IEEEtran}
\bibliography{bibliography}

\vfill

\begin{acronym}
	\acro{AES}{Advanced Encryption Standard}
	\acro{AES-NI}{Advanced Encryption Standard New Instructions}
	\acro{ANN}{Artificial Neural Network}
	\acro{API}{Application Programming Interface}
	\acro{AUC}{Area Under Curve}
	\acro{BBL}{Basic Block}
	\acro{CBC}{Cipher Block Chaining}
	\acro{CPU}{Central Processing Unit}
	\acro{CNN}{Convolutional Neural Network}
	\acro{DCNN}{Dynamic Convolutional Neural Network}
	\acro{DFG}{Data Flow Graph}
	\acro{ELF}{Executable and Linkable Format}
	\acro{GPU}{Graphics Processing Unit}
	\acro{IV}{Initialization Vector}
	\acro{JIT}{Just-In-Time}
	\acro{MD5}{Message Digest 5}
	\acro{NLP}{Natural Language Processing}
	\acro{PE}{Portable Executable}
	\acro{RC4}{Rivest Cipher 4}
	\acro{RNN}{Recurrent Neural Network}
	\acro{RSA}{Rivest-Shamir-Adleman}
	\acro{SPN}{Substitution-Permutation Network}
	\acro{SSE}{Streaming SIMD Extensions}
	\acro{SSL}{Secure Sockets Layer}
\end{acronym}

\end{document}

%% file: ent-blow.tex
\begin{tikzpicture}[scale=0.5]

\begin{axis}[
    xtick={0,1000,2000,3000,4000,5000},
    legend pos=north west,
    ymajorgrids=true,
    grid style=dashed,
]

\addplot[color=blue]
coordinates {
(0,0)(1,0)(2,1)(3,9)(4,0)(5,0)(6,0)(7,0)(8,0)(9,0)(10,0)(11,2)(12,0)(13,0)(14,2)(15,2)(16,0)(17,0)(18,0)(19,0)(20,0)(21,0)(22,0)(23,0)(24,0)(25,0)(26,9)(27,0)(28,0)(29,0)(30,0)(31,0)(32,0)(33,0)(34,4)(35,0)(36,0)(37,21)(38,13)(39,4)(40,1)(41,0)(42,0)(43,0)(44,0)(45,1)(46,4)(47,0)(48,0)(49,3)(50,0)(51,0)(52,0)(53,0)(54,2)(55,3)(56,0)(57,1)(58,0)(59,0)(60,0)(61,2)(62,1)(63,0)(64,0)(65,16)(66,0)(67,0)(68,0)(69,0)(70,0)(71,0)(72,8)(73,0)(74,2)(75,4)(76,0)(77,0)(78,0)(79,1)(80,1)(81,3)(82,2)(83,15)(84,3)(85,0)(86,0)(87,7)(88,0)(89,3)(90,2)(91,0)(92,2)(93,0)(94,0)(95,0)(96,1)(97,0)(98,0)(99,1)(100,0)(101,0)(102,0)(103,0)(104,0)(105,0)(106,0)(107,0)(108,0)(109,0)(110,0)(111,0)(112,0)(113,5)(114,0)(115,0)(116,0)(117,0)(118,7)(119,0)(120,0)(121,0)(122,0)(123,0)(124,0)(125,0)(126,0)(127,10)(128,0)(129,1)(130,0)(131,0)(132,0)(133,0)(134,0)(135,0)(136,0)(137,0)(138,0)(139,0)(140,0)(141,3)(142,0)(143,0)(144,0)(145,0)(146,0)(147,0)(148,0)(149,0)(150,0)(151,1)(152,0)(153,0)(154,0)(155,0)(156,0)(157,0)(158,0)(159,2)(160,1)(161,0)(162,0)(163,0)(164,0)(165,0)(166,0)(167,0)(168,1)(169,0)(170,3)(171,0)(172,0)(173,2)(174,0)(175,0)(176,0)(177,0)(178,0)(179,0)(180,2)(181,0)(182,1)(183,0)(184,1)(185,0)(186,0)(187,2)(188,2)(189,0)(190,2)(191,2)(192,0)(193,0)(194,0)(195,2)(196,5)(197,0)(198,0)(199,0)(200,0)(201,5)(202,8)(203,0)(204,0)(205,0)(206,0)(207,0)(208,0)(209,0)(210,0)(211,0)(212,0)(213,0)(214,0)(215,0)(216,0)(217,0)(218,0)(219,14)(220,0)(221,0)(222,1)(223,0)(224,0)(225,0)(226,0)(227,0)(228,0)(229,0)(230,1)(231,1)(232,1)(233,0)(234,1)(235,6)(236,0)(237,6)(238,3)(239,0)(240,1)(241,1)(242,0)(243,0)(244,0)(245,0)(246,0)(247,0)(248,0)(249,0)(250,3)(251,2)(252,0)(253,0)(254,2)(255,0)(256,0)(257,0)(258,0)(259,0)(260,3)(261,23)(262,1)(263,2)(264,0)(265,10)(266,1)(267,0)(268,0)(269,1)(270,1)(271,3)(272,0)(273,0)(274,2)(275,2)(276,18)(277,1)(278,0)(279,0)(280,1)(281,0)(282,10)(283,1)(284,0)(285,5)(286,0)(287,0)(288,12)(289,15)(290,0)(291,0)(292,0)(293,4)(294,0)(295,0)(296,29)(297,0)(298,0)(299,2)(300,0)(301,384)(302,0)(303,0)(304,0)(305,0)(306,0)(307,0)(308,0)(309,0)(310,2)(311,24)(312,0)(313,0)(314,0)(315,0)(316,0)(317,0)(318,0)(319,0)(320,0)(321,0)(322,0)(323,1)(324,0)(325,0)(326,2)(327,0)(328,0)(329,2)(330,0)(331,0)(332,1)(333,3)(334,0)(335,0)(336,0)(337,0)(338,0)(339,0)(340,1)(341,0)(342,5)(343,3)(344,0)(345,0)(346,29)(347,0)(348,301)(349,0)(350,0)(351,0)(352,1)(353,2)(354,1)(355,1)(356,0)(357,0)(358,0)(359,0)(360,3)(361,0)(362,0)(363,3)(364,0)(365,0)(366,0)(367,0)(368,0)(369,0)(370,0)(371,0)(372,0)(373,0)(374,1)(375,0)(376,0)(377,0)(378,0)(379,0)(380,0)(381,0)(382,0)(383,0)(384,0)(385,0)(386,0)(387,0)(388,0)(389,0)(390,0)(391,7)(392,0)(393,0)(394,1)(395,0)(396,51)(397,0)(398,0)(399,0)(400,0)(401,8)(402,0)(403,3)(404,0)(405,0)(406,0)(407,0)(408,0)(409,0)(410,0)(411,0)(412,0)(413,0)(414,0)(415,0)(416,0)(417,0)(418,0)(419,0)(420,0)(421,0)(422,0)(423,0)(424,0)(425,0)(426,0)(427,0)(428,0)(429,0)(430,0)(431,0)(432,0)(433,15)(434,0)(435,0)(436,0)(437,0)(438,0)(439,0)(440,0)(441,1)(442,4)(443,0)(444,5)(445,0)(446,0)(447,0)(448,3)(449,4)(450,0)(451,0)(452,1)(453,0)(454,0)(455,0)(456,1)(457,0)(458,0)(459,0)(460,0)(461,0)(462,3)(463,0)(464,0)(465,0)(466,0)(467,1)(468,2)(469,2)(470,3)(471,0)(472,0)(473,0)(474,0)(475,4)(476,0)(477,0)(478,0)(479,1)(480,1)(481,0)(482,0)(483,0)(484,0)(485,0)(486,0)(487,2)(488,0)(489,2)(490,0)(491,0)(492,0)(493,0)(494,0)(495,0)(496,0)(497,0)(498,2)(499,14)(500,1)(501,133)(502,0)(503,6)(504,6)(505,0)(506,0)(507,0)(508,0)(509,9)(510,0)(511,0)(512,0)(513,0)(514,0)(515,0)(516,10)(517,12)(518,0)(519,0)(520,0)(521,1)(522,1)(523,1)(524,1)(525,1)(526,1)(527,1)(528,0)(529,0)(530,0)(531,0)(532,0)(533,0)(534,0)(535,0)(536,0)(537,0)(538,0)(539,2)(540,0)(541,0)(542,1)(543,0)(544,0)(545,8)(546,0)(547,5)(548,0)(549,0)(550,2)(551,5)(552,11)(553,0)(554,0)(555,0)(556,3)(557,0)(558,0)(559,5)(560,1)(561,0)(562,0)(563,0)(564,0)(565,0)(566,17)(567,0)(568,0)(569,10)(570,0)(571,0)(572,0)(573,0)(574,0)(575,0)(576,0)(577,0)(578,0)(579,0)(580,0)(581,0)(582,0)(583,0)(584,1)(585,0)(586,0)(587,0)(588,0)(589,0)(590,0)(591,0)(592,0)(593,0)(594,0)(595,0)(596,0)(597,1)(598,0)(599,0)(600,0)(601,0)(602,0)(603,0)(604,0)(605,0)(606,0)(607,0)(608,0)(609,0)(610,0)(611,0)(612,0)(613,2)(614,0)(615,0)(616,0)(617,0)(618,0)(619,0)(620,0)(621,0)(622,0)(623,0)(624,0)(625,0)(626,0)(627,0)(628,0)(629,0)(630,0)(631,0)(632,0)(633,0)(634,0)(635,0)(636,0)(637,2)(638,3)(639,2)(640,0)(641,0)(642,0)(643,0)(644,1)(645,0)(646,1)(647,2)(648,0)(649,0)(650,0)(651,0)(652,0)(653,0)(654,2)(655,0)(656,0)(657,0)(658,1)(659,0)(660,0)(661,0)(662,0)(663,1)(664,12)(665,0)(666,0)(667,0)(668,1)(669,2)(670,1)(671,0)(672,0)(673,0)(674,0)(675,0)(676,0)(677,0)(678,0)(679,0)(680,0)(681,0)(682,0)(683,0)(684,1)(685,1)(686,0)(687,1)(688,0)(689,11)(690,0)(691,0)(692,0)(693,5)(694,0)(695,10)(696,0)(697,0)(698,0)(699,0)(700,0)(701,0)(702,0)(703,0)(704,0)(705,0)(706,0)(707,0)(708,0)(709,0)(710,0)(711,4)(712,0)(713,0)(714,0)(715,0)(716,0)(717,0)(718,0)(719,0)(720,2)(721,0)(722,3)(723,0)(724,0)(725,0)(726,0)(727,2)(728,0)(729,0)(730,2)(731,2)(732,0)(733,0)(734,0)(735,0)(736,0)(737,0)(738,14)(739,0)(740,2)(741,3)(742,0)(743,0)(744,2)(745,0)(746,0)(747,0)(748,0)(749,0)(750,2)(751,0)(752,0)(753,0)(754,2)(755,0)(756,0)(757,0)(758,0)(759,0)(760,0)(761,0)(762,0)(763,0)(764,0)(765,0)(766,0)(767,2)(768,0)(769,0)(770,0)(771,0)(772,0)(773,0)(774,0)(775,0)(776,0)(777,0)(778,0)(779,0)(780,0)(781,0)(782,1)(783,0)(784,0)(785,0)(786,0)(787,1)(788,0)(789,0)(790,0)(791,0)(792,0)(793,4)(794,0)(795,0)(796,0)(797,0)(798,0)(799,0)(800,0)(801,0)(802,0)(803,0)(804,0)(805,0)(806,0)(807,0)(808,0)(809,1)(810,0)(811,0)(812,0)(813,2)(814,0)(815,0)(816,0)(817,4)(818,0)(819,0)(820,0)(821,0)(822,0)(823,0)(824,0)(825,0)(826,0)(827,0)(828,0)(829,0)(830,0)(831,0)(832,0)(833,0)(834,0)(835,1)(836,0)(837,0)(838,0)(839,0)(840,0)(841,0)(842,0)(843,0)(844,0)(845,0)(846,0)(847,3)(848,1)(849,0)(850,0)(851,0)(852,0)(853,0)(854,18)(855,0)(856,0)(857,0)(858,0)(859,0)(860,0)(861,0)(862,0)(863,0)(864,0)(865,0)(866,0)(867,0)(868,0)(869,0)(870,0)(871,0)(872,0)(873,2)(874,0)(875,0)(876,0)(877,0)(878,0)(879,0)(880,0)(881,0)(882,0)(883,2)(884,0)(885,0)(886,0)(887,0)(888,0)(889,0)(890,0)(891,0)(892,0)(893,0)(894,0)(895,0)(896,2)(897,0)(898,0)(899,5)(900,0)(901,0)(902,0)(903,0)(904,0)(905,0)(906,0)(907,1)(908,3)(909,0)(910,0)(911,2)(912,0)(913,0)(914,2)(915,0)(916,0)(917,0)(918,0)(919,0)(920,6)(921,2)(922,2)(923,2)(924,0)(925,2)(926,4)(927,1)(928,2)(929,0)(930,0)(931,0)(932,2)(933,11)(934,0)(935,0)(936,0)(937,0)(938,0)(939,0)(940,0)(941,0)(942,2)(943,0)(944,0)(945,0)(946,0)(947,1)(948,0)(949,1)(950,0)(951,7)(952,0)(953,0)(954,0)(955,0)(956,0)(957,2)(958,0)(959,1)(960,0)(961,0)(962,0)(963,0)(964,3)(965,0)(966,0)(967,0)(968,0)(969,4)(970,0)(971,0)(972,11)(973,1)(974,0)(975,6)(976,0)(977,0)(978,0)(979,0)(980,0)(981,1)(982,1)(983,0)(984,0)(985,0)(986,0)(987,2)(988,4)(989,0)(990,0)(991,0)(992,0)(993,0)(994,0)(995,0)(996,0)(997,0)(998,14)(999,0)(1000,0)(1001,9)(1002,0)(1003,0)(1004,0)(1005,1)(1006,0)(1007,0)(1008,0)(1009,7)(1010,0)(1011,0)(1012,0)(1013,0)(1014,0)(1015,0)(1016,0)(1017,0)(1018,0)(1019,26)(1020,0)(1021,0)(1022,0)(1023,0)(1024,0)(1025,0)(1026,0)(1027,0)(1028,1)(1029,0)(1030,0)(1031,0)(1032,0)(1033,5)(1034,0)(1035,0)(1036,0)(1037,0)(1038,0)(1039,0)(1040,0)(1041,0)(1042,0)(1043,0)(1044,0)(1045,1)(1046,4)(1047,0)(1048,25)(1049,0)(1050,0)(1051,1)(1052,0)(1053,0)(1054,18)(1055,0)(1056,12)(1057,0)(1058,0)(1059,0)(1060,0)(1061,1)(1062,0)(1063,0)(1064,0)(1065,0)(1066,2)(1067,1)(1068,0)(1069,0)(1070,0)(1071,0)(1072,12)(1073,0)(1074,0)(1075,0)(1076,5)(1077,23)(1078,2)(1079,0)(1080,0)(1081,0)(1082,0)(1083,0)(1084,1)(1085,0)(1086,0)(1087,0)(1088,0)(1089,0)(1090,0)(1091,0)(1092,16)(1093,16)(1094,0)(1095,0)(1096,0)(1097,0)(1098,0)(1099,0)(1100,0)(1101,0)(1102,0)(1103,3)(1104,0)(1105,0)(1106,0)(1107,0)(1108,0)(1109,12)(1110,6)(1111,6)(1112,0)(1113,0)(1114,0)(1115,108)(1116,0)(1117,0)(1118,0)(1119,0)(1120,0)(1121,0)(1122,0)(1123,0)(1124,0)(1125,0)(1126,0)(1127,0)(1128,3)(1129,0)(1130,0)(1131,0)(1132,0)(1133,5)(1134,0)(1135,0)(1136,0)(1137,0)(1138,0)(1139,5)(1140,0)(1141,2)(1142,0)(1143,0)(1144,3)(1145,0)(1146,0)(1147,0)(1148,0)(1149,8)(1150,0)(1151,0)(1152,0)(1153,0)(1154,0)(1155,0)(1156,0)(1157,0)(1158,7)(1159,1)(1160,0)(1161,0)(1162,0)(1163,11)(1164,0)(1165,0)(1166,0)(1167,0)(1168,0)(1169,0)(1170,1188)(1171,0)(1172,0)(1173,0)(1174,2)(1175,1)(1176,0)(1177,0)(1178,0)(1179,0)(1180,0)(1181,0)(1182,0)(1183,646)(1184,186)(1185,3)(1186,0)(1187,0)(1188,24)(1189,21)(1190,67)(1191,13)(1192,492)(1193,3)(1194,0)(1195,0)(1196,0)(1197,0)(1198,0)(1199,3)(1200,0)(1201,188)(1202,171)(1203,0)(1204,0)(1205,0)(1206,0)(1207,0)(1208,0)(1209,0)(1210,0)(1211,0)(1212,0)(1213,0)(1214,0)(1215,5)(1216,0)(1217,0)(1218,0)(1219,0)(1220,0)(1221,2)(1222,0)(1223,1)(1224,39)(1225,0)(1226,0)(1227,0)(1228,0)(1229,0)(1230,13)(1231,0)(1232,0)(1233,12)(1234,0)(1235,0)(1236,175)(1237,0)(1238,47)(1239,1110)(1240,1)(1241,174)(1242,782)(1243,1855)(1244,21)(1245,2)(1246,4)(1247,254)(1248,0)(1249,1086)(1250,344)(1251,36)(1252,7)(1253,163)(1254,206)(1255,0)(1256,0)(1257,0)(1258,0)(1259,0)(1260,34)(1261,4)(1262,10)(1263,0)(1264,0)(1265,112)(1266,1)(1267,7)(1268,7)(1269,7)(1270,0)(1271,0)(1272,369)(1273,0)(1274,46)(1275,842)(1276,28)(1277,189)(1278,0)(1279,55)(1280,1)(1281,56)(1282,0)(1283,0)(1284,0)(1285,15)(1286,15)(1287,0)(1288,3)(1289,89)(1290,0)(1291,2)(1292,5)(1293,0)(1294,28)(1295,0)(1296,91)(1297,0)(1298,0)(1299,5)(1300,67)(1301,4)(1302,394)(1303,0)(1304,51)(1305,0)(1306,69)(1307,0)(1308,0)(1309,46)(1310,6)(1311,0)(1312,28)(1313,0)(1314,0)(1315,0)(1316,23)(1317,0)(1318,0)(1319,0)(1320,0)(1321,0)(1322,2)(1323,0)(1324,0)(1325,20)(1326,41)(1327,54)(1328,99)(1329,93)(1330,42)(1331,1)(1332,19)(1333,1)(1334,24)(1335,38)(1336,86)(1337,60)(1338,1)(1339,0)(1340,0)(1341,76)(1342,0)(1343,58)(1344,0)(1345,0)(1346,0)(1347,0)(1348,0)(1349,0)(1350,0)(1351,2)(1352,596)(1353,68)(1354,19)(1355,164)(1356,0)(1357,2)(1358,2)(1359,0)(1360,0)(1361,0)(1362,0)(1363,70)(1364,51)(1365,0)(1366,3)(1367,0)(1368,0)(1369,1)(1370,3)(1371,1)(1372,15)(1373,225)(1374,45)(1375,9)(1376,21)(1377,0)(1378,3)(1379,49)(1380,16)(1381,387)(1382,96)(1383,1)(1384,0)(1385,0)(1386,3)(1387,1)(1388,69)(1389,0)(1390,0)(1391,0)(1392,0)(1393,0)(1394,0)(1395,4)(1396,0)(1397,0)(1398,0)(1399,0)(1400,0)(1401,0)(1402,0)(1403,0)(1404,0)(1405,0)(1406,0)(1407,0)(1408,2)(1409,0)(1410,15)(1411,779)(1412,583)(1413,51)(1414,0)(1415,11)(1416,1446)(1417,8)(1418,14)(1419,0)(1420,28)(1421,0)(1422,0)(1423,0)(1424,29)(1425,0)(1426,0)(1427,0)(1428,0)(1429,0)(1430,0)(1431,0)(1432,0)(1433,0)(1434,0)(1435,2)(1436,0)(1437,2)(1438,0)(1439,7)(1440,0)(1441,11)(1442,42)(1443,7)(1444,0)(1445,4)(1446,15)(1447,0)(1448,0)(1449,0)(1450,0)(1451,0)(1452,0)(1453,0)(1454,0)(1455,0)(1456,0)(1457,0)(1458,0)(1459,0)(1460,0)(1461,3)(1462,4)(1463,0)(1464,0)(1465,0)(1466,6)(1467,19)(1468,0)(1469,0)(1470,3)(1471,0)(1472,0)(1473,0)(1474,5)(1475,38)(1476,992)(1477,15)(1478,174)(1479,0)(1480,0)(1481,20)(1482,5)(1483,0)(1484,6)(1485,0)(1486,10)(1487,0)(1488,17)(1489,0)(1490,0)(1491,0)(1492,0)(1493,0)(1494,0)(1495,0)(1496,0)(1497,0)(1498,0)(1499,0)(1500,436)(1501,0)(1502,0)(1503,0)(1504,0)(1505,0)(1506,0)(1507,0)(1508,0)(1509,0)(1510,0)(1511,7)(1512,20)(1513,0)(1514,89)(1515,0)(1516,8)(1517,26)(1518,69)(1519,0)(1520,0)(1521,0)(1522,224)(1523,16)(1524,31)(1525,0)(1526,0)(1527,0)(1528,4)(1529,0)(1530,0)(1531,13)(1532,5)(1533,889)(1534,0)(1535,0)(1536,0)(1537,0)(1538,0)(1539,23)(1540,0)(1541,0)(1542,0)(1543,99)(1544,0)(1545,9)(1546,0)(1547,0)(1548,0)(1549,0)(1550,0)(1551,0)(1552,0)(1553,0)(1554,0)(1555,3)(1556,0)(1557,0)(1558,0)(1559,0)(1560,17)(1561,0)(1562,1)(1563,0)(1564,1)(1565,0)(1566,0)(1567,7657)(1568,1)(1569,0)(1570,0)(1571,0)(1572,14)(1573,0)(1574,51)(1575,0)(1576,0)(1577,686)(1578,0)(1579,0)(1580,785)(1581,601)(1582,421)(1583,586)(1584,392)(1585,52)(1586,2)(1587,0)(1588,0)(1589,0)(1590,0)(1591,124)(1592,536)(1593,537)(1594,517)(1595,0)(1596,2)(1597,0)(1598,0)(1599,0)(1600,0)(1601,0)(1602,0)(1603,0)(1604,0)(1605,0)(1606,0)(1607,0)(1608,0)(1609,0)(1610,21)(1611,0)(1612,1)(1613,32)(1614,0)(1615,0)(1616,0)(1617,0)(1618,0)(1619,0)(1620,0)(1621,2)(1622,0)(1623,370)(1624,0)(1625,31)(1626,0)(1627,0)(1628,0)(1629,0)(1630,0)(1631,0)(1632,0)(1633,0)(1634,0)(1635,0)(1636,0)(1637,0)(1638,0)(1639,0)(1640,0)(1641,0)(1642,5)(1643,1)(1644,0)(1645,0)(1646,0)(1647,1)(1648,0)(1649,0)(1650,0)(1651,0)(1652,0)(1653,1)(1654,33)(1655,0)(1656,0)(1657,38)(1658,0)(1659,0)(1660,3)(1661,0)(1662,0)(1663,0)(1664,0)(1665,0)(1666,0)(1667,0)(1668,0)(1669,0)(1670,4)(1671,0)(1672,0)(1673,0)(1674,0)(1675,0)(1676,0)(1677,8)(1678,0)(1679,0)(1680,1)(1681,0)(1682,0)(1683,0)(1684,0)(1685,0)(1686,0)(1687,0)(1688,0)(1689,0)(1690,0)(1691,0)(1692,2)(1693,0)(1694,0)(1695,0)(1696,0)(1697,0)(1698,0)(1699,5)(1700,21)(1701,7)(1702,1)(1703,0)(1704,0)(1705,0)(1706,4)(1707,3)(1708,0)(1709,0)(1710,0)(1711,0)(1712,9)(1713,0)(1714,0)(1715,10)(1716,0)(1717,3)(1718,5)(1719,0)(1720,0)(1721,0)(1722,0)(1723,9)(1724,2)(1725,10)(1726,0)(1727,0)(1728,0)(1729,0)(1730,0)(1731,0)(1732,0)(1733,1)(1734,0)(1735,0)(1736,4)(1737,4)(1738,0)(1739,8)(1740,0)(1741,7)(1742,0)(1743,0)(1744,0)(1745,0)(1746,0)(1747,0)(1748,0)(1749,0)(1750,0)(1751,0)(1752,0)(1753,0)(1754,0)(1755,0)(1756,0)(1757,0)(1758,0)(1759,0)(1760,0)(1761,0)(1762,0)(1763,0)(1764,0)(1765,0)(1766,0)(1767,3)(1768,0)(1769,0)(1770,0)(1771,3)(1772,0)(1773,0)(1774,0)(1775,0)(1776,0)(1777,0)(1778,0)(1779,0)(1780,0)(1781,0)(1782,0)(1783,0)(1784,0)(1785,0)(1786,0)(1787,0)(1788,0)(1789,0)(1790,0)(1791,0)(1792,0)(1793,0)(1794,0)(1795,7)(1796,0)(1797,0)(1798,1)(1799,2)(1800,0)(1801,0)(1802,0)(1803,0)(1804,0)(1805,1)(1806,0)(1807,2)(1808,13)(1809,0)(1810,0)(1811,3)(1812,1)(1813,0)(1814,0)(1815,0)(1816,0)(1817,0)(1818,5)(1819,27)(1820,0)(1821,0)(1822,0)(1823,0)(1824,0)(1825,0)(1826,0)(1827,0)(1828,0)(1829,0)(1830,0)(1831,0)(1832,0)(1833,1)(1834,0)(1835,0)(1836,0)(1837,0)(1838,0)(1839,0)(1840,7)(1841,5)(1842,0)(1843,0)(1844,0)(1845,0)(1846,0)(1847,13)(1848,0)(1849,1)(1850,0)(1851,0)(1852,0)(1853,6)(1854,4)(1855,0)(1856,0)(1857,6)(1858,5)(1859,0)(1860,0)(1861,0)(1862,0)(1863,17)(1864,4)(1865,8)(1866,0)(1867,0)(1868,6)(1869,0)(1870,0)(1871,0)(1872,0)(1873,0)(1874,0)(1875,0)(1876,0)(1877,0)(1878,0)(1879,548)(1880,0)(1881,0)(1882,0)(1883,0)(1884,0)(1885,0)(1886,0)(1887,0)(1888,0)(1889,0)(1890,0)(1891,0)(1892,2554)(1893,0)(1894,0)(1895,16)(1896,21)(1897,0)(1898,1)(1899,0)(1900,0)(1901,0)(1902,0)(1903,0)(1904,11)(1905,6)(1906,1)(1907,1)(1908,0)(1909,0)(1910,15)(1911,28)(1912,0)(1913,0)(1914,0)(1915,0)(1916,1)(1917,0)(1918,0)(1919,0)(1920,0)(1921,0)(1922,0)(1923,8)(1924,3)(1925,0)(1926,0)(1927,0)(1928,0)(1929,0)(1930,0)(1931,0)(1932,0)(1933,0)(1934,0)(1935,0)(1936,0)(1937,0)(1938,7)(1939,1)(1940,0)(1941,0)(1942,0)(1943,0)(1944,0)(1945,0)(1946,0)(1947,0)(1948,0)(1949,0)(1950,0)(1951,0)(1952,0)(1953,96)(1954,0)(1955,0)(1956,0)(1957,2)(1958,0)(1959,28)(1960,0)(1961,0)(1962,0)(1963,0)(1964,0)(1965,0)(1966,7)(1967,0)(1968,9)(1969,0)(1970,3)(1971,0)(1972,0)(1973,0)(1974,0)(1975,0)(1976,0)(1977,66)(1978,0)(1979,1)(1980,0)(1981,1)(1982,1)(1983,0)(1984,1)(1985,1)(1986,0)(1987,0)(1988,1)(1989,0)(1990,0)(1991,0)(1992,0)(1993,0)(1994,384)(1995,0)(1996,0)(1997,0)(1998,0)(1999,0)(2000,8)(2001,97)(2002,48)(2003,3)(2004,0)(2005,0)(2006,3)(2007,0)(2008,1)(2009,19)(2010,2)(2011,0)(2012,2)(2013,129)(2014,2)(2015,0)(2016,0)(2017,0)(2018,2)(2019,0)(2020,10)(2021,0)(2022,0)(2023,209)(2024,0)(2025,0)(2026,0)(2027,0)(2028,0)(2029,6)(2030,0)(2031,4)(2032,35)(2033,85)(2034,0)(2035,2)(2036,0)(2037,0)(2038,0)(2039,0)(2040,0)(2041,0)(2042,1)(2043,89)(2044,15)(2045,1)(2046,36)(2047,47)(2048,0)(2049,3)(2050,0)(2051,15)(2052,15)(2053,0)(2054,0)(2055,13)(2056,1)(2057,190)(2058,95)(2059,0)(2060,0)(2061,93)(2062,0)(2063,0)(2064,1)(2065,56)(2066,15)(2067,0)(2068,0)(2069,0)(2070,99)(2071,160)(2072,0)(2073,157)(2074,178)(2075,32)(2076,84)(2077,60)(2078,60)(2079,14)(2080,14)(2081,361)(2082,42)(2083,14)(2084,0)(2085,22)(2086,0)(2087,0)(2088,14)(2089,14)(2090,0)(2091,14)(2092,15)(2093,2)(2094,4)(2095,2)(2096,0)(2097,1)(2098,0)(2099,0)(2100,6)(2101,4)(2102,0)(2103,196)(2104,0)(2105,0)(2106,0)(2107,327)(2108,0)(2109,16)(2110,0)(2111,0)(2112,0)(2113,0)(2114,14)(2115,0)(2116,2)(2117,1)(2118,0)(2119,0)(2120,0)(2121,0)(2122,4)(2123,0)(2124,0)(2125,0)(2126,0)(2127,0)(2128,7)(2129,0)(2130,0)(2131,2)(2132,0)(2133,2)(2134,2)(2135,0)(2136,0)(2137,0)(2138,3)(2139,14)(2140,0)(2141,0)(2142,0)(2143,0)(2144,0)(2145,0)(2146,0)(2147,0)(2148,0)(2149,0)(2150,0)(2151,0)(2152,0)(2153,1)(2154,0)(2155,0)(2156,0)(2157,0)(2158,0)(2159,0)(2160,0)(2161,0)(2162,0)(2163,0)(2164,0)(2165,0)(2166,0)(2167,0)(2168,0)(2169,0)(2170,0)(2171,0)(2172,0)(2173,0)(2174,0)(2175,0)(2176,1)(2177,0)(2178,0)(2179,0)(2180,0)(2181,0)(2182,0)(2183,0)(2184,0)(2185,0)(2186,0)(2187,0)(2188,0)(2189,0)(2190,5)(2191,0)(2192,0)(2193,0)(2194,0)(2195,0)(2196,253)(2197,16)(2198,0)(2199,0)(2200,0)(2201,0)(2202,0)(2203,5)(2204,3)(2205,0)(2206,0)(2207,0)(2208,0)(2209,1)(2210,3)(2211,0)(2212,0)(2213,0)(2214,0)(2215,0)(2216,0)(2217,0)(2218,1)(2219,0)(2220,0)(2221,0)(2222,0)(2223,0)(2224,0)(2225,0)(2226,0)(2227,2)(2228,0)(2229,3)(2230,0)(2231,0)(2232,0)(2233,0)(2234,0)(2235,5)(2236,0)(2237,5)(2238,0)(2239,0)(2240,0)(2241,0)(2242,0)(2243,0)(2244,4)(2245,0)(2246,0)(2247,0)(2248,0)(2249,0)(2250,3)(2251,3)(2252,0)(2253,0)(2254,0)(2255,0)(2256,1)(2257,0)(2258,0)(2259,0)(2260,2)(2261,1)(2262,1)(2263,0)(2264,554)(2265,160)(2266,162)(2267,64)(2268,195)(2269,489)(2270,0)(2271,0)(2272,0)(2273,16)(2274,0)(2275,16)(2276,0)(2277,0)(2278,0)(2279,49)(2280,0)(2281,48)(2282,0)(2283,17)(2284,142)(2285,210)(2286,0)(2287,7)(2288,48)(2289,146)(2290,0)(2291,13)(2292,226)(2293,32)(2294,62)(2295,5)(2296,324)(2297,45)(2298,63)(2299,0)(2300,0)(2301,0)(2302,132)(2303,8)(2304,4)(2305,6)(2306,0)(2307,0)(2308,1)(2309,0)(2310,0)(2311,0)(2312,0)(2313,2)(2314,2)(2315,8)(2316,10)(2317,65)(2318,260)(2319,65)(2320,195)(2321,0)(2322,34)(2323,0)(2324,204)(2325,497)(2326,33)(2327,21)(2328,66)(2329,0)(2330,93)(2331,316)(2332,62)(2333,3)(2334,0)(2335,0)(2336,0)(2337,2)(2338,2)(2339,2)(2340,4)(2341,7)(2342,0)(2343,2)(2344,5)(2345,0)(2346,0)(2347,0)(2348,6)(2349,0)(2350,0)(2351,1)(2352,0)(2353,1)(2354,14)(2355,0)(2356,0)(2357,3)(2358,0)(2359,0)(2360,0)(2361,0)(2362,0)(2363,0)(2364,0)(2365,0)(2366,0)(2367,0)(2368,0)(2369,0)(2370,0)(2371,0)(2372,0)(2373,1)(2374,4)(2375,0)(2376,0)(2377,0)(2378,2)(2379,0)(2380,0)(2381,0)(2382,0)(2383,1)(2384,0)(2385,0)(2386,3)(2387,6)(2388,0)(2389,0)(2390,0)(2391,0)(2392,0)(2393,1)(2394,0)(2395,0)(2396,1)(2397,0)(2398,0)(2399,0)(2400,5)(2401,0)(2402,0)(2403,0)(2404,0)(2405,0)(2406,0)(2407,0)(2408,0)(2409,0)(2410,9)(2411,0)(2412,0)(2413,0)(2414,0)(2415,1)(2416,0)(2417,0)(2418,0)(2419,0)(2420,0)(2421,0)(2422,0)(2423,3)(2424,4)(2425,0)(2426,0)(2427,0)(2428,0)(2429,0)(2430,0)(2431,4)(2432,0)(2433,0)(2434,0)(2435,0)(2436,0)(2437,0)(2438,0)(2439,0)(2440,0)(2441,0)(2442,1)(2443,7)(2444,0)(2445,0)(2446,0)(2447,14)(2448,0)(2449,0)(2450,0)(2451,0)(2452,0)
};
\addlegendentry{Blowfish}
 
\end{axis}
\end{tikzpicture}%

%% file: ent-rc4.tex
\begin{tikzpicture}[scale=0.5]

\begin{axis}[
    xtick={0,1000,2000,3000,4000,5000},
    legend pos=north east,
    ymajorgrids=true,
    grid style=dashed,
]

\addplot[color=blue]
coordinates {
	(0,0)(1,0)(2,1)(3,9)(4,0)(5,0)(6,0)(7,0)(8,0)(9,0)(10,0)(11,2)(12,0)(13,0)(14,2)(15,2)(16,0)(17,0)(18,0)(19,0)(20,0)(21,0)(22,0)(23,0)(24,0)(25,0)(26,9)(27,0)(28,0)(29,0)(30,0)(31,0)(32,0)(33,0)(34,4)(35,0)(36,0)(37,20)(38,13)(39,4)(40,1)(41,0)(42,0)(43,0)(44,0)(45,1)(46,4)(47,0)(48,0)(49,3)(50,0)(51,0)(52,0)(53,0)(54,2)(55,3)(56,0)(57,1)(58,0)(59,0)(60,0)(61,2)(62,1)(63,0)(64,0)(65,16)(66,0)(67,0)(68,0)(69,0)(70,0)(71,0)(72,8)(73,0)(74,3)(75,4)(76,0)(77,0)(78,0)(79,0)(80,1)(81,3)(82,2)(83,15)(84,3)(85,0)(86,0)(87,7)(88,0)(89,3)(90,2)(91,0)(92,2)(93,0)(94,0)(95,0)(96,1)(97,0)(98,0)(99,1)(100,0)(101,0)(102,0)(103,0)(104,0)(105,0)(106,0)(107,0)(108,0)(109,0)(110,0)(111,0)(112,0)(113,3)(114,0)(115,0)(116,0)(117,0)(118,7)(119,0)(120,0)(121,0)(122,0)(123,0)(124,0)(125,0)(126,0)(127,10)(128,0)(129,1)(130,0)(131,0)(132,0)(133,0)(134,0)(135,0)(136,0)(137,0)(138,0)(139,0)(140,0)(141,3)(142,0)(143,0)(144,0)(145,0)(146,0)(147,3)(148,0)(149,0)(150,0)(151,1)(152,0)(153,0)(154,0)(155,0)(156,0)(157,0)(158,0)(159,2)(160,1)(161,0)(162,0)(163,0)(164,0)(165,0)(166,0)(167,0)(168,1)(169,0)(170,3)(171,0)(172,0)(173,2)(174,0)(175,0)(176,0)(177,0)(178,0)(179,0)(180,2)(181,0)(182,1)(183,0)(184,1)(185,0)(186,0)(187,2)(188,2)(189,0)(190,2)(191,2)(192,0)(193,0)(194,0)(195,2)(196,5)(197,0)(198,0)(199,0)(200,0)(201,5)(202,8)(203,0)(204,0)(205,0)(206,0)(207,0)(208,0)(209,0)(210,0)(211,0)(212,0)(213,0)(214,0)(215,0)(216,0)(217,0)(218,0)(219,14)(220,0)(221,0)(222,1)(223,0)(224,0)(225,0)(226,0)(227,0)(228,0)(229,0)(230,1)(231,1)(232,1)(233,0)(234,1)(235,6)(236,0)(237,1)(238,3)(239,0)(240,1)(241,1)(242,0)(243,0)(244,0)(245,0)(246,0)(247,0)(248,0)(249,0)(250,3)(251,2)(252,0)(253,0)(254,2)(255,0)(256,0)(257,0)(258,0)(259,0)(260,3)(261,23)(262,1)(263,2)(264,0)(265,10)(266,1)(267,0)(268,0)(269,1)(270,1)(271,3)(272,0)(273,0)(274,10)(275,2)(276,18)(277,1)(278,0)(279,0)(280,1)(281,0)(282,10)(283,1)(284,0)(285,5)(286,0)(287,0)(288,12)(289,15)(290,0)(291,0)(292,0)(293,4)(294,0)(295,0)(296,29)(297,0)(298,0)(299,1)(300,0)(301,400)(302,0)(303,0)(304,0)(305,0)(306,0)(307,0)(308,0)(309,0)(310,22)(311,0)(312,2)(313,0)(314,0)(315,0)(316,0)(317,0)(318,0)(319,0)(320,0)(321,0)(322,0)(323,1)(324,0)(325,0)(326,12)(327,0)(328,0)(329,1)(330,0)(331,0)(332,0)(333,3)(334,0)(335,0)(336,0)(337,0)(338,0)(339,0)(340,0)(341,0)(342,5)(343,3)(344,0)(345,0)(346,46)(347,0)(348,288)(349,0)(350,0)(351,0)(352,1)(353,2)(354,0)(355,0)(356,0)(357,0)(358,0)(359,0)(360,0)(361,0)(362,0)(363,5)(364,0)(365,0)(366,0)(367,0)(368,0)(369,0)(370,0)(371,0)(372,0)(373,0)(374,1)(375,0)(376,0)(377,0)(378,0)(379,0)(380,0)(381,0)(382,0)(383,0)(384,0)(385,0)(386,0)(387,0)(388,0)(389,0)(390,0)(391,7)(392,0)(393,0)(394,1)(395,0)(396,35)(397,0)(398,0)(399,0)(400,0)(401,8)(402,0)(403,3)(404,0)(405,0)(406,0)(407,0)(408,0)(409,0)(410,0)(411,0)(412,0)(413,0)(414,0)(415,0)(416,0)(417,0)(418,0)(419,0)(420,0)(421,0)(422,0)(423,0)(424,0)(425,0)(426,0)(427,0)(428,0)(429,0)(430,0)(431,0)(432,0)(433,15)(434,0)(435,0)(436,0)(437,0)(438,0)(439,0)(440,0)(441,1)(442,6)(443,0)(444,1)(445,2)(446,0)(447,0)(448,3)(449,3)(450,1)(451,0)(452,0)(453,0)(454,0)(455,0)(456,1)(457,0)(458,1)(459,0)(460,0)(461,0)(462,4)(463,0)(464,0)(465,0)(466,0)(467,1)(468,2)(469,2)(470,4)(471,0)(472,0)(473,0)(474,0)(475,4)(476,0)(477,0)(478,0)(479,5)(480,2)(481,0)(482,0)(483,0)(484,0)(485,0)(486,0)(487,2)(488,0)(489,1)(490,0)(491,1)(492,0)(493,0)(494,0)(495,0)(496,0)(497,0)(498,2)(499,2)(500,1)(501,134)(502,0)(503,0)(504,0)(505,0)(506,0)(507,0)(508,0)(509,9)(510,0)(511,0)(512,0)(513,0)(514,0)(515,0)(516,27)(517,8)(518,0)(519,0)(520,0)(521,0)(522,1)(523,1)(524,1)(525,1)(526,1)(527,1)(528,0)(529,0)(530,0)(531,0)(532,0)(533,0)(534,0)(535,0)(536,0)(537,0)(538,0)(539,8)(540,0)(541,0)(542,1)(543,0)(544,0)(545,8)(546,0)(547,5)(548,0)(549,0)(550,2)(551,0)(552,0)(553,5)(554,0)(555,5)(556,10)(557,0)(558,0)(559,0)(560,3)(561,0)(562,0)(563,0)(564,0)(565,0)(566,0)(567,0)(568,0)(569,4)(570,0)(571,0)(572,0)(573,0)(574,0)(575,0)(576,0)(577,0)(578,2)(579,0)(580,1)(581,1)(582,0)(583,0)(584,1)(585,0)(586,0)(587,0)(588,0)(589,0)(590,0)(591,0)(592,1)(593,0)(594,0)(595,0)(596,0)(597,0)(598,0)(599,0)(600,0)(601,0)(602,0)(603,1)(604,0)(605,0)(606,0)(607,0)(608,2)(609,6)(610,0)(611,0)(612,0)(613,0)(614,0)(615,1)(616,0)(617,0)(618,0)(619,0)(620,0)(621,0)(622,0)(623,0)(624,0)(625,0)(626,0)(627,0)(628,0)(629,0)(630,0)(631,0)(632,0)(633,0)(634,0)(635,0)(636,0)(637,0)(638,3)(639,0)(640,1)(641,3)(642,0)(643,1)(644,0)(645,0)(646,0)(647,0)(648,0)(649,0)(650,6)(651,0)(652,0)(653,0)(654,2)(655,0)(656,0)(657,0)(658,0)(659,0)(660,0)(661,0)(662,0)(663,0)(664,0)(665,0)(666,0)(667,0)(668,0)(669,0)(670,0)(671,0)(672,1)(673,0)(674,0)(675,0)(676,1)(677,0)(678,0)(679,0)(680,2)(681,5)(682,0)(683,0)(684,0)(685,0)(686,1)(687,2)(688,1)(689,0)(690,0)(691,0)(692,0)(693,1)(694,0)(695,1)(696,0)(697,0)(698,0)(699,0)(700,0)(701,0)(702,0)(703,0)(704,0)(705,0)(706,0)(707,1)(708,0)(709,0)(710,0)(711,0)(712,0)(713,4)(714,1)(715,0)(716,10)(717,6)(718,0)(719,0)(720,0)(721,5)(722,0)(723,1)(724,0)(725,0)(726,0)(727,0)(728,0)(729,1)(730,0)(731,0)(732,0)(733,1)(734,0)(735,2)(736,0)(737,0)(738,0)(739,2)(740,0)(741,0)(742,7)(743,2)(744,0)(745,0)(746,0)(747,0)(748,3)(749,1)(750,1)(751,2)(752,0)(753,0)(754,0)(755,6)(756,0)(757,0)(758,0)(759,0)(760,0)(761,2)(762,0)(763,0)(764,0)(765,0)(766,0)(767,0)(768,0)(769,0)(770,0)(771,3)(772,0)(773,0)(774,0)(775,0)(776,0)(777,0)(778,0)(779,0)(780,0)(781,0)(782,1)(783,1)(784,0)(785,0)(786,0)(787,0)(788,0)(789,0)(790,0)(791,0)(792,0)(793,1)(794,0)(795,0)(796,0)(797,0)(798,0)(799,0)(800,0)(801,0)(802,0)(803,0)(804,0)(805,0)(806,0)(807,0)(808,0)(809,0)(810,0)(811,0)(812,0)(813,0)(814,0)(815,0)(816,0)(817,0)(818,0)(819,0)(820,0)(821,0)(822,0)(823,0)(824,0)(825,0)(826,0)(827,0)(828,1)(829,0)(830,0)(831,0)(832,0)(833,0)(834,0)(835,0)(836,0)(837,0)(838,0)(839,0)(840,0)(841,0)(842,0)(843,0)(844,0)(845,0)(846,0)(847,0)(848,0)(849,0)(850,0)(851,0)(852,0)(853,18)(854,1)(855,0)(856,0)(857,2)(858,0)(859,0)(860,0)(861,0)(862,0)(863,0)(864,0)(865,0)(866,0)(867,0)(868,3)(869,0)(870,0)(871,0)(872,1)(873,0)(874,0)(875,0)(876,0)(877,0)(878,0)(879,0)(880,1)(881,1)(882,0)(883,0)(884,0)(885,0)(886,0)(887,0)(888,0)(889,0)(890,0)(891,0)(892,11)(893,0)(894,0)(895,0)(896,0)(897,6)(898,0)(899,0)(900,0)(901,0)(902,0)(903,0)(904,3)(905,0)(906,0)(907,1)(908,0)(909,0)(910,0)(911,4)(912,0)(913,0)(914,25)(915,0)(916,0)(917,0)(918,0)(919,0)(920,0)(921,0)(922,0)(923,0)(924,0)(925,0)(926,0)(927,0)(928,15)(929,0)(930,0)(931,0)(932,0)(933,0)(934,0)(935,0)(936,4)(937,1)(938,3)(939,0)(940,0)(941,0)(942,0)(943,0)(944,0)(945,0)(946,0)(947,0)(948,0)(949,0)(950,7)(951,1)(952,0)(953,0)(954,0)(955,0)(956,0)(957,0)(958,0)(959,0)(960,0)(961,0)(962,0)(963,0)(964,0)(965,2)(966,0)(967,0)(968,0)(969,0)(970,0)(971,1)(972,0)(973,1)(974,0)(975,7)(976,0)(977,0)(978,2)(979,0)(980,2)(981,0)(982,1)(983,0)(984,0)(985,0)(986,0)(987,3)(988,0)(989,0)(990,0)(991,0)(992,4)(993,0)(994,0)(995,11)(996,6)(997,1)(998,0)(999,1)(1000,0)(1001,0)(1002,0)(1003,0)(1004,1)(1005,1)(1006,0)(1007,0)(1008,0)(1009,0)(1010,2)(1011,4)(1012,0)(1013,0)(1014,0)(1015,0)(1016,0)(1017,0)(1018,0)(1019,0)(1020,0)(1021,14)(1022,0)(1023,0)(1024,9)(1025,0)(1026,0)(1027,0)(1028,0)(1029,0)(1030,0)(1031,0)(1032,6)(1033,7)(1034,0)(1035,0)(1036,0)(1037,0)(1038,0)(1039,0)(1040,0)(1041,0)(1042,0)(1043,0)(1044,0)(1045,0)(1046,2)(1047,8)(1048,0)(1049,0)(1050,0)(1051,0)(1052,0)(1053,0)(1054,0)(1055,0)(1056,27)(1057,0)(1058,0)(1059,0)(1060,0)(1061,0)(1062,3)(1063,16)(1064,0)(1065,1)(1066,4)(1067,0)(1068,25)(1069,0)(1070,0)(1071,1)(1072,0)(1073,3)(1074,0)(1075,0)(1076,0)(1077,0)(1078,1)(1079,0)(1080,0)(1081,0)(1082,0)(1083,2)(1084,0)(1085,0)(1086,16)(1087,0)(1088,1)(1089,0)(1090,0)(1091,0)(1092,0)(1093,0)(1094,5)(1095,2)(1096,0)(1097,0)(1098,0)(1099,0)(1100,0)(1101,0)(1102,0)(1103,0)(1104,0)(1105,0)(1106,0)(1107,0)(1108,0)(1109,0)(1110,0)(1111,0)(1112,0)(1113,16)(1114,42)(1115,0)(1116,0)(1117,0)(1118,0)(1119,0)(1120,0)(1121,0)(1122,0)(1123,0)(1124,0)(1125,0)(1126,0)(1127,3)(1128,0)(1129,3)(1130,0)(1131,0)(1132,0)(1133,0)(1134,0)(1135,12)(1136,6)(1137,6)(1138,0)(1139,0)(1140,0)(1141,108)(1142,11)(1143,0)(1144,0)(1145,0)(1146,0)(1147,0)(1148,0)(1149,0)(1150,0)(1151,0)(1152,0)(1153,0)(1154,0)(1155,0)(1156,0)(1157,0)(1158,0)(1159,0)(1160,0)(1161,0)(1162,3)(1163,0)(1164,0)(1165,0)(1166,0)(1167,5)(1168,0)(1169,0)(1170,0)(1171,0)(1172,0)(1173,5)(1174,0)(1175,11)(1176,0)(1177,3)(1178,0)(1179,0)(1180,0)(1181,0)(1182,7)(1183,0)(1184,0)(1185,0)(1186,0)(1187,0)(1188,0)(1189,0)(1190,0)(1191,7)(1192,1)(1193,1)(1194,0)(1195,8)(1196,10)(1197,1)(1198,0)(1199,0)(1200,0)(1201,0)(1202,0)(1203,1188)(1204,0)(1205,0)(1206,0)(1207,2)(1208,1)(1209,0)(1210,0)(1211,0)(1212,0)(1213,56)(1214,0)(1215,0)(1216,649)(1217,186)(1218,3)(1219,0)(1220,0)(1221,85)(1222,21)(1223,56)(1224,13)(1225,440)(1226,3)(1227,0)(1228,0)(1229,0)(1230,0)(1231,0)(1232,1)(1233,0)(1234,80)(1235,8)(1236,0)(1237,0)(1238,0)(1239,0)(1240,0)(1241,0)(1242,0)(1243,0)(1244,0)(1245,0)(1246,0)(1247,0)(1248,5)(1249,0)(1250,0)(1251,0)(1252,0)(1253,0)(1254,2)(1255,0)(1256,26)(1257,39)(1258,0)(1259,0)(1260,0)(1261,0)(1262,0)(1263,13)(1264,0)(1265,0)(1266,12)(1267,0)(1268,0)(1269,175)(1270,0)(1271,48)(1272,1111)(1273,0)(1274,174)(1275,782)(1276,1815)(1277,21)(1278,7)(1279,0)(1280,254)(1281,5)(1282,1081)(1283,340)(1284,57)(1285,0)(1286,163)(1287,202)(1288,0)(1289,0)(1290,0)(1291,0)(1292,0)(1293,34)(1294,0)(1295,10)(1296,0)(1297,0)(1298,113)(1299,1)(1300,7)(1301,8)(1302,7)(1303,0)(1304,0)(1305,369)(1306,0)(1307,46)(1308,917)(1309,1)(1310,203)(1311,0)(1312,21)(1313,28)(1314,62)(1315,14)(1316,0)(1317,0)(1318,0)(1319,0)(1320,13)(1321,5)(1322,0)(1323,147)(1324,78)(1325,0)(1326,4)(1327,91)(1328,0)(1329,0)(1330,81)(1331,2)(1332,0)(1333,0)(1334,35)(1335,0)(1336,388)(1337,0)(1338,50)(1339,0)(1340,96)(1341,0)(1342,0)(1343,29)(1344,77)(1345,0)(1346,28)(1347,0)(1348,0)(1349,0)(1350,25)(1351,0)(1352,0)(1353,0)(1354,0)(1355,0)(1356,2)(1357,0)(1358,31)(1359,20)(1360,41)(1361,54)(1362,99)(1363,85)(1364,40)(1365,0)(1366,19)(1367,10)(1368,120)(1369,39)(1370,26)(1371,158)(1372,23)(1373,0)(1374,0)(1375,76)(1376,0)(1377,58)(1378,0)(1379,0)(1380,0)(1381,0)(1382,0)(1383,2)(1384,596)(1385,68)(1386,19)(1387,148)(1388,0)(1389,1)(1390,2)(1391,9)(1392,0)(1393,0)(1394,0)(1395,70)(1396,23)(1397,0)(1398,3)(1399,0)(1400,0)(1401,1)(1402,2)(1403,1)(1404,1)(1405,215)(1406,14)(1407,0)(1408,21)(1409,0)(1410,6)(1411,24)(1412,2)(1413,429)(1414,96)(1415,1)(1416,0)(1417,0)(1418,3)(1419,1)(1420,69)(1421,0)(1422,0)(1423,0)(1424,0)(1425,0)(1426,0)(1427,4)(1428,8)(1429,0)(1430,0)(1431,0)(1432,0)(1433,0)(1434,0)(1435,0)(1436,0)(1437,0)(1438,0)(1439,0)(1440,2)(1441,0)(1442,15)(1443,632)(1444,583)(1445,55)(1446,6)(1447,6)(1448,1288)(1449,0)(1450,27)(1451,0)(1452,36)(1453,0)(1454,0)(1455,0)(1456,30)(1457,0)(1458,0)(1459,0)(1460,0)(1461,0)(1462,0)(1463,0)(1464,0)(1465,0)(1466,0)(1467,2)(1468,0)(1469,2)(1470,0)(1471,7)(1472,0)(1473,8)(1474,0)(1475,0)(1476,5)(1477,1)(1478,10)(1479,0)(1480,0)(1481,0)(1482,0)(1483,0)(1484,0)(1485,0)(1486,0)(1487,0)(1488,0)(1489,0)(1490,0)(1491,3)(1492,7)(1493,0)(1494,0)(1495,0)(1496,6)(1497,10)(1498,1)(1499,0)(1500,3)(1501,0)(1502,4)(1503,0)(1504,5)(1505,38)(1506,996)(1507,12)(1508,135)(1509,0)(1510,0)(1511,12)(1512,2)(1513,0)(1514,5)(1515,0)(1516,10)(1517,0)(1518,10)(1519,0)(1520,0)(1521,0)(1522,0)(1523,0)(1524,0)(1525,0)(1526,0)(1527,26)(1528,0)(1529,0)(1530,0)(1531,406)(1532,0)(1533,0)(1534,0)(1535,0)(1536,0)(1537,0)(1538,0)(1539,0)(1540,0)(1541,0)(1542,0)(1543,0)(1544,5)(1545,10)(1546,0)(1547,89)(1548,4)(1549,0)(1550,27)(1551,12)(1552,2)(1553,28)(1554,0)(1555,251)(1556,16)(1557,32)(1558,6)(1559,0)(1560,0)(1561,4)(1562,0)(1563,0)(1564,0)(1565,10)(1566,6)(1567,981)(1568,0)(1569,0)(1570,0)(1571,0)(1572,0)(1573,0)(1574,27)(1575,0)(1576,0)(1577,0)(1578,0)(1579,0)(1580,0)(1581,9)(1582,0)(1583,0)(1584,0)(1585,0)(1586,0)(1587,0)(1588,0)(1589,0)(1590,0)(1591,3)(1592,0)(1593,0)(1594,0)(1595,0)(1596,11)(1597,0)(1598,1)(1599,0)(1600,0)(1601,0)(1602,0)(1603,7657)(1604,0)(1605,0)(1606,0)(1607,5)(1608,7)(1609,0)(1610,0)(1611,0)(1612,0)(1613,686)(1614,0)(1615,0)(1616,776)(1617,584)(1618,396)(1619,586)(1620,392)(1621,52)(1622,12)(1623,0)(1624,0)(1625,0)(1626,0)(1627,78)(1628,536)(1629,537)(1630,566)(1631,0)(1632,0)(1633,0)(1634,9)(1635,0)(1636,0)(1637,0)(1638,0)(1639,0)(1640,0)(1641,0)(1642,5)(1643,4)(1644,13)(1645,0)(1646,0)(1647,32)(1648,0)(1649,0)(1650,0)(1651,0)(1652,0)(1653,0)(1654,0)(1655,63)(1656,1)(1657,38)(1658,0)(1659,0)(1660,2)(1661,0)(1662,173)(1663,211)(1664,0)(1665,37)(1666,0)(1667,68)(1668,84)(1669,115)(1670,4)(1671,2)(1672,2)(1673,0)(1674,0)(1675,0)(1676,0)(1677,0)(1678,0)(1679,0)(1680,0)(1681,0)(1682,2)(1683,99)(1684,1)(1685,0)(1686,0)(1687,0)(1688,5)(1689,0)(1690,0)(1691,7)(1692,0)(1693,0)(1694,0)(1695,0)(1696,0)(1697,3)(1698,0)(1699,0)(1700,0)(1701,0)(1702,0)(1703,0)(1704,0)(1705,0)(1706,12)(1707,1)(1708,0)(1709,0)(1710,0)(1711,0)(1712,0)(1713,0)(1714,0)(1715,8)(1716,0)(1717,0)(1718,1)(1719,0)(1720,0)(1721,0)(1722,0)(1723,0)(1724,0)(1725,0)(1726,0)(1727,0)(1728,0)(1729,0)(1730,0)(1731,0)(1732,2)(1733,27)(1734,0)(1735,3)(1736,7)(1737,19)(1738,0)(1739,0)(1740,0)(1741,0)(1742,0)(1743,0)(1744,1)(1745,0)(1746,0)(1747,9)(1748,0)(1749,0)(1750,10)(1751,0)(1752,3)(1753,5)(1754,0)(1755,0)(1756,0)(1757,7)(1758,2)(1759,9)(1760,0)(1761,0)(1762,0)(1763,0)(1764,0)(1765,4)(1766,1)(1767,0)(1768,0)(1769,4)(1770,4)(1771,0)(1772,8)(1773,0)(1774,7)(1775,0)(1776,0)(1777,0)(1778,0)(1779,0)(1780,0)(1781,0)(1782,0)(1783,0)(1784,0)(1785,0)(1786,0)(1787,0)(1788,0)(1789,0)(1790,0)(1791,0)(1792,0)(1793,0)(1794,0)(1795,0)(1796,0)(1797,0)(1798,0)(1799,0)(1800,3)(1801,0)(1802,0)(1803,0)(1804,3)(1805,0)(1806,0)(1807,0)(1808,0)(1809,0)(1810,0)(1811,0)(1812,0)(1813,0)(1814,0)(1815,0)(1816,0)(1817,0)(1818,0)(1819,0)(1820,0)(1821,0)(1822,0)(1823,0)(1824,0)(1825,0)(1826,0)(1827,0)(1828,7)(1829,0)(1830,0)(1831,1)(1832,2)(1833,0)(1834,0)(1835,0)(1836,0)(1837,0)(1838,1)(1839,0)(1840,2)(1841,11)(1842,0)(1843,0)(1844,2)(1845,2)(1846,0)(1847,71)(1848,15)(1849,0)(1850,0)(1851,0)(1852,5)(1853,27)(1854,0)(1855,0)(1856,0)(1857,0)(1858,0)(1859,0)(1860,0)(1861,0)(1862,0)(1863,0)(1864,0)(1865,0)(1866,0)(1867,1)(1868,0)(1869,0)(1870,0)(1871,0)(1872,0)(1873,0)(1874,7)(1875,2)(1876,0)(1877,0)(1878,0)(1879,1)(1880,0)(1881,0)(1882,0)(1883,0)(1884,0)(1885,0)(1886,0)(1887,11)(1888,4)(1889,0)(1890,0)(1891,6)(1892,5)(1893,0)(1894,0)(1895,0)(1896,0)(1897,20)(1898,6)(1899,2)(1900,0)(1901,0)(1902,5)(1903,7)(1904,2)(1905,0)(1906,0)(1907,0)(1908,0)(1909,0)(1910,0)(1911,1)(1912,0)(1913,9)(1914,16)(1915,1)(1916,0)(1917,0)(1918,1)(1919,0)(1920,0)(1921,0)(1922,0)(1923,0)(1924,0)(1925,0)(1926,0)(1927,0)(1928,0)(1929,0)(1930,0)(1931,0)(1932,0)(1933,0)(1934,1)(1935,0)(1936,0)(1937,0)(1938,0)(1939,0)(1940,0)(1941,0)(1942,0)(1943,0)(1944,0)(1945,0)(1946,0)(1947,0)(1948,0)(1949,0)(1950,0)(1951,0)(1952,0)(1953,0)(1954,0)(1955,0)(1956,0)(1957,1)(1958,0)(1959,0)(1960,0)(1961,0)(1962,0)(1963,0)(1964,0)(1965,0)(1966,0)(1967,0)(1968,0)(1969,0)(1970,0)(1971,5)(1972,0)(1973,0)(1974,0)(1975,0)(1976,253)(1977,16)(1978,0)(1979,0)(1980,2)(1981,0)(1982,0)(1983,0)(1984,0)(1985,0)(1986,0)(1987,5)(1988,3)(1989,0)(1990,0)(1991,0)(1992,0)(1993,1)(1994,3)(1995,0)(1996,0)(1997,0)(1998,0)(1999,0)(2000,0)(2001,0)(2002,1)(2003,0)(2004,0)(2005,0)(2006,0)(2007,0)(2008,0)(2009,0)(2010,0)(2011,2)(2012,0)(2013,3)(2014,0)(2015,0)(2016,0)(2017,0)(2018,20)(2019,4)(2020,9)(2021,1)(2022,0)(2023,0)(2024,10)(2025,6)(2026,3)(2027,0)(2028,4)(2029,0)(2030,0)(2031,0)(2032,0)(2033,0)(2034,0)(2035,6)(2036,1)(2037,1)(2038,7)(2039,1)(2040,0)(2041,0)(2042,0)(2043,0)(2044,0)(2045,0)(2046,0)(2047,0)(2048,2)(2049,1)(2050,2)(2051,0)(2052,0)(2053,0)(2054,0)(2055,1)(2056,14)(2057,0)(2058,0)(2059,0)(2060,0)(2061,0)(2062,0)(2063,0)(2064,0)(2065,0)(2066,0)(2067,0)(2068,0)(2069,0)(2070,1)(2071,255)(2072,1)(2073,509)(2074,4)(2075,7)(2076,0)(2077,0)(2078,0)(2079,2)(2080,7)(2081,0)(2082,0)(2083,0)(2084,0)(2085,0)(2086,0)(2087,0)(2088,0)(2089,5)(2090,0)(2091,0)(2092,0)(2093,2)(2094,0)(2095,0)(2096,0)(2097,2)(2098,41)(2099,31)(2100,33)(2101,0)(2102,0)(2103,1)(2104,0)(2105,0)(2106,0)(2107,15)(2108,0)(2109,0)(2110,0)(2111,2)(2112,1)(2113,1)(2114,5)(2115,0)(2116,3)(2117,5)(2118,0)(2119,5)(2120,0)(2121,5)(2122,0)(2123,0)(2124,0)(2125,0)(2126,0)(2127,0)(2128,0)(2129,0)(2130,5)(2131,0)(2132,0)(2133,4)(2134,0)(2135,0)(2136,0)(2137,0)(2138,0)(2139,0)(2140,0)(2141,0)(2142,128)(2143,4)(2144,0)(2145,0)(2146,0)(2147,5)(2148,0)(2149,0)(2150,0)(2151,0)(2152,0)(2153,0)(2154,0)(2155,0)(2156,0)(2157,0)(2158,0)(2159,0)(2160,0)(2161,0)(2162,1)(2163,0)(2164,0)(2165,5)(2166,0)(2167,0)(2168,21)(2169,0)(2170,0)(2171,0)(2172,0)(2173,1)(2174,0)(2175,6)(2176,0)(2177,0)(2178,0)(2179,0)(2180,1)(2181,0)(2182,0)(2183,0)(2184,0)(2185,0)(2186,0)(2187,7)(2188,0)(2189,9)(2190,1)(2191,1)(2192,2)(2193,0)(2194,0)(2195,0)(2196,0)(2197,0)(2198,0)(2199,0)(2200,0)(2201,0)(2202,0)(2203,0)(2204,0)(2205,0)(2206,0)(2207,0)(2208,4)(2209,0)(2210,0)(2211,0)(2212,0)(2213,1)(2214,0)(2215,0)(2216,0)(2217,0)(2218,0)(2219,0)(2220,0)(2221,0)(2222,0)(2223,0)(2224,0)(2225,56)(2226,29)(2227,14)(2228,3)(2229,0)(2230,0)(2231,3)(2232,0)(2233,1)(2234,0)(2235,2)(2236,0)(2237,2)(2238,0)(2239,2)(2240,0)(2241,0)(2242,0)(2243,463)(2244,79)(2245,10)(2246,0)(2247,0)(2248,521)(2249,0)(2250,0)(2251,0)(2252,0)(2253,0)(2254,6)(2255,0)(2256,4)(2257,1252)(2258,85)(2259,0)(2260,2)(2261,17)(2262,0)(2263,0)(2264,0)(2265,0)(2266,0)(2267,1)(2268,292)(2269,149)(2270,34)(2271,42)(2272,0)(2273,0)(2274,3)(2275,0)(2276,15)(2277,15)(2278,0)(2279,0)(2280,17)(2281,356)(2282,190)(2283,301)(2284,0)(2285,0)(2286,93)(2287,53)(2288,0)(2289,209)(2290,56)(2291,15)(2292,0)(2293,0)(2294,0)(2295,176)(2296,160)(2297,0)(2298,157)(2299,178)(2300,259)(2301,84)(2302,60)(2303,60)(2304,14)(2305,14)(2306,28)(2307,42)(2308,14)(2309,0)(2310,28)(2311,0)(2312,0)(2313,108)(2314,14)(2315,0)(2316,14)(2317,15)(2318,0)(2319,0)(2320,0)(2321,0)(2322,6)(2323,4)(2324,0)(2325,0)(2326,0)(2327,0)(2328,0)(2329,0)(2330,0)(2331,16)(2332,0)(2333,0)(2334,0)(2335,0)(2336,45)(2337,0)(2338,0)(2339,0)(2340,0)(2341,0)(2342,3)(2343,0)(2344,0)(2345,5)(2346,0)(2347,0)(2348,2)(2349,0)(2350,2)(2351,2)(2352,0)(2353,0)(2354,0)(2355,0)(2356,0)(2357,0)(2358,0)(2359,0)(2360,2)(2361,0)(2362,0)(2363,0)(2364,3)(2365,1)(2366,0)(2367,0)(2368,5)(2369,0)(2370,5)(2371,0)(2372,0)(2373,0)(2374,0)(2375,0)(2376,0)(2377,4)(2378,0)(2379,0)(2380,0)(2381,0)(2382,0)(2383,3)(2384,3)(2385,0)(2386,0)(2387,0)(2388,0)(2389,1)(2390,0)(2391,0)(2392,0)(2393,2)(2394,0)(2395,0)(2396,394)(2397,64)(2398,64)(2399,16)(2400,31)(2401,134)(2402,16)(2403,16)(2404,48)(2405,32)(2406,32)(2407,75)(2408,42)(2409,129)(2410,32)(2411,8)(2412,41)(2413,0)(2414,0)(2415,0)(2416,175)(2417,32)(2418,0)(2419,0)(2420,2)(2421,135)(2422,33)(2423,37)(2424,0)(2425,137)(2426,0)(2427,46)(2428,0)(2429,0)(2430,2)(2431,15)(2432,4)(2433,0)(2434,0)(2435,0)(2436,0)(2437,0)(2438,0)(2439,1)(2440,0)(2441,0)(2442,0)(2443,0)(2444,2)(2445,2)(2446,7)(2447,65)(2448,130)(2449,33)(2450,0)(2451,31)(2452,93)(2453,12)(2454,0)(2455,0)(2456,66)(2457,0)(2458,0)(2459,69)(2460,69)(2461,0)(2462,12)(2463,0)(2464,315)(2465,12)(2466,292)(2467,0)(2468,0)(2469,0)(2470,8)(2471,0)(2472,0)(2473,8)(2474,0)(2475,4)(2476,7)(2477,0)(2478,2)(2479,5)(2480,0)(2481,0)(2482,0)(2483,6)(2484,0)(2485,0)(2486,1)(2487,0)(2488,1)(2489,14)(2490,0)(2491,0)(2492,3)(2493,0)(2494,0)(2495,0)(2496,0)(2497,0)(2498,0)(2499,0)(2500,0)(2501,0)(2502,0)(2503,0)(2504,0)(2505,0)(2506,0)(2507,0)(2508,0)(2509,0)(2510,5)(2511,0)(2512,0)(2513,0)(2514,0)(2515,0)(2516,2)(2517,0)(2518,0)(2519,0)(2520,0)(2521,0)(2522,1)(2523,0)(2524,0)(2525,3)(2526,6)(2527,0)(2528,0)(2529,0)(2530,0)(2531,0)(2532,1)(2533,0)(2534,0)(2535,1)(2536,0)(2537,0)(2538,0)(2539,5)(2540,0)(2541,0)(2542,0)(2543,0)(2544,0)(2545,0)(2546,0)(2547,0)(2548,0)(2549,9)(2550,0)(2551,0)(2552,0)(2553,0)(2554,1)(2555,0)(2556,0)(2557,0)(2558,0)(2559,3)(2560,4)(2561,0)(2562,0)(2563,0)(2564,0)(2565,0)(2566,0)(2567,4)(2568,0)(2569,0)(2570,0)(2571,0)(2572,0)(2573,0)(2574,0)(2575,0)(2576,0)(2577,0)(2578,1)(2579,7)(2580,0)(2581,0)(2582,0)(2583,14)(2584,0)(2585,0)(2586,0)(2587,0)(2588,0)
};
\addlegendentry{RC4}

\end{axis}
\end{tikzpicture}

%% file: ent-md5.tex
\begin{tikzpicture}[scale=0.5]

\begin{axis}[
    xtick={0,1000,2000,3000,4000,5000},
    legend pos=north west,
    ymajorgrids=true,
    grid style=dashed,
]

\addplot[color=blue]
coordinates {
(0,0)(1,0)(2,1)(3,9)(4,0)(5,0)(6,0)(7,0)(8,0)(9,0)(10,0)(11,2)(12,0)(13,0)(14,2)(15,2)(16,0)(17,0)(18,0)(19,0)(20,0)(21,0)(22,0)(23,0)(24,0)(25,0)(26,9)(27,0)(28,0)(29,0)(30,0)(31,0)(32,0)(33,0)(34,4)(35,0)(36,0)(37,20)(38,13)(39,4)(40,1)(41,0)(42,0)(43,0)(44,0)(45,1)(46,4)(47,0)(48,0)(49,3)(50,0)(51,0)(52,0)(53,0)(54,2)(55,3)(56,0)(57,1)(58,0)(59,0)(60,0)(61,2)(62,1)(63,0)(64,0)(65,16)(66,0)(67,0)(68,0)(69,0)(70,0)(71,0)(72,8)(73,1)(74,6)(75,4)(76,0)(77,0)(78,0)(79,0)(80,1)(81,3)(82,2)(83,15)(84,3)(85,0)(86,0)(87,7)(88,0)(89,3)(90,2)(91,0)(92,2)(93,0)(94,0)(95,0)(96,1)(97,0)(98,0)(99,1)(100,0)(101,0)(102,0)(103,0)(104,0)(105,0)(106,0)(107,0)(108,0)(109,0)(110,0)(111,0)(112,0)(113,3)(114,0)(115,0)(116,0)(117,0)(118,7)(119,0)(120,0)(121,0)(122,0)(123,0)(124,0)(125,0)(126,0)(127,10)(128,0)(129,1)(130,0)(131,0)(132,0)(133,0)(134,0)(135,0)(136,0)(137,0)(138,0)(139,0)(140,0)(141,3)(142,0)(143,0)(144,0)(145,0)(146,0)(147,5)(148,2)(149,0)(150,0)(151,1)(152,0)(153,1)(154,6)(155,0)(156,1)(157,0)(158,0)(159,2)(160,1)(161,0)(162,0)(163,0)(164,0)(165,0)(166,0)(167,0)(168,1)(169,0)(170,3)(171,0)(172,0)(173,2)(174,0)(175,0)(176,0)(177,0)(178,0)(179,0)(180,2)(181,0)(182,1)(183,0)(184,1)(185,0)(186,0)(187,2)(188,2)(189,0)(190,2)(191,2)(192,0)(193,0)(194,0)(195,2)(196,5)(197,0)(198,2)(199,0)(200,0)(201,5)(202,8)(203,0)(204,0)(205,0)(206,0)(207,0)(208,0)(209,0)(210,0)(211,0)(212,0)(213,0)(214,0)(215,0)(216,0)(217,0)(218,0)(219,14)(220,0)(221,0)(222,1)(223,0)(224,0)(225,0)(226,0)(227,0)(228,0)(229,0)(230,1)(231,1)(232,1)(233,0)(234,1)(235,6)(236,0)(237,0)(238,3)(239,0)(240,1)(241,1)(242,0)(243,0)(244,0)(245,0)(246,0)(247,0)(248,0)(249,0)(250,3)(251,2)(252,0)(253,0)(254,2)(255,0)(256,0)(257,0)(258,0)(259,0)(260,3)(261,23)(262,1)(263,2)(264,0)(265,10)(266,1)(267,0)(268,0)(269,1)(270,1)(271,3)(272,0)(273,0)(274,0)(275,7)(276,20)(277,1)(278,0)(279,0)(280,1)(281,0)(282,10)(283,1)(284,0)(285,5)(286,0)(287,0)(288,12)(289,15)(290,0)(291,0)(292,0)(293,4)(294,0)(295,0)(296,29)(297,0)(298,0)(299,1)(300,0)(301,384)(302,0)(303,0)(304,0)(305,0)(306,0)(307,0)(308,0)(309,0)(310,0)(311,2)(312,12)(313,0)(314,0)(315,2)(316,0)(317,0)(318,0)(319,1)(320,0)(321,0)(322,0)(323,0)(324,0)(325,0)(326,0)(327,0)(328,0)(329,3)(330,0)(331,0)(332,0)(333,0)(334,0)(335,0)(336,0)(337,0)(338,5)(339,3)(340,0)(341,0)(342,37)(343,0)(344,289)(345,4)(346,0)(347,0)(348,0)(349,0)(350,0)(351,0)(352,0)(353,0)(354,0)(355,3)(356,0)(357,0)(358,0)(359,0)(360,0)(361,0)(362,0)(363,0)(364,0)(365,0)(366,1)(367,0)(368,0)(369,0)(370,0)(371,0)(372,0)(373,0)(374,0)(375,0)(376,0)(377,0)(378,0)(379,0)(380,0)(381,0)(382,0)(383,0)(384,7)(385,0)(386,0)(387,0)(388,1)(389,29)(390,0)(391,0)(392,0)(393,0)(394,8)(395,0)(396,3)(397,0)(398,0)(399,0)(400,0)(401,0)(402,0)(403,0)(404,0)(405,0)(406,0)(407,0)(408,0)(409,0)(410,0)(411,0)(412,0)(413,0)(414,0)(415,0)(416,0)(417,0)(418,0)(419,0)(420,0)(421,0)(422,0)(423,0)(424,0)(425,15)(426,0)(427,0)(428,0)(429,0)(430,0)(431,0)(432,0)(433,1)(434,3)(435,0)(436,0)(437,0)(438,1)(439,2)(440,0)(441,0)(442,3)(443,6)(444,0)(445,0)(446,0)(447,0)(448,0)(449,0)(450,1)(451,0)(452,0)(453,0)(454,0)(455,0)(456,4)(457,0)(458,0)(459,0)(460,0)(461,1)(462,2)(463,2)(464,3)(465,0)(466,0)(467,0)(468,0)(469,4)(470,0)(471,0)(472,0)(473,5)(474,2)(475,0)(476,0)(477,0)(478,0)(479,0)(480,0)(481,2)(482,0)(483,0)(484,1)(485,0)(486,0)(487,0)(488,0)(489,0)(490,0)(491,0)(492,0)(493,2)(494,10)(495,1)(496,133)(497,0)(498,9)(499,0)(500,0)(501,0)(502,0)(503,42)(504,0)(505,1)(506,0)(507,0)(508,0)(509,0)(510,11)(511,9)(512,0)(513,0)(514,0)(515,0)(516,1)(517,1)(518,1)(519,1)(520,1)(521,1)(522,0)(523,0)(524,0)(525,0)(526,0)(527,0)(528,0)(529,0)(530,0)(531,0)(532,7)(533,1)(534,0)(535,0)(536,1)(537,0)(538,1)(539,9)(540,0)(541,5)(542,0)(543,0)(544,3)(545,0)(546,0)(547,0)(548,4)(549,3)(550,0)(551,0)(552,0)(553,1)(554,0)(555,0)(556,0)(557,0)(558,0)(559,0)(560,6)(561,0)(562,0)(563,0)(564,0)(565,0)(566,0)(567,0)(568,0)(569,0)(570,0)(571,0)(572,0)(573,0)(574,0)(575,0)(576,0)(577,0)(578,1)(579,0)(580,0)(581,0)(582,1)(583,0)(584,0)(585,0)(586,0)(587,0)(588,0)(589,0)(590,0)(591,3)(592,1)(593,0)(594,0)(595,0)(596,0)(597,1)(598,0)(599,0)(600,0)(601,0)(602,0)(603,0)(604,0)(605,0)(606,0)(607,0)(608,0)(609,0)(610,0)(611,5)(612,0)(613,0)(614,1)(615,0)(616,0)(617,0)(618,1)(619,0)(620,0)(621,0)(622,0)(623,0)(624,0)(625,0)(626,0)(627,0)(628,0)(629,0)(630,0)(631,0)(632,0)(633,0)(634,0)(635,0)(636,0)(637,0)(638,1)(639,1)(640,0)(641,0)(642,0)(643,1)(644,0)(645,0)(646,0)(647,0)(648,7)(649,0)(650,0)(651,0)(652,0)(653,2)(654,0)(655,0)(656,0)(657,0)(658,0)(659,0)(660,0)(661,0)(662,3)(663,0)(664,0)(665,0)(666,0)(667,0)(668,0)(669,0)(670,2)(671,0)(672,0)(673,1)(674,0)(675,1)(676,6)(677,0)(678,1)(679,2)(680,1)(681,0)(682,0)(683,0)(684,0)(685,0)(686,0)(687,0)(688,0)(689,0)(690,0)(691,0)(692,0)(693,0)(694,0)(695,0)(696,0)(697,0)(698,0)(699,18)(700,0)(701,0)(702,0)(703,0)(704,0)(705,0)(706,0)(707,0)(708,0)(709,0)(710,4)(711,0)(712,0)(713,0)(714,0)(715,0)(716,0)(717,0)(718,0)(719,0)(720,0)(721,0)(722,0)(723,0)(724,2)(725,0)(726,0)(727,0)(728,0)(729,0)(730,2)(731,0)(732,0)(733,2)(734,2)(735,0)(736,0)(737,0)(738,0)(739,6)(740,0)(741,5)(742,1)(743,0)(744,1)(745,1)(746,0)(747,0)(748,1)(749,0)(750,0)(751,2)(752,0)(753,0)(754,0)(755,0)(756,1)(757,0)(758,0)(759,0)(760,0)(761,0)(762,0)(763,0)(764,0)(765,0)(766,0)(767,1)(768,0)(769,0)(770,0)(771,0)(772,0)(773,0)(774,0)(775,0)(776,0)(777,0)(778,0)(779,0)(780,2)(781,0)(782,0)(783,0)(784,1)(785,0)(786,0)(787,0)(788,0)(789,0)(790,0)(791,1)(792,0)(793,1)(794,0)(795,0)(796,0)(797,0)(798,0)(799,1)(800,0)(801,0)(802,0)(803,0)(804,0)(805,1)(806,0)(807,1)(808,0)(809,0)(810,0)(811,0)(812,0)(813,0)(814,0)(815,0)(816,0)(817,0)(818,0)(819,0)(820,0)(821,4)(822,0)(823,0)(824,0)(825,0)(826,0)(827,0)(828,0)(829,11)(830,0)(831,0)(832,0)(833,1)(834,0)(835,0)(836,0)(837,1)(838,0)(839,0)(840,0)(841,0)(842,0)(843,0)(844,0)(845,0)(846,0)(847,0)(848,0)(849,0)(850,0)(851,0)(852,0)(853,0)(854,4)(855,0)(856,0)(857,0)(858,0)(859,0)(860,0)(861,0)(862,0)(863,0)(864,0)(865,0)(866,0)(867,0)(868,0)(869,0)(870,0)(871,0)(872,0)(873,0)(874,2)(875,0)(876,4)(877,0)(878,0)(879,0)(880,3)(881,4)(882,0)(883,0)(884,18)(885,0)(886,0)(887,0)(888,0)(889,0)(890,0)(891,0)(892,0)(893,0)(894,0)(895,0)(896,0)(897,2)(898,0)(899,0)(900,0)(901,0)(902,0)(903,0)(904,0)(905,0)(906,16)(907,4)(908,0)(909,0)(910,0)(911,0)(912,0)(913,2)(914,1)(915,0)(916,0)(917,0)(918,0)(919,0)(920,14)(921,0)(922,2)(923,0)(924,1)(925,1)(926,2)(927,0)(928,0)(929,9)(930,1)(931,3)(932,6)(933,0)(934,0)(935,0)(936,0)(937,0)(938,0)(939,0)(940,0)(941,0)(942,0)(943,1)(944,0)(945,1)(946,0)(947,7)(948,0)(949,0)(950,0)(951,0)(952,2)(953,0)(954,2)(955,0)(956,1)(957,0)(958,0)(959,0)(960,0)(961,3)(962,0)(963,0)(964,0)(965,0)(966,4)(967,0)(968,0)(969,16)(970,1)(971,0)(972,0)(973,0)(974,0)(975,0)(976,0)(977,0)(978,1)(979,1)(980,0)(981,0)(982,0)(983,0)(984,2)(985,4)(986,0)(987,0)(988,0)(989,0)(990,0)(991,0)(992,0)(993,0)(994,0)(995,14)(996,0)(997,0)(998,9)(999,0)(1000,0)(1001,0)(1002,1)(1003,0)(1004,0)(1005,0)(1006,0)(1007,0)(1008,0)(1009,7)(1010,0)(1011,0)(1012,0)(1013,0)(1014,0)(1015,0)(1016,0)(1017,0)(1018,0)(1019,2)(1020,0)(1021,0)(1022,0)(1023,0)(1024,0)(1025,0)(1026,0)(1027,0)(1028,0)(1029,0)(1030,0)(1031,0)(1032,2)(1033,0)(1034,0)(1035,3)(1036,0)(1037,0)(1038,4)(1039,0)(1040,25)(1041,0)(1042,0)(1043,1)(1044,0)(1045,0)(1046,0)(1047,0)(1048,11)(1049,0)(1050,0)(1051,0)(1052,1)(1053,0)(1054,0)(1055,0)(1056,0)(1057,3)(1058,0)(1059,0)(1060,0)(1061,0)(1062,0)(1063,0)(1064,0)(1065,0)(1066,0)(1067,21)(1068,18)(1069,2)(1070,0)(1071,0)(1072,0)(1073,0)(1074,0)(1075,0)(1076,0)(1077,0)(1078,0)(1079,0)(1080,0)(1081,0)(1082,0)(1083,0)(1084,0)(1085,0)(1086,0)(1087,40)(1088,0)(1089,0)(1090,0)(1091,0)(1092,0)(1093,0)(1094,0)(1095,0)(1096,0)(1097,3)(1098,0)(1099,0)(1100,0)(1101,0)(1102,0)(1103,12)(1104,6)(1105,6)(1106,0)(1107,0)(1108,0)(1109,108)(1110,0)(1111,16)(1112,0)(1113,2)(1114,0)(1115,0)(1116,24)(1117,0)(1118,0)(1119,0)(1120,0)(1121,0)(1122,0)(1123,0)(1124,0)(1125,0)(1126,0)(1127,0)(1128,0)(1129,0)(1130,0)(1131,0)(1132,0)(1133,3)(1134,0)(1135,0)(1136,0)(1137,0)(1138,5)(1139,0)(1140,0)(1141,0)(1142,0)(1143,0)(1144,5)(1145,0)(1146,0)(1147,0)(1148,3)(1149,0)(1150,0)(1151,0)(1152,0)(1153,7)(1154,0)(1155,0)(1156,0)(1157,0)(1158,0)(1159,0)(1160,0)(1161,0)(1162,7)(1163,1)(1164,0)(1165,0)(1166,0)(1167,10)(1168,0)(1169,0)(1170,0)(1171,0)(1172,0)(1173,0)(1174,1188)(1175,0)(1176,0)(1177,0)(1178,3)(1179,1)(1180,0)(1181,0)(1182,0)(1183,0)(1184,0)(1185,0)(1186,0)(1187,1372)(1188,11)(1189,3)(1190,0)(1191,0)(1192,20)(1193,21)(1194,0)(1195,55)(1196,0)(1197,3)(1198,0)(1199,0)(1200,0)(1201,0)(1202,0)(1203,0)(1204,2)(1205,996)(1206,8)(1207,0)(1208,0)(1209,0)(1210,0)(1211,0)(1212,0)(1213,0)(1214,0)(1215,0)(1216,0)(1217,0)(1218,0)(1219,0)(1220,0)(1221,5)(1222,0)(1223,0)(1224,0)(1225,0)(1226,0)(1227,2)(1228,0)(1229,0)(1230,35)(1231,0)(1232,0)(1233,0)(1234,0)(1235,0)(1236,13)(1237,0)(1238,0)(1239,12)(1240,0)(1241,4)(1242,189)(1243,2)(1244,46)(1245,955)(1246,2)(1247,174)(1248,203)(1249,2465)(1250,21)(1251,2)(1252,0)(1253,258)(1254,10)(1255,108)(1256,355)(1257,36)(1258,0)(1259,163)(1260,661)(1261,0)(1262,0)(1263,0)(1264,0)(1265,0)(1266,34)(1267,0)(1268,10)(1269,0)(1270,1)(1271,112)(1272,1)(1273,7)(1274,7)(1275,7)(1276,0)(1277,0)(1278,163)(1279,0)(1280,46)(1281,62)(1282,0)(1283,189)(1284,0)(1285,22)(1286,44)(1287,647)(1288,5)(1289,0)(1290,0)(1291,0)(1292,0)(1293,13)(1294,5)(1295,0)(1296,0)(1297,20)(1298,75)(1299,0)(1300,0)(1301,797)(1302,0)(1303,0)(1304,70)(1305,0)(1306,0)(1307,0)(1308,35)(1309,0)(1310,355)(1311,0)(1312,71)(1313,0)(1314,57)(1315,0)(1316,0)(1317,25)(1318,2)(1319,0)(1320,0)(1321,31)(1322,0)(1323,0)(1324,0)(1325,0)(1326,25)(1327,0)(1328,0)(1329,0)(1330,0)(1331,14)(1332,0)(1333,0)(1334,0)(1335,25)(1336,41)(1337,135)(1338,99)(1339,85)(1340,19)(1341,0)(1342,19)(1343,0)(1344,20)(1345,38)(1346,195)(1347,71)(1348,6)(1349,1)(1350,0)(1351,76)(1352,0)(1353,872)(1354,0)(1355,0)(1356,0)(1357,0)(1358,0)(1359,0)(1360,0)(1361,56)(1362,19)(1363,343)(1364,19)(1365,147)(1366,0)(1367,6)(1368,2)(1369,0)(1370,0)(1371,0)(1372,0)(1373,1)(1374,15)(1375,0)(1376,2096)(1377,0)(1378,0)(1379,1)(1380,2)(1381,1)(1382,1)(1383,6)(1384,7)(1385,6)(1386,28)(1387,0)(1388,17)(1389,25)(1390,0)(1391,3)(1392,1)(1393,1)(1394,0)(1395,0)(1396,3)(1397,1)(1398,1)(1399,0)(1400,0)(1401,0)(1402,0)(1403,0)(1404,0)(1405,4)(1406,0)(1407,0)(1408,0)(1409,0)(1410,0)(1411,0)(1412,0)(1413,0)(1414,0)(1415,0)(1416,0)(1417,0)(1418,2)(1419,0)(1420,15)(1421,1115)(1422,4)(1423,91)(1424,194)(1425,588)(1426,0)(1427,0)(1428,10)(1429,0)(1430,11)(1431,0)(1432,0)(1433,4)(1434,9)(1435,0)(1436,0)(1437,0)(1438,0)(1439,0)(1440,0)(1441,0)(1442,0)(1443,0)(1444,0)(1445,2)(1446,0)(1447,2)(1448,0)(1449,7)(1450,0)(1451,18)(1452,0)(1453,0)(1454,0)(1455,591)(1456,37)(1457,0)(1458,0)(1459,0)(1460,0)(1461,0)(1462,8)(1463,0)(1464,0)(1465,0)(1466,0)(1467,0)(1468,0)(1469,0)(1470,0)(1471,3)(1472,4)(1473,0)(1474,0)(1475,0)(1476,14)(1477,16)(1478,0)(1479,0)(1480,4)(1481,0)(1482,0)(1483,0)(1484,5)(1485,1203)(1486,432)(1487,12)(1488,213)(1489,2)(1490,0)(1491,12)(1492,43)(1493,0)(1494,76)(1495,0)(1496,16)(1497,0)(1498,18)(1499,0)(1500,0)(1501,0)(1502,0)(1503,0)(1504,0)(1505,0)(1506,0)(1507,10)(1508,0)(1509,21)(1510,63)(1511,0)(1512,0)(1513,0)(1514,0)(1515,0)(1516,0)(1517,234)(1518,19)(1519,10)(1520,456)(1521,456)(1522,0)(1523,0)(1524,0)(1525,0)(1526,0)(1527,0)(1528,0)(1529,0)(1530,0)(1531,0)(1532,9)(1533,0)(1534,0)(1535,0)(1536,0)(1537,0)(1538,7)(1539,0)(1540,1)(1541,1)(1542,0)(1543,7657)(1544,0)(1545,83)(1546,0)(1547,6)(1548,22)(1549,0)(1550,0)(1551,0)(1552,0)(1553,2)(1554,0)(1555,70)(1556,0)(1557,136)(1558,1)(1559,0)(1560,0)(1561,0)(1562,196)(1563,134)(1564,2)(1565,0)(1566,2)(1567,576)(1568,0)(1569,688)(1570,0)(1571,0)(1572,0)(1573,0)(1574,0)(1575,84)(1576,0)(1577,42)(1578,1)(1579,48)(1580,0)(1581,13)(1582,119)(1583,113)(1584,178)(1585,27)(1586,67)(1587,452)(1588,31)(1589,49)(1590,84)(1591,128)(1592,18)(1593,14)(1594,0)(1595,72)(1596,0)(1597,0)(1598,0)(1599,0)(1600,2)(1601,0)(1602,283)(1603,28)(1604,0)(1605,0)(1606,0)(1607,0)(1608,0)(1609,0)(1610,0)(1611,0)(1612,0)(1613,0)(1614,0)(1615,0)(1616,0)(1617,105)(1618,0)(1619,0)(1620,0)(1621,0)(1622,0)(1623,0)(1624,0)(1625,3)(1626,0)(1627,0)(1628,0)(1629,0)(1630,0)(1631,0)(1632,0)(1633,12)(1634,0)(1635,0)(1636,2)(1637,0)(1638,0)(1639,0)(1640,0)(1641,26)(1642,20)(1643,26)(1644,0)(1645,8)(1646,0)(1647,0)(1648,0)(1649,0)(1650,0)(1651,0)(1652,0)(1653,0)(1654,0)(1655,0)(1656,0)(1657,0)(1658,0)(1659,0)(1660,8)(1661,0)(1662,0)(1663,1)(1664,0)(1665,0)(1666,0)(1667,0)(1668,0)(1669,0)(1670,0)(1671,0)(1672,0)(1673,0)(1674,1)(1675,0)(1676,0)(1677,0)(1678,0)(1679,0)(1680,0)(1681,8)(1682,20)(1683,47)(1684,10)(1685,31)(1686,7)(1687,0)(1688,0)(1689,0)(1690,13)(1691,6)(1692,4)(1693,3)(1694,0)(1695,0)(1696,0)(1697,0)(1698,0)(1699,0)(1700,0)(1701,0)(1702,8)(1703,0)(1704,0)(1705,10)(1706,0)(1707,3)(1708,5)(1709,0)(1710,0)(1711,3)(1712,1)(1713,0)(1714,1)(1715,6)(1716,0)(1717,0)(1718,0)(1719,0)(1720,0)(1721,0)(1722,10)(1723,0)(1724,0)(1725,0)(1726,1)(1727,0)(1728,0)(1729,4)(1730,4)(1731,0)(1732,8)(1733,0)(1734,7)(1735,0)(1736,0)(1737,0)(1738,0)(1739,0)(1740,0)(1741,0)(1742,0)(1743,0)(1744,0)(1745,0)(1746,0)(1747,0)(1748,0)(1749,0)(1750,0)(1751,0)(1752,0)(1753,0)(1754,0)(1755,0)(1756,0)(1757,0)(1758,0)(1759,0)(1760,3)(1761,0)(1762,0)(1763,0)(1764,3)(1765,0)(1766,0)(1767,0)(1768,0)(1769,0)(1770,0)(1771,0)(1772,0)(1773,0)(1774,0)(1775,0)(1776,0)(1777,0)(1778,0)(1779,0)(1780,0)(1781,0)(1782,0)(1783,0)(1784,0)(1785,0)(1786,0)(1787,0)(1788,7)(1789,0)(1790,0)(1791,1)(1792,2)(1793,0)(1794,0)(1795,0)(1796,0)(1797,0)(1798,1)(1799,0)(1800,2)(1801,13)(1802,0)(1803,0)(1804,3)(1805,1)(1806,1)(1807,0)(1808,0)(1809,0)(1810,5)(1811,27)(1812,0)(1813,0)(1814,0)(1815,0)(1816,0)(1817,0)(1818,0)(1819,0)(1820,0)(1821,0)(1822,0)(1823,0)(1824,0)(1825,1)(1826,0)(1827,0)(1828,0)(1829,0)(1830,0)(1831,0)(1832,7)(1833,2)(1834,1)(1835,0)(1836,0)(1837,0)(1838,0)(1839,0)(1840,0)(1841,0)(1842,0)(1843,0)(1844,0)(1845,0)(1846,0)(1847,0)(1848,0)(1849,6)(1850,4)(1851,0)(1852,0)(1853,6)(1854,5)(1855,0)(1856,0)(1857,14)(1858,1)(1859,25)(1860,0)(1861,0)(1862,1)(1863,0)(1864,0)(1865,0)(1866,0)(1867,1)(1868,0)(1869,17)(1870,3)(1871,0)(1872,0)(1873,0)(1874,0)(1875,0)(1876,0)(1877,15)(1878,1)(1879,2)(1880,0)(1881,0)(1882,0)(1883,4)(1884,1)(1885,4)(1886,0)(1887,0)(1888,3)(1889,0)(1890,0)(1891,23)(1892,0)(1893,0)(1894,3)(1895,12)(1896,0)(1897,12)(1898,0)(1899,0)(1900,3)(1901,2)(1902,0)(1903,7)(1904,0)(1905,1)(1906,0)(1907,4)(1908,2)(1909,0)(1910,0)(1911,0)(1912,0)(1913,0)(1914,0)(1915,0)(1916,0)(1917,0)(1918,0)(1919,0)(1920,0)(1921,0)(1922,0)(1923,0)(1924,0)(1925,0)(1926,0)(1927,0)(1928,0)(1929,0)(1930,0)(1931,0)(1932,0)(1933,0)(1934,0)(1935,0)(1936,0)(1937,0)(1938,0)(1939,0)(1940,0)(1941,0)(1942,0)(1943,0)(1944,0)(1945,0)(1946,0)(1947,0)(1948,0)(1949,0)(1950,0)(1951,0)(1952,0)(1953,0)(1954,0)(1955,0)(1956,0)(1957,0)(1958,0)(1959,0)(1960,0)(1961,0)(1962,0)(1963,0)(1964,0)(1965,0)(1966,0)(1967,0)(1968,0)(1969,0)(1970,0)(1971,0)(1972,0)(1973,0)(1974,0)(1975,0)(1976,0)(1977,0)(1978,0)(1979,0)(1980,0)(1981,0)(1982,0)(1983,0)(1984,16)(1985,0)(1986,0)(1987,0)(1988,0)(1989,0)(1990,0)(1991,0)(1992,0)(1993,0)(1994,0)(1995,0)(1996,7)(1997,0)(1998,9)(1999,0)(2000,0)(2001,6)(2002,0)(2003,0)(2004,0)(2005,0)(2006,0)(2007,0)(2008,0)(2009,0)(2010,0)(2011,0)(2012,0)(2013,0)(2014,0)(2015,0)(2016,0)(2017,0)(2018,0)(2019,0)(2020,0)(2021,0)(2022,0)(2023,0)(2024,6)(2025,0)(2026,3)(2027,0)(2028,0)(2029,0)(2030,0)(2031,1)(2032,0)(2033,0)(2034,0)(2035,0)(2036,0)(2037,0)(2038,0)(2039,0)(2040,0)(2041,0)(2042,0)(2043,8)(2044,29)(2045,14)(2046,3)(2047,0)(2048,0)(2049,3)(2050,0)(2051,1)(2052,0)(2053,2)(2054,0)(2055,2)(2056,0)(2057,2)(2058,0)(2059,0)(2060,0)(2061,2)(2062,0)(2063,10)(2064,0)(2065,0)(2066,209)(2067,0)(2068,0)(2069,0)(2070,0)(2071,0)(2072,6)(2073,0)(2074,4)(2075,35)(2076,85)(2077,0)(2078,2)(2079,0)(2080,0)(2081,0)(2082,0)(2083,0)(2084,0)(2085,1)(2086,89)(2087,15)(2088,1)(2089,36)(2090,0)(2091,0)(2092,3)(2093,0)(2094,15)(2095,15)(2096,0)(2097,0)(2098,13)(2099,1)(2100,190)(2101,95)(2102,0)(2103,0)(2104,93)(2105,0)(2106,0)(2107,1)(2108,56)(2109,15)(2110,0)(2111,0)(2112,0)(2113,99)(2114,158)(2115,0)(2116,157)(2117,178)(2118,0)(2119,84)(2120,60)(2121,60)(2122,14)(2123,14)(2124,28)(2125,42)(2126,14)(2127,0)(2128,22)(2129,0)(2130,0)(2131,14)(2132,14)(2133,0)(2134,14)(2135,15)(2136,2)(2137,4)(2138,2)(2139,0)(2140,0)(2141,0)(2142,0)(2143,6)(2144,4)(2145,0)(2146,0)(2147,0)(2148,0)(2149,0)(2150,0)(2151,0)(2152,16)(2153,0)(2154,0)(2155,0)(2156,0)(2157,14)(2158,0)(2159,0)(2160,0)(2161,1)(2162,2)(2163,0)(2164,0)(2165,0)(2166,0)(2167,0)(2168,0)(2169,0)(2170,0)(2171,0)(2172,0)(2173,0)(2174,5)(2175,0)(2176,0)(2177,2)(2178,0)(2179,2)(2180,2)(2181,0)(2182,0)(2183,0)(2184,3)(2185,15)(2186,0)(2187,0)(2188,0)(2189,0)(2190,0)(2191,0)(2192,0)(2193,0)(2194,0)(2195,0)(2196,1)(2197,0)(2198,0)(2199,0)(2200,0)(2201,0)(2202,0)(2203,0)(2204,0)(2205,0)(2206,0)(2207,0)(2208,0)(2209,0)(2210,0)(2211,0)(2212,0)(2213,0)(2214,0)(2215,0)(2216,0)(2217,0)(2218,0)(2219,1)(2220,0)(2221,0)(2222,0)(2223,0)(2224,0)(2225,0)(2226,0)(2227,0)(2228,0)(2229,0)(2230,0)(2231,0)(2232,0)(2233,5)(2234,0)(2235,0)(2236,0)(2237,0)(2238,0)(2239,253)(2240,16)(2241,0)(2242,0)(2243,0)(2244,0)(2245,0)(2246,5)(2247,3)(2248,0)(2249,0)(2250,0)(2251,0)(2252,1)(2253,3)(2254,0)(2255,0)(2256,0)(2257,0)(2258,0)(2259,0)(2260,0)(2261,1)(2262,0)(2263,0)(2264,0)(2265,0)(2266,0)(2267,0)(2268,0)(2269,0)(2270,2)(2271,0)(2272,3)(2273,0)(2274,0)(2275,0)(2276,0)(2277,0)(2278,5)(2279,0)(2280,5)(2281,0)(2282,0)(2283,0)(2284,0)(2285,0)(2286,0)(2287,4)(2288,0)(2289,0)(2290,0)(2291,0)(2292,0)(2293,3)(2294,3)(2295,0)(2296,0)(2297,0)(2298,0)(2299,0)(2300,0)(2301,0)(2302,0)(2303,2)(2304,5)(2305,0)(2306,0)(2307,0)(2308,6)(2309,0)(2310,0)(2311,1)(2312,0)(2313,1)(2314,14)(2315,0)(2316,0)(2317,3)(2318,0)(2319,0)(2320,0)(2321,0)(2322,0)(2323,0)(2324,0)(2325,0)(2326,0)(2327,0)(2328,0)(2329,0)(2330,0)(2331,0)(2332,0)(2333,0)(2334,0)(2335,0)(2336,2)(2337,3)(2338,1)(2339,0)(2340,0)(2341,0)(2342,0)(2343,2)(2344,0)(2345,0)(2346,0)(2347,0)(2348,0)(2349,1)(2350,0)(2351,0)(2352,3)(2353,6)(2354,0)(2355,0)(2356,0)(2357,0)(2358,0)(2359,1)(2360,0)(2361,0)(2362,1)(2363,0)(2364,0)(2365,0)(2366,5)(2367,0)(2368,0)(2369,0)(2370,0)(2371,0)(2372,0)(2373,0)(2374,0)(2375,0)(2376,9)(2377,0)(2378,0)(2379,0)(2380,0)(2381,1)(2382,0)(2383,0)(2384,0)(2385,0)(2386,0)(2387,0)(2388,0)(2389,0)(2390,3)(2391,4)(2392,0)(2393,0)(2394,0)(2395,0)(2396,0)(2397,0)(2398,4)(2399,0)(2400,0)(2401,0)(2402,0)(2403,0)(2404,0)(2405,0)(2406,0)(2407,0)(2408,0)(2409,1)(2410,7)(2411,0)(2412,0)(2413,0)(2414,14)(2415,0)(2416,0)(2417,0)(2418,0)(2419,0)
};
\addlegendentry{MD5}
 
\end{axis}
\end{tikzpicture}%

%% file: ent-rsa.tex
\begin{tikzpicture}[scale=0.5]

\begin{axis}[
    xtick={0,1000,2000,3000,4000,5000},
    legend pos=north east,
    ymajorgrids=true,
    grid style=dashed,
]

\addplot[color=blue]
coordinates {
	(0,0)(1,0)(2,1)(3,9)(4,0)(5,0)(6,0)(7,0)(8,0)(9,0)(10,0)(11,2)(12,0)(13,0)(14,2)(15,2)(16,0)(17,0)(18,0)(19,0)(20,0)(21,0)(22,0)(23,0)(24,0)(25,0)(26,9)(27,0)(28,0)(29,0)(30,0)(31,0)(32,0)(33,0)(34,4)(35,0)(36,0)(37,20)(38,13)(39,4)(40,1)(41,0)(42,0)(43,0)(44,0)(45,1)(46,4)(47,0)(48,0)(49,3)(50,0)(51,0)(52,0)(53,0)(54,2)(55,3)(56,0)(57,1)(58,0)(59,0)(60,0)(61,2)(62,1)(63,0)(64,0)(65,16)(66,0)(67,0)(68,0)(69,0)(70,0)(71,0)(72,8)(73,0)(74,8)(75,4)(76,0)(77,0)(78,0)(79,1)(80,6)(81,3)(82,2)(83,15)(84,3)(85,0)(86,0)(87,7)(88,0)(89,3)(90,2)(91,0)(92,2)(93,0)(94,0)(95,0)(96,1)(97,0)(98,0)(99,1)(100,0)(101,0)(102,0)(103,0)(104,0)(105,0)(106,0)(107,0)(108,0)(109,0)(110,0)(111,0)(112,0)(113,3)(114,0)(115,0)(116,0)(117,0)(118,7)(119,0)(120,0)(121,0)(122,0)(123,0)(124,0)(125,0)(126,0)(127,10)(128,0)(129,1)(130,0)(131,0)(132,0)(133,0)(134,0)(135,0)(136,0)(137,0)(138,0)(139,0)(140,0)(141,8)(142,0)(143,0)(144,0)(145,0)(146,0)(147,2)(148,0)(149,0)(150,0)(151,1)(152,0)(153,0)(154,0)(155,0)(156,2)(157,0)(158,0)(159,2)(160,1)(161,0)(162,0)(163,0)(164,0)(165,0)(166,0)(167,0)(168,1)(169,0)(170,3)(171,0)(172,0)(173,2)(174,0)(175,0)(176,0)(177,0)(178,0)(179,0)(180,2)(181,0)(182,1)(183,0)(184,1)(185,0)(186,0)(187,2)(188,2)(189,0)(190,2)(191,2)(192,0)(193,0)(194,0)(195,2)(196,5)(197,0)(198,0)(199,0)(200,1)(201,5)(202,8)(203,0)(204,0)(205,0)(206,0)(207,0)(208,0)(209,0)(210,0)(211,0)(212,0)(213,0)(214,0)(215,0)(216,0)(217,0)(218,0)(219,14)(220,0)(221,0)(222,1)(223,0)(224,0)(225,0)(226,0)(227,0)(228,0)(229,0)(230,1)(231,1)(232,1)(233,0)(234,1)(235,6)(236,0)(237,0)(238,3)(239,0)(240,1)(241,1)(242,0)(243,0)(244,0)(245,0)(246,0)(247,0)(248,0)(249,0)(250,3)(251,2)(252,0)(253,0)(254,2)(255,0)(256,0)(257,0)(258,0)(259,0)(260,3)(261,23)(262,1)(263,2)(264,0)(265,10)(266,1)(267,0)(268,0)(269,1)(270,1)(271,3)(272,0)(273,0)(274,14)(275,2)(276,18)(277,1)(278,0)(279,0)(280,1)(281,0)(282,10)(283,1)(284,0)(285,5)(286,0)(287,0)(288,12)(289,15)(290,0)(291,0)(292,0)(293,4)(294,0)(295,0)(296,29)(297,0)(298,0)(299,1)(300,0)(301,5)(302,0)(303,0)(304,0)(305,0)(306,3)(307,0)(308,0)(309,0)(310,0)(311,0)(312,0)(313,0)(314,1)(315,5)(316,0)(317,0)(318,4)(319,1)(320,0)(321,0)(322,0)(323,0)(324,0)(325,0)(326,0)(327,0)(328,0)(329,3)(330,0)(331,0)(332,0)(333,0)(334,0)(335,0)(336,0)(337,0)(338,5)(339,3)(340,0)(341,0)(342,33)(343,0)(344,302)(345,4)(346,0)(347,0)(348,0)(349,0)(350,0)(351,0)(352,0)(353,0)(354,0)(355,3)(356,0)(357,0)(358,0)(359,0)(360,0)(361,0)(362,0)(363,0)(364,0)(365,0)(366,1)(367,0)(368,0)(369,0)(370,0)(371,0)(372,0)(373,0)(374,0)(375,0)(376,0)(377,0)(378,0)(379,0)(380,0)(381,0)(382,0)(383,0)(384,7)(385,0)(386,0)(387,0)(388,1)(389,25)(390,0)(391,0)(392,0)(393,0)(394,8)(395,0)(396,3)(397,0)(398,0)(399,0)(400,0)(401,0)(402,0)(403,0)(404,0)(405,0)(406,0)(407,0)(408,0)(409,0)(410,0)(411,0)(412,0)(413,0)(414,0)(415,0)(416,0)(417,0)(418,0)(419,0)(420,0)(421,0)(422,0)(423,0)(424,0)(425,15)(426,0)(427,0)(428,0)(429,0)(430,0)(431,0)(432,0)(433,1)(434,4)(435,0)(436,0)(437,0)(438,0)(439,0)(440,0)(441,0)(442,3)(443,5)(444,0)(445,0)(446,0)(447,0)(448,0)(449,0)(450,1)(451,0)(452,0)(453,0)(454,0)(455,0)(456,3)(457,0)(458,0)(459,0)(460,0)(461,1)(462,2)(463,2)(464,3)(465,0)(466,0)(467,0)(468,0)(469,4)(470,0)(471,0)(472,0)(473,1)(474,1)(475,0)(476,0)(477,0)(478,0)(479,0)(480,0)(481,2)(482,0)(483,0)(484,1)(485,0)(486,0)(487,0)(488,0)(489,0)(490,0)(491,0)(492,0)(493,2)(494,0)(495,1)(496,133)(497,0)(498,13)(499,0)(500,0)(501,0)(502,0)(503,9)(504,0)(505,0)(506,0)(507,0)(508,0)(509,0)(510,10)(511,8)(512,0)(513,0)(514,0)(515,0)(516,1)(517,1)(518,1)(519,1)(520,1)(521,1)(522,0)(523,0)(524,0)(525,0)(526,0)(527,0)(528,0)(529,0)(530,0)(531,0)(532,0)(533,1)(534,0)(535,0)(536,1)(537,0)(538,1)(539,9)(540,0)(541,5)(542,0)(543,0)(544,3)(545,0)(546,1)(547,0)(548,0)(549,1)(550,0)(551,0)(552,0)(553,4)(554,0)(555,0)(556,0)(557,0)(558,0)(559,0)(560,0)(561,0)(562,0)(563,0)(564,0)(565,0)(566,0)(567,0)(568,0)(569,0)(570,5)(571,0)(572,0)(573,1)(574,0)(575,0)(576,0)(577,0)(578,1)(579,1)(580,0)(581,0)(582,0)(583,0)(584,0)(585,0)(586,4)(587,0)(588,0)(589,2)(590,0)(591,0)(592,0)(593,0)(594,0)(595,0)(596,0)(597,0)(598,0)(599,0)(600,0)(601,0)(602,0)(603,0)(604,0)(605,0)(606,1)(607,0)(608,3)(609,0)(610,1)(611,0)(612,26)(613,0)(614,0)(615,1)(616,0)(617,0)(618,0)(619,1)(620,0)(621,0)(622,0)(623,0)(624,0)(625,0)(626,0)(627,0)(628,0)(629,0)(630,0)(631,0)(632,0)(633,0)(634,0)(635,0)(636,0)(637,0)(638,0)(639,0)(640,0)(641,0)(642,0)(643,2)(644,0)(645,0)(646,0)(647,0)(648,0)(649,0)(650,0)(651,0)(652,0)(653,0)(654,0)(655,1)(656,0)(657,2)(658,0)(659,0)(660,1)(661,0)(662,0)(663,0)(664,0)(665,0)(666,0)(667,1)(668,0)(669,0)(670,0)(671,1)(672,1)(673,2)(674,7)(675,0)(676,0)(677,0)(678,1)(679,2)(680,1)(681,0)(682,0)(683,0)(684,0)(685,0)(686,1)(687,0)(688,0)(689,1)(690,1)(691,0)(692,0)(693,0)(694,0)(695,14)(696,0)(697,0)(698,0)(699,36)(700,0)(701,0)(702,0)(703,0)(704,0)(705,0)(706,0)(707,0)(708,0)(709,2)(710,0)(711,0)(712,0)(713,0)(714,0)(715,0)(716,1)(717,0)(718,1)(719,2)(720,0)(721,0)(722,0)(723,0)(724,0)(725,2)(726,0)(727,0)(728,6)(729,2)(730,0)(731,0)(732,0)(733,0)(734,0)(735,0)(736,1)(737,5)(738,7)(739,0)(740,10)(741,0)(742,2)(743,0)(744,4)(745,0)(746,0)(747,0)(748,2)(749,35)(750,0)(751,0)(752,0)(753,0)(754,0)(755,0)(756,0)(757,0)(758,0)(759,0)(760,0)(761,0)(762,1)(763,0)(764,1)(765,0)(766,0)(767,0)(768,0)(769,0)(770,0)(771,0)(772,0)(773,0)(774,0)(775,0)(776,0)(777,0)(778,0)(779,0)(780,2)(781,0)(782,0)(783,0)(784,0)(785,0)(786,0)(787,0)(788,0)(789,0)(790,0)(791,0)(792,0)(793,0)(794,0)(795,0)(796,0)(797,0)(798,0)(799,0)(800,0)(801,0)(802,0)(803,0)(804,0)(805,0)(806,0)(807,0)(808,0)(809,0)(810,0)(811,0)(812,0)(813,0)(814,1)(815,0)(816,0)(817,0)(818,0)(819,0)(820,0)(821,0)(822,1)(823,0)(824,0)(825,0)(826,0)(827,0)(828,0)(829,0)(830,0)(831,0)(832,0)(833,0)(834,0)(835,0)(836,0)(837,0)(838,0)(839,0)(840,0)(841,0)(842,0)(843,1)(844,2)(845,2)(846,0)(847,0)(848,0)(849,5)(850,0)(851,0)(852,0)(853,0)(854,0)(855,0)(856,2)(857,0)(858,0)(859,0)(860,0)(861,0)(862,0)(863,2)(864,0)(865,0)(866,0)(867,0)(868,1)(869,2)(870,0)(871,0)(872,1)(873,0)(874,0)(875,0)(876,0)(877,0)(878,3)(879,0)(880,0)(881,0)(882,0)(883,0)(884,0)(885,0)(886,0)(887,0)(888,0)(889,0)(890,0)(891,2)(892,0)(893,0)(894,0)(895,0)(896,3)(897,0)(898,0)(899,0)(900,0)(901,0)(902,2)(903,0)(904,0)(905,0)(906,0)(907,0)(908,0)(909,0)(910,0)(911,3)(912,0)(913,1)(914,0)(915,0)(916,0)(917,1)(918,0)(919,0)(920,0)(921,0)(922,0)(923,0)(924,0)(925,0)(926,0)(927,0)(928,0)(929,0)(930,0)(931,0)(932,0)(933,0)(934,0)(935,0)(936,0)(937,0)(938,0)(939,0)(940,0)(941,0)(942,0)(943,0)(944,3)(945,3)(946,0)(947,0)(948,4)(949,0)(950,0)(951,0)(952,3)(953,0)(954,0)(955,0)(956,0)(957,1)(958,1)(959,1)(960,1)(961,0)(962,0)(963,0)(964,0)(965,0)(966,0)(967,0)(968,0)(969,0)(970,0)(971,0)(972,0)(973,0)(974,0)(975,0)(976,0)(977,0)(978,0)(979,0)(980,0)(981,1)(982,0)(983,1)(984,0)(985,7)(986,0)(987,0)(988,0)(989,0)(990,0)(991,0)(992,0)(993,0)(994,2)(995,0)(996,1)(997,0)(998,0)(999,0)(1000,0)(1001,0)(1002,3)(1003,0)(1004,0)(1005,0)(1006,0)(1007,4)(1008,0)(1009,0)(1010,11)(1011,1)(1012,0)(1013,0)(1014,0)(1015,0)(1016,0)(1017,0)(1018,0)(1019,1)(1020,1)(1021,0)(1022,0)(1023,0)(1024,0)(1025,2)(1026,4)(1027,0)(1028,0)(1029,0)(1030,0)(1031,0)(1032,0)(1033,0)(1034,0)(1035,0)(1036,14)(1037,0)(1038,0)(1039,9)(1040,1)(1041,0)(1042,0)(1043,2)(1044,1)(1045,7)(1046,0)(1047,0)(1048,0)(1049,0)(1050,0)(1051,0)(1052,0)(1053,0)(1054,0)(1055,6)(1056,0)(1057,0)(1058,0)(1059,0)(1060,0)(1061,0)(1062,0)(1063,26)(1064,0)(1065,0)(1066,0)(1067,1)(1068,0)(1069,0)(1070,0)(1071,0)(1072,18)(1073,0)(1074,1)(1075,4)(1076,0)(1077,25)(1078,0)(1079,0)(1080,1)(1081,0)(1082,5)(1083,0)(1084,0)(1085,0)(1086,1)(1087,0)(1088,0)(1089,0)(1090,0)(1091,2)(1092,19)(1093,0)(1094,0)(1095,0)(1096,0)(1097,0)(1098,0)(1099,0)(1100,10)(1101,0)(1102,0)(1103,0)(1104,0)(1105,0)(1106,0)(1107,0)(1108,0)(1109,0)(1110,0)(1111,0)(1112,0)(1113,0)(1114,0)(1115,0)(1116,0)(1117,0)(1118,0)(1119,0)(1120,0)(1121,0)(1122,0)(1123,24)(1124,32)(1125,0)(1126,0)(1127,0)(1128,0)(1129,0)(1130,0)(1131,0)(1132,0)(1133,0)(1134,0)(1135,0)(1136,0)(1137,3)(1138,0)(1139,0)(1140,0)(1141,0)(1142,0)(1143,12)(1144,6)(1145,6)(1146,0)(1147,0)(1148,0)(1149,108)(1150,8)(1151,0)(1152,2)(1153,0)(1154,0)(1155,0)(1156,0)(1157,0)(1158,0)(1159,0)(1160,0)(1161,0)(1162,0)(1163,0)(1164,0)(1165,0)(1166,0)(1167,0)(1168,3)(1169,0)(1170,0)(1171,0)(1172,0)(1173,5)(1174,0)(1175,0)(1176,0)(1177,0)(1178,0)(1179,5)(1180,1)(1181,0)(1182,0)(1183,3)(1184,0)(1185,0)(1186,0)(1187,0)(1188,7)(1189,0)(1190,0)(1191,0)(1192,0)(1193,0)(1194,0)(1195,0)(1196,1)(1197,7)(1198,1)(1199,0)(1200,0)(1201,8)(1202,10)(1203,0)(1204,0)(1205,0)(1206,0)(1207,0)(1208,0)(1209,1188)(1210,0)(1211,1)(1212,0)(1213,2)(1214,1)(1215,0)(1216,0)(1217,0)(1218,0)(1219,7)(1220,0)(1221,0)(1222,744)(1223,11)(1224,3)(1225,0)(1226,0)(1227,50)(1228,23)(1229,0)(1230,7)(1231,20)(1232,3)(1233,0)(1234,0)(1235,0)(1236,0)(1237,0)(1238,5)(1239,32)(1240,25)(1241,8)(1242,0)(1243,0)(1244,0)(1245,0)(1246,0)(1247,0)(1248,0)(1249,0)(1250,0)(1251,0)(1252,0)(1253,0)(1254,12)(1255,0)(1256,0)(1257,0)(1258,0)(1259,0)(1260,2)(1261,192)(1262,0)(1263,35)(1264,0)(1265,0)(1266,0)(1267,0)(1268,0)(1269,13)(1270,0)(1271,0)(1272,12)(1273,537)(1274,42)(1275,175)(1276,3)(1277,46)(1278,1103)(1279,785)(1280,174)(1281,223)(1282,1632)(1283,21)(1284,2)(1285,0)(1286,1053)(1287,0)(1288,128)(1289,1121)(1290,40)(1291,4)(1292,163)(1293,578)(1294,0)(1295,0)(1296,0)(1297,0)(1298,0)(1299,34)(1300,0)(1301,11)(1302,0)(1303,41)(1304,112)(1305,8)(1306,7)(1307,7)(1308,7)(1309,0)(1310,0)(1311,163)(1312,0)(1313,46)(1314,2246)(1315,8)(1316,194)(1317,1)(1318,65)(1319,6)(1320,660)(1321,2)(1322,0)(1323,0)(1324,15)(1325,5)(1326,0)(1327,136)(1328,77)(1329,0)(1330,7)(1331,7)(1332,0)(1333,0)(1334,71)(1335,3)(1336,0)(1337,0)(1338,250)(1339,0)(1340,355)(1341,1)(1342,37)(1343,0)(1344,116)(1345,0)(1346,0)(1347,25)(1348,31)(1349,0)(1350,28)(1351,0)(1352,0)(1353,0)(1354,33)(1355,0)(1356,0)(1357,0)(1358,0)(1359,0)(1360,0)(1361,0)(1362,212)(1363,56)(1364,41)(1365,54)(1366,99)(1367,87)(1368,20)(1369,4)(1370,24)(1371,1)(1372,59)(1373,40)(1374,0)(1375,131)(1376,971)(1377,1)(1378,4)(1379,135)(1380,0)(1381,58)(1382,0)(1383,0)(1384,0)(1385,0)(1386,0)(1387,194)(1388,99)(1389,3)(1390,47)(1391,147)(1392,0)(1393,3)(1394,2)(1395,0)(1396,0)(1397,0)(1398,1030)(1399,0)(1400,1)(1401,22)(1402,0)(1403,3)(1404,0)(1405,0)(1406,1)(1407,355)(1408,1)(1409,601)(1410,14)(1411,29)(1412,172)(1413,308)(1414,0)(1415,69)(1416,115)(1417,2)(1418,3)(1419,1)(1420,26)(1421,0)(1422,0)(1423,4)(1424,1)(1425,1)(1426,0)(1427,0)(1428,0)(1429,0)(1430,0)(1431,0)(1432,13)(1433,0)(1434,0)(1435,0)(1436,0)(1437,0)(1438,0)(1439,0)(1440,0)(1441,0)(1442,0)(1443,0)(1444,0)(1445,2)(1446,0)(1447,15)(1448,553)(1449,16)(1450,123)(1451,0)(1452,19)(1453,74)(1454,7)(1455,41)(1456,0)(1457,15)(1458,0)(1459,0)(1460,33)(1461,5)(1462,0)(1463,10)(1464,0)(1465,0)(1466,0)(1467,0)(1468,0)(1469,0)(1470,0)(1471,0)(1472,2)(1473,0)(1474,2)(1475,0)(1476,7)(1477,32)(1478,22)(1479,13)(1480,6)(1481,0)(1482,0)(1483,302)(1484,0)(1485,0)(1486,0)(1487,0)(1488,0)(1489,0)(1490,384)(1491,5)(1492,0)(1493,0)(1494,0)(1495,0)(1496,0)(1497,0)(1498,3)(1499,4)(1500,0)(1501,0)(1502,0)(1503,6)(1504,15)(1505,0)(1506,0)(1507,3)(1508,0)(1509,0)(1510,0)(1511,6)(1512,217)(1513,19)(1514,19)(1515,134)(1516,0)(1517,0)(1518,12)(1519,0)(1520,0)(1521,5)(1522,0)(1523,10)(1524,29)(1525,8)(1526,0)(1527,0)(1528,0)(1529,0)(1530,0)(1531,0)(1532,0)(1533,4)(1534,0)(1535,0)(1536,25)(1537,0)(1538,0)(1539,0)(1540,0)(1541,0)(1542,0)(1543,0)(1544,0)(1545,10)(1546,3)(1547,10)(1548,0)(1549,76)(1550,0)(1551,219)(1552,22)(1553,5)(1554,92)(1555,0)(1556,2)(1557,4)(1558,17)(1559,477)(1560,50)(1561,0)(1562,0)(1563,0)(1564,0)(1565,0)(1566,0)(1567,0)(1568,0)(1569,0)(1570,0)(1571,0)(1572,0)(1573,0)(1574,0)(1575,0)(1576,0)(1577,0)(1578,0)(1579,0)(1580,3)(1581,2)(1582,0)(1583,0)(1584,0)(1585,0)(1586,0)(1587,0)(1588,0)(1589,11)(1590,0)(1591,0)(1592,0)(1593,0)(1594,7657)(1595,0)(1596,0)(1597,0)(1598,366)(1599,2)(1600,48)(1601,0)(1602,0)(1603,781)(1604,195)(1605,394)(1606,973)(1607,2)(1608,0)(1609,0)(1610,0)(1611,0)(1612,6)(1613,387)(1614,0)(1615,0)(1616,0)(1617,0)(1618,0)(1619,0)(1620,0)(1621,0)(1622,0)(1623,0)(1624,119)(1625,0)(1626,0)(1627,0)(1628,0)(1629,0)(1630,51)(1631,1)(1632,0)(1633,2)(1634,0)(1635,56)(1636,121)(1637,236)(1638,136)(1639,31)(1640,47)(1641,70)(1642,0)(1643,0)(1644,0)(1645,0)(1646,0)(1647,0)(1648,0)(1649,0)(1650,0)(1651,2)(1652,0)(1653,99)(1654,0)(1655,0)(1656,0)(1657,3)(1658,0)(1659,0)(1660,6)(1661,0)(1662,0)(1663,0)(1664,0)(1665,1)(1666,0)(1667,0)(1668,0)(1669,0)(1670,0)(1671,0)(1672,0)(1673,0)(1674,5)(1675,0)(1676,0)(1677,8)(1678,0)(1679,0)(1680,1)(1681,0)(1682,0)(1683,0)(1684,0)(1685,0)(1686,0)(1687,0)(1688,0)(1689,0)(1690,0)(1691,0)(1692,0)(1693,2)(1694,1)(1695,0)(1696,0)(1697,0)(1698,1)(1699,2)(1700,1)(1701,12)(1702,0)(1703,7)(1704,19)(1705,39)(1706,0)(1707,0)(1708,0)(1709,2)(1710,0)(1711,0)(1712,0)(1713,0)(1714,0)(1715,0)(1716,0)(1717,0)(1718,0)(1719,0)(1720,0)(1721,9)(1722,0)(1723,0)(1724,10)(1725,0)(1726,3)(1727,5)(1728,0)(1729,0)(1730,0)(1731,0)(1732,6)(1733,3)(1734,40)(1735,7)(1736,0)(1737,17)(1738,2)(1739,0)(1740,0)(1741,0)(1742,0)(1743,0)(1744,1)(1745,0)(1746,8)(1747,2)(1748,0)(1749,0)(1750,1)(1751,0)(1752,0)(1753,4)(1754,4)(1755,0)(1756,8)(1757,0)(1758,7)(1759,0)(1760,0)(1761,0)(1762,0)(1763,0)(1764,0)(1765,0)(1766,0)(1767,0)(1768,0)(1769,0)(1770,0)(1771,0)(1772,0)(1773,0)(1774,0)(1775,0)(1776,0)(1777,0)(1778,0)(1779,0)(1780,0)(1781,0)(1782,0)(1783,0)(1784,3)(1785,0)(1786,0)(1787,0)(1788,3)(1789,0)(1790,0)(1791,0)(1792,0)(1793,0)(1794,0)(1795,0)(1796,0)(1797,0)(1798,0)(1799,0)(1800,0)(1801,0)(1802,0)(1803,0)(1804,0)(1805,0)(1806,0)(1807,0)(1808,0)(1809,0)(1810,0)(1811,0)(1812,7)(1813,0)(1814,0)(1815,1)(1816,2)(1817,0)(1818,0)(1819,0)(1820,0)(1821,0)(1822,1)(1823,0)(1824,2)(1825,9)(1826,0)(1827,0)(1828,2)(1829,0)(1830,3)(1831,1)(1832,3)(1833,1)(1834,0)(1835,0)(1836,0)(1837,5)(1838,0)(1839,0)(1840,0)(1841,0)(1842,0)(1843,0)(1844,0)(1845,0)(1846,0)(1847,0)(1848,0)(1849,0)(1850,0)(1851,0)(1852,1)(1853,0)(1854,0)(1855,0)(1856,0)(1857,0)(1858,0)(1859,7)(1860,11)(1861,0)(1862,0)(1863,0)(1864,0)(1865,0)(1866,2)(1867,0)(1868,2)(1869,0)(1870,0)(1871,0)(1872,8)(1873,4)(1874,0)(1875,0)(1876,6)(1877,5)(1878,0)(1879,0)(1880,76)(1881,5)(1882,0)(1883,0)(1884,0)(1885,0)(1886,2)(1887,2)(1888,0)(1889,0)(1890,2)(1891,11)(1892,110)(1893,0)(1894,98)(1895,0)(1896,0)(1897,0)(1898,0)(1899,0)(1900,1)(1901,0)(1902,0)(1903,0)(1904,0)(1905,0)(1906,4)(1907,2)(1908,0)(1909,0)(1910,1)(1911,0)(1912,0)(1913,0)(1914,2)(1915,0)(1916,0)(1917,0)(1918,0)(1919,10)(1920,21)(1921,3)(1922,0)(1923,0)(1924,0)(1925,0)(1926,0)(1927,0)(1928,0)(1929,0)(1930,0)(1931,1)(1932,0)(1933,0)(1934,0)(1935,0)(1936,0)(1937,0)(1938,0)(1939,0)(1940,0)(1941,0)(1942,0)(1943,0)(1944,0)(1945,0)(1946,0)(1947,0)(1948,0)(1949,0)(1950,0)(1951,0)(1952,0)(1953,0)(1954,1)(1955,0)(1956,0)(1957,0)(1958,0)(1959,0)(1960,0)(1961,0)(1962,0)(1963,0)(1964,0)(1965,0)(1966,0)(1967,0)(1968,5)(1969,0)(1970,2)(1971,0)(1972,0)(1973,253)(1974,16)(1975,0)(1976,0)(1977,0)(1978,0)(1979,0)(1980,0)(1981,0)(1982,0)(1983,0)(1984,5)(1985,3)(1986,0)(1987,0)(1988,0)(1989,0)(1990,1)(1991,3)(1992,0)(1993,0)(1994,0)(1995,0)(1996,0)(1997,0)(1998,0)(1999,1)(2000,0)(2001,0)(2002,0)(2003,0)(2004,0)(2005,0)(2006,0)(2007,0)(2008,2)(2009,0)(2010,3)(2011,0)(2012,0)(2013,0)(2014,0)(2015,15)(2016,2)(2017,0)(2018,0)(2019,0)(2020,12)(2021,0)(2022,2)(2023,199)(2024,2)(2025,4)(2026,24)(2027,0)(2028,0)(2029,1)(2030,0)(2031,6)(2032,6)(2033,3)(2034,34)(2035,58)(2036,7)(2037,3)(2038,0)(2039,0)(2040,2)(2041,0)(2042,0)(2043,0)(2044,0)(2045,0)(2046,0)(2047,1)(2048,17)(2049,0)(2050,14)(2051,0)(2052,0)(2053,22)(2054,9)(2055,0)(2056,0)(2057,2)(2058,82)(2059,17)(2060,18)(2061,19)(2062,7)(2063,0)(2064,0)(2065,0)(2066,1)(2067,3)(2068,29)(2069,3)(2070,108)(2071,0)(2072,7)(2073,2)(2074,10)(2075,0)(2076,0)(2077,0)(2078,2)(2079,2)(2080,1)(2081,8)(2082,0)(2083,82)(2084,0)(2085,0)(2086,14)(2087,0)(2088,0)(2089,7)(2090,0)(2091,21)(2092,7)(2093,7)(2094,2)(2095,1)(2096,0)(2097,13)(2098,2)(2099,26)(2100,0)(2101,0)(2102,0)(2103,0)(2104,0)(2105,0)(2106,50)(2107,0)(2108,0)(2109,1)(2110,0)(2111,0)(2112,0)(2113,0)(2114,0)(2115,0)(2116,0)(2117,0)(2118,0)(2119,0)(2120,4)(2121,33)(2122,9)(2123,0)(2124,15)(2125,7)(2126,25)(2127,10)(2128,0)(2129,28)(2130,14)(2131,0)(2132,5)(2133,0)(2134,11)(2135,3)(2136,0)(2137,0)(2138,0)(2139,3)(2140,0)(2141,0)(2142,0)(2143,4)(2144,0)(2145,0)(2146,4)(2147,3)(2148,0)(2149,9)(2150,0)(2151,5)(2152,1)(2153,0)(2154,0)(2155,2)(2156,0)(2157,0)(2158,0)(2159,0)(2160,3)(2161,0)(2162,0)(2163,0)(2164,1)(2165,0)(2166,0)(2167,0)(2168,14)(2169,0)(2170,0)(2171,2)(2172,2)(2173,6)(2174,9)(2175,0)(2176,0)(2177,3)(2178,0)(2179,0)(2180,9)(2181,3)(2182,0)(2183,0)(2184,0)(2185,0)(2186,0)(2187,0)(2188,0)(2189,72)(2190,4)(2191,3)(2192,0)(2193,1)(2194,0)(2195,1)(2196,0)(2197,0)(2198,13)(2199,0)(2200,0)(2201,18)(2202,0)(2203,0)(2204,0)(2205,0)(2206,0)(2207,4)(2208,0)(2209,0)(2210,0)(2211,0)(2212,27)(2213,0)(2214,10)(2215,0)(2216,0)(2217,0)(2218,1)(2219,0)(2220,0)(2221,1)(2222,12)(2223,0)(2224,0)(2225,0)(2226,19)(2227,3)(2228,0)(2229,0)(2230,0)(2231,4)(2232,0)(2233,0)(2234,0)(2235,9)(2236,6)(2237,340)(2238,6)(2239,0)(2240,3)(2241,0)(2242,6)(2243,0)(2244,2)(2245,0)(2246,0)(2247,0)(2248,4)(2249,0)(2250,5)(2251,6)(2252,0)(2253,0)(2254,0)(2255,0)(2256,0)(2257,0)(2258,0)(2259,0)(2260,5)(2261,0)(2262,0)(2263,0)(2264,0)(2265,0)(2266,0)(2267,0)(2268,2)(2269,0)(2270,0)(2271,0)(2272,1)(2273,0)(2274,3)(2275,0)(2276,0)(2277,0)(2278,0)(2279,0)(2280,2)(2281,1)(2282,0)(2283,0)(2284,0)(2285,0)(2286,0)(2287,0)(2288,0)(2289,3)(2290,0)(2291,0)(2292,0)(2293,0)(2294,0)(2295,2)(2296,0)(2297,0)(2298,0)(2299,9)(2300,0)(2301,0)(2302,1)(2303,1)(2304,62)(2305,0)(2306,3)(2307,0)(2308,0)(2309,13)(2310,7)(2311,0)(2312,1)(2313,0)(2314,0)(2315,0)(2316,0)(2317,27)(2318,3)(2319,0)(2320,10)(2321,0)(2322,0)(2323,8)(2324,5)(2325,0)(2326,0)(2327,0)(2328,0)(2329,1)(2330,0)(2331,0)(2332,0)(2333,0)(2334,0)(2335,0)(2336,0)(2337,0)(2338,0)(2339,0)(2340,0)(2341,0)(2342,0)(2343,0)(2344,0)(2345,0)(2346,8)(2347,0)(2348,0)(2349,0)(2350,0)(2351,0)(2352,0)(2353,0)(2354,0)(2355,0)(2356,0)(2357,0)(2358,0)(2359,0)(2360,0)(2361,0)(2362,64)(2363,0)(2364,0)(2365,0)(2366,0)(2367,0)(2368,0)(2369,0)(2370,0)(2371,0)(2372,0)(2373,0)(2374,44)(2375,2)(2376,0)(2377,0)(2378,0)(2379,0)(2380,0)(2381,0)(2382,0)(2383,0)(2384,331)(2385,8)(2386,0)(2387,0)(2388,0)(2389,0)(2390,158)(2391,79)(2392,0)(2393,0)(2394,0)(2395,0)(2396,0)(2397,0)(2398,0)(2399,1)(2400,0)(2401,0)(2402,4)(2403,0)(2404,0)(2405,4)(2406,8)(2407,0)(2408,0)(2409,0)(2410,5)(2411,0)(2412,0)(2413,0)(2414,0)(2415,0)(2416,0)(2417,0)(2418,0)(2419,8)(2420,2)(2421,0)(2422,0)(2423,62)(2424,0)(2425,10)(2426,5)(2427,0)(2428,0)(2429,0)(2430,0)(2431,1)(2432,0)(2433,0)(2434,0)(2435,3)(2436,0)(2437,3)(2438,0)(2439,0)(2440,7)(2441,0)(2442,1)(2443,78)(2444,0)(2445,0)(2446,7)(2447,13)(2448,7)(2449,0)(2450,0)(2451,0)(2452,1)(2453,1)(2454,0)(2455,0)(2456,0)(2457,0)(2458,0)(2459,1)(2460,1)(2461,5)(2462,1)(2463,0)(2464,1)(2465,0)(2466,1)(2467,3)(2468,0)(2469,1)(2470,1)(2471,13)(2472,5)(2473,0)(2474,1)(2475,0)(2476,7)(2477,0)(2478,0)(2479,0)(2480,6)(2481,0)(2482,5)(2483,0)(2484,0)(2485,0)(2486,0)(2487,0)(2488,0)(2489,0)(2490,14)(2491,0)(2492,6)(2493,0)(2494,0)(2495,1)(2496,0)(2497,0)(2498,0)(2499,0)(2500,0)(2501,0)(2502,0)(2503,0)(2504,10)(2505,0)(2506,0)(2507,0)(2508,0)(2509,0)(2510,0)(2511,0)(2512,0)(2513,0)(2514,0)(2515,0)(2516,0)(2517,0)(2518,0)(2519,0)(2520,0)(2521,0)(2522,4)(2523,0)(2524,0)(2525,0)(2526,0)(2527,0)(2528,0)(2529,4)(2530,1)(2531,8)(2532,62)(2533,44)(2534,0)(2535,0)(2536,0)(2537,0)(2538,1)(2539,0)(2540,3)(2541,0)(2542,0)(2543,0)(2544,0)(2545,0)(2546,0)(2547,1)(2548,0)(2549,3)(2550,0)(2551,0)(2552,2)(2553,0)(2554,1)(2555,0)(2556,0)(2557,0)(2558,0)(2559,0)(2560,0)(2561,0)(2562,0)(2563,0)(2564,1)(2565,0)(2566,32)(2567,0)(2568,0)(2569,89)(2570,0)(2571,1)(2572,3)(2573,0)(2574,1)(2575,0)(2576,0)(2577,8)(2578,0)(2579,0)(2580,0)(2581,1)(2582,0)(2583,0)(2584,0)(2585,0)(2586,0)(2587,0)(2588,0)(2589,0)(2590,0)(2591,0)(2592,0)(2593,0)(2594,0)(2595,0)(2596,0)(2597,0)(2598,0)(2599,0)(2600,0)(2601,0)(2602,1)(2603,0)(2604,1)(2605,0)(2606,0)(2607,0)(2608,0)(2609,0)(2610,0)(2611,0)(2612,4)(2613,0)(2614,0)(2615,0)(2616,5)(2617,7)(2618,0)(2619,3)(2620,18)(2621,0)(2622,0)(2623,0)(2624,0)(2625,0)(2626,1)(2627,0)(2628,3)(2629,5)(2630,7)(2631,4)(2632,0)(2633,0)(2634,0)(2635,0)(2636,0)(2637,0)(2638,0)(2639,1)(2640,2)(2641,2)(2642,0)(2643,0)(2644,0)(2645,0)(2646,0)(2647,0)(2648,0)(2649,0)(2650,5)(2651,6)(2652,0)(2653,2)(2654,11)(2655,0)(2656,1)(2657,9)(2658,0)(2659,0)(2660,4)(2661,0)(2662,1)(2663,2)(2664,0)(2665,0)(2666,0)(2667,4)(2668,0)(2669,0)(2670,0)(2671,0)(2672,0)(2673,0)(2674,0)(2675,5)(2676,0)(2677,0)(2678,5)(2679,0)(2680,0)(2681,0)(2682,0)(2683,0)(2684,0)(2685,0)(2686,0)(2687,0)(2688,4)(2689,0)(2690,0)(2691,0)(2692,1)(2693,0)(2694,0)(2695,0)(2696,8)(2697,0)(2698,0)(2699,0)(2700,1)(2701,47)(2702,3)(2703,1)(2704,0)(2705,0)(2706,0)(2707,1)(2708,4)(2709,3)(2710,1)(2711,0)(2712,0)(2713,1)(2714,0)(2715,1)(2716,5)(2717,3)(2718,7)(2719,9)(2720,0)(2721,1)(2722,0)(2723,0)(2724,0)(2725,1)(2726,0)(2727,1)(2728,18)(2729,0)(2730,0)(2731,0)(2732,0)(2733,0)(2734,3)(2735,5)(2736,0)(2737,1)(2738,0)(2739,0)(2740,6)(2741,0)(2742,0)(2743,5)(2744,0)(2745,0)(2746,1)(2747,0)(2748,1)(2749,101)(2750,49)(2751,2)(2752,1)(2753,1)(2754,0)(2755,0)(2756,0)(2757,6)(2758,0)(2759,0)(2760,0)(2761,0)(2762,0)(2763,0)(2764,0)(2765,0)(2766,0)(2767,0)(2768,0)(2769,0)(2770,0)(2771,0)(2772,0)(2773,0)(2774,0)(2775,0)(2776,0)(2777,0)(2778,1)(2779,0)(2780,0)(2781,7)(2782,6)(2783,0)(2784,2)(2785,9)(2786,2)(2787,0)(2788,0)(2789,0)(2790,0)(2791,1)(2792,0)(2793,0)(2794,0)(2795,5)(2796,0)(2797,0)(2798,2)(2799,0)(2800,0)(2801,5)(2802,0)(2803,15)(2804,0)(2805,0)(2806,0)(2807,4)(2808,0)(2809,0)(2810,0)(2811,4)(2812,0)(2813,0)(2814,0)(2815,0)(2816,0)(2817,2)(2818,1)(2819,73)(2820,2)(2821,7)(2822,14)(2823,0)(2824,9)(2825,0)(2826,0)(2827,8)(2828,0)(2829,0)(2830,0)(2831,0)(2832,0)(2833,0)(2834,0)(2835,0)(2836,2)(2837,0)(2838,0)(2839,0)(2840,1)(2841,14)(2842,8)(2843,0)(2844,0)(2845,0)(2846,0)(2847,0)(2848,0)(2849,0)(2850,0)(2851,0)(2852,0)(2853,0)(2854,0)(2855,0)(2856,0)(2857,0)(2858,4)(2859,0)(2860,6)(2861,2)(2862,0)(2863,0)(2864,0)(2865,0)(2866,0)(2867,0)(2868,0)(2869,4)(2870,0)(2871,3)(2872,0)(2873,1)(2874,0)(2875,1)(2876,1)(2877,6)(2878,0)(2879,0)(2880,0)(2881,0)(2882,3)(2883,0)(2884,0)(2885,0)(2886,2)(2887,9)(2888,0)(2889,2)(2890,0)(2891,0)(2892,0)(2893,0)(2894,2)(2895,0)(2896,3)(2897,6)(2898,0)(2899,1)(2900,0)(2901,0)(2902,0)(2903,0)(2904,17)(2905,4)(2906,5)(2907,1)(2908,2)(2909,0)(2910,0)(2911,0)(2912,0)(2913,1)(2914,6)(2915,11)(2916,4)(2917,1)(2918,0)(2919,2)(2920,0)(2921,0)(2922,0)(2923,0)(2924,1)(2925,0)(2926,2)(2927,0)(2928,0)(2929,0)(2930,0)(2931,0)(2932,0)(2933,0)(2934,0)(2935,0)(2936,0)(2937,0)(2938,0)(2939,1)(2940,1)(2941,0)(2942,0)(2943,0)(2944,0)(2945,0)(2946,2)(2947,3)(2948,4)(2949,0)(2950,0)(2951,1)(2952,1)(2953,0)(2954,0)(2955,1)(2956,14)(2957,6)(2958,4)(2959,0)(2960,0)(2961,4)(2962,5)(2963,0)(2964,0)(2965,0)(2966,0)(2967,3)(2968,1)(2969,32)(2970,0)(2971,0)(2972,0)(2973,0)(2974,0)(2975,0)(2976,0)(2977,0)(2978,0)(2979,1)(2980,0)(2981,1)(2982,0)(2983,0)(2984,1)(2985,0)(2986,4)(2987,0)(2988,1)(2989,0)(2990,0)(2991,0)(2992,0)(2993,10)(2994,0)(2995,1)(2996,0)(2997,0)(2998,0)(2999,0)(3000,0)(3001,0)(3002,0)(3003,0)(3004,0)(3005,0)(3006,0)(3007,1)(3008,5)(3009,5)(3010,0)(3011,0)(3012,0)(3013,0)(3014,0)(3015,0)(3016,0)(3017,0)(3018,0)(3019,0)(3020,5)(3021,1)(3022,9)(3023,0)(3024,0)(3025,0)(3026,0)(3027,0)(3028,0)(3029,0)(3030,0)(3031,0)(3032,0)(3033,0)(3034,63)(3035,521)(3036,17)(3037,28)(3038,0)(3039,20)(3040,0)(3041,0)(3042,0)(3043,0)(3044,0)(3045,0)(3046,0)(3047,0)(3048,3)(3049,5)(3050,1)(3051,0)(3052,25)(3053,4)(3054,1)(3055,0)(3056,0)(3057,0)(3058,11)(3059,2)(3060,1)(3061,0)(3062,0)(3063,0)(3064,0)(3065,0)(3066,0)(3067,0)(3068,0)(3069,0)(3070,0)(3071,0)(3072,3)(3073,0)(3074,0)(3075,0)(3076,0)(3077,0)(3078,0)(3079,0)(3080,0)(3081,4)(3082,2)(3083,1)(3084,0)(3085,0)(3086,0)(3087,0)(3088,2)(3089,0)(3090,0)(3091,6)(3092,0)(3093,0)(3094,0)(3095,0)(3096,71)(3097,1)(3098,0)(3099,0)(3100,3)(3101,0)(3102,3)(3103,0)(3104,0)(3105,0)(3106,2)(3107,0)(3108,0)(3109,0)(3110,0)(3111,0)(3112,1)(3113,8)(3114,0)(3115,0)(3116,0)(3117,0)(3118,0)(3119,3)(3120,0)(3121,0)(3122,1)(3123,0)(3124,16)(3125,5)(3126,0)(3127,0)(3128,0)(3129,5)(3130,0)(3131,0)(3132,0)(3133,10)(3134,4)(3135,0)(3136,1)(3137,0)(3138,0)(3139,0)(3140,0)(3141,6)(3142,0)(3143,0)(3144,0)(3145,0)(3146,6)(3147,0)(3148,2)(3149,49)(3150,3)(3151,2)(3152,2)(3153,0)(3154,0)(3155,5)(3156,9)(3157,0)(3158,0)(3159,21)(3160,0)(3161,19)(3162,0)(3163,0)(3164,2)(3165,0)(3166,0)(3167,36)(3168,0)(3169,0)(3170,4)(3171,0)(3172,0)(3173,0)(3174,0)(3175,0)(3176,1)(3177,0)(3178,0)(3179,0)(3180,0)(3181,7)(3182,1)(3183,0)(3184,0)(3185,11)(3186,25)(3187,0)(3188,0)(3189,0)(3190,8)(3191,1)(3192,0)(3193,0)(3194,0)(3195,2)(3196,12)(3197,0)(3198,0)(3199,0)(3200,0)(3201,0)(3202,2)(3203,0)(3204,0)(3205,0)(3206,0)(3207,0)(3208,0)(3209,0)(3210,0)(3211,0)(3212,0)(3213,1)(3214,0)(3215,0)(3216,8)(3217,1)(3218,0)(3219,0)(3220,5)(3221,11)(3222,1)(3223,0)(3224,0)(3225,0)(3226,51)(3227,0)(3228,56)(3229,0)(3230,1)(3231,0)(3232,0)(3233,0)(3234,4)(3235,0)(3236,0)(3237,0)(3238,4)(3239,1)(3240,0)(3241,12)(3242,2)(3243,0)(3244,409)(3245,0)(3246,0)(3247,4)(3248,0)(3249,1)(3250,0)(3251,0)(3252,4)(3253,5)(3254,6)(3255,8)(3256,0)(3257,0)(3258,210)(3259,46)(3260,0)(3261,1)(3262,0)(3263,2)(3264,1)(3265,0)(3266,604)(3267,49)(3268,0)(3269,0)(3270,0)(3271,0)(3272,5)(3273,0)(3274,0)(3275,10)(3276,2)(3277,0)(3278,0)(3279,0)(3280,6)(3281,0)(3282,12)(3283,0)(3284,0)(3285,0)(3286,0)(3287,0)(3288,0)(3289,12)(3290,0)(3291,0)(3292,0)(3293,0)(3294,0)(3295,12)(3296,0)(3297,0)(3298,4)(3299,0)(3300,0)(3301,0)(3302,0)(3303,0)(3304,0)(3305,0)(3306,3)(3307,2)(3308,0)(3309,0)(3310,0)(3311,0)(3312,0)(3313,15)(3314,0)(3315,0)(3316,0)(3317,0)(3318,0)(3319,0)(3320,0)(3321,0)(3322,0)(3323,1)(3324,0)(3325,0)(3326,0)(3327,0)(3328,12)(3329,0)(3330,0)(3331,0)(3332,0)(3333,0)(3334,0)(3335,2)(3336,0)(3337,0)(3338,0)(3339,0)(3340,0)(3341,0)(3342,0)(3343,0)(3344,2)(3345,1)(3346,0)(3347,0)(3348,0)(3349,0)(3350,0)(3351,0)(3352,0)(3353,0)(3354,0)(3355,0)(3356,0)(3357,0)(3358,0)(3359,0)(3360,0)(3361,0)(3362,0)(3363,0)(3364,0)(3365,0)(3366,971)(3367,20)(3368,0)(3369,0)(3370,68)(3371,0)(3372,18)(3373,2)(3374,0)(3375,196)(3376,0)(3377,147)(3378,24)(3379,4)(3380,52)(3381,77)(3382,48)(3383,1048)(3384,566)(3385,146)(3386,98)(3387,101)(3388,0)(3389,2)(3390,0)(3391,0)(3392,1)(3393,0)(3394,0)(3395,0)(3396,0)(3397,0)(3398,0)(3399,0)(3400,0)(3401,0)(3402,2)(3403,0)(3404,0)(3405,0)(3406,1)(3407,0)(3408,0)(3409,0)(3410,4)(3411,0)(3412,0)(3413,0)(3414,0)(3415,0)(3416,0)(3417,0)(3418,0)(3419,6)(3420,0)(3421,1)(3422,0)(3423,0)(3424,0)(3425,0)(3426,0)(3427,0)(3428,0)(3429,0)(3430,0)(3431,0)(3432,0)(3433,0)(3434,0)(3435,0)(3436,0)(3437,0)(3438,0)(3439,0)(3440,0)(3441,0)(3442,0)(3443,1)(3444,0)(3445,2)(3446,0)(3447,5)(3448,20)(3449,0)(3450,0)(3451,0)(3452,0)(3453,0)(3454,4)(3455,1)(3456,0)(3457,0)(3458,1018)(3459,10)(3460,0)(3461,0)(3462,0)(3463,915)(3464,42)(3465,2)(3466,0)(3467,110)(3468,20)(3469,0)(3470,0)(3471,4)(3472,0)(3473,30)(3474,0)(3475,0)(3476,0)(3477,16)(3478,400)(3479,5)(3480,0)(3481,0)(3482,0)(3483,17)(3484,0)(3485,291)(3486,19)(3487,0)(3488,436)(3489,43)(3490,1)(3491,0)(3492,0)(3493,0)(3494,0)(3495,53)(3496,0)(3497,798)(3498,0)(3499,2)(3500,29)(3501,86)(3502,29)(3503,0)(3504,0)(3505,0)(3506,1)(3507,52)(3508,0)(3509,1)(3510,0)(3511,1)(3512,2)(3513,0)(3514,0)(3515,0)(3516,0)(3517,0)(3518,0)(3519,0)(3520,0)(3521,1)(3522,0)(3523,0)(3524,0)(3525,0)(3526,8)(3527,0)(3528,0)(3529,0)(3530,5)(3531,40)(3532,0)(3533,1)(3534,90)(3535,0)(3536,1)(3537,55)(3538,5)(3539,23)(3540,324)(3541,0)(3542,3)(3543,0)(3544,0)(3545,4)(3546,13)(3547,12)(3548,0)(3549,0)(3550,0)(3551,0)(3552,0)(3553,0)(3554,4)(3555,0)(3556,0)(3557,0)(3558,16)(3559,1)(3560,0)(3561,52)(3562,9)(3563,0)(3564,1)(3565,0)(3566,0)(3567,1)(3568,0)(3569,0)(3570,0)(3571,0)(3572,0)(3573,0)(3574,0)(3575,0)(3576,0)(3577,0)(3578,0)(3579,0)(3580,0)(3581,0)(3582,0)(3583,0)(3584,5)(3585,0)(3586,0)(3587,0)(3588,0)(3589,0)(3590,3)(3591,0)(3592,0)(3593,100)(3594,0)(3595,0)(3596,0)(3597,0)(3598,0)(3599,0)(3600,0)(3601,0)(3602,0)(3603,0)(3604,0)(3605,0)(3606,1004)(3607,6)(3608,0)(3609,0)(3610,0)(3611,0)(3612,0)(3613,5)(3614,0)(3615,0)(3616,0)(3617,0)(3618,0)(3619,0)(3620,0)(3621,0)(3622,0)(3623,1)(3624,0)(3625,0)(3626,0)(3627,0)(3628,0)(3629,4)(3630,7)(3631,29)(3632,373)(3633,5)(3634,0)(3635,0)(3636,0)(3637,0)(3638,0)(3639,0)(3640,0)(3641,0)(3642,3)(3643,4)(3644,0)(3645,0)(3646,0)(3647,504)(3648,0)(3649,0)(3650,139)(3651,299)(3652,0)(3653,49)(3654,7)(3655,103)(3656,2)(3657,0)(3658,0)(3659,0)(3660,0)(3661,0)(3662,3)(3663,0)(3664,0)(3665,7)(3666,0)(3667,0)(3668,1)(3669,0)(3670,0)(3671,0)(3672,0)(3673,0)(3674,138)(3675,50)(3676,200)(3677,169)(3678,302)(3679,257)(3680,50)(3681,0)(3682,337)(3683,107)(3684,50)(3685,0)(3686,0)(3687,0)(3688,100)(3689,0)(3690,0)(3691,161)(3692,0)(3693,0)(3694,1)(3695,0)(3696,0)(3697,1)(3698,1)(3699,0)(3700,0)(3701,0)(3702,0)(3703,0)(3704,0)(3705,0)(3706,1)(3707,0)(3708,0)(3709,0)(3710,61)(3711,22)(3712,3)(3713,2)(3714,7)(3715,0)(3716,0)(3717,0)(3718,0)(3719,9)(3720,35)(3721,0)(3722,0)(3723,0)(3724,18)(3725,0)(3726,0)(3727,0)(3728,4)(3729,0)(3730,0)(3731,0)(3732,0)(3733,0)(3734,0)(3735,0)(3736,0)(3737,0)(3738,0)(3739,2)(3740,0)(3741,2)(3742,0)(3743,0)(3744,0)(3745,0)(3746,2)(3747,0)(3748,0)(3749,0)(3750,0)(3751,0)(3752,0)(3753,0)(3754,0)(3755,160)(3756,0)(3757,5)(3758,6)(3759,1)(3760,9)(3761,8)(3762,0)(3763,1)(3764,18)(3765,279)(3766,34)(3767,3)(3768,0)(3769,0)(3770,3)(3771,0)(3772,0)(3773,0)(3774,0)(3775,0)(3776,0)(3777,0)(3778,113)(3779,34)(3780,34)(3781,0)(3782,420)(3783,2)(3784,0)(3785,0)(3786,0)(3787,5)(3788,0)(3789,0)(3790,0)(3791,0)(3792,0)(3793,0)(3794,0)(3795,5)(3796,0)(3797,0)(3798,1)(3799,0)(3800,0)(3801,0)(3802,0)(3803,0)(3804,12)(3805,5)(3806,0)(3807,0)(3808,0)(3809,0)(3810,0)(3811,0)(3812,0)(3813,2)(3814,1)(3815,0)(3816,0)(3817,0)(3818,0)(3819,0)(3820,1)(3821,0)(3822,0)(3823,8)(3824,0)(3825,1)(3826,0)(3827,0)(3828,0)(3829,1)(3830,0)(3831,6)(3832,0)(3833,0)(3834,2)(3835,1)(3836,3)(3837,17)(3838,0)(3839,0)(3840,0)(3841,0)(3842,0)(3843,0)(3844,0)(3845,0)(3846,0)(3847,0)(3848,0)(3849,0)(3850,0)(3851,0)(3852,0)(3853,2)(3854,0)(3855,0)(3856,2)(3857,0)(3858,0)(3859,0)(3860,0)(3861,0)(3862,0)(3863,2)(3864,0)(3865,0)(3866,0)(3867,0)(3868,0)(3869,0)(3870,0)(3871,0)(3872,0)(3873,0)(3874,0)(3875,0)(3876,0)(3877,0)(3878,0)(3879,0)(3880,0)(3881,0)(3882,0)(3883,2)(3884,2)(3885,6)(3886,16)(3887,5)(3888,9)(3889,0)(3890,0)(3891,1)(3892,0)(3893,1)(3894,3)(3895,0)(3896,9)(3897,0)(3898,12)(3899,0)(3900,20)(3901,2)(3902,0)(3903,0)(3904,0)(3905,0)(3906,2)(3907,0)(3908,0)(3909,134)(3910,115)(3911,4)(3912,0)(3913,0)(3914,8)(3915,4)(3916,0)(3917,30)(3918,77)(3919,1)(3920,1)(3921,19)(3922,0)(3923,15)(3924,0)(3925,3)(3926,30)(3927,0)(3928,4)(3929,10)(3930,0)(3931,3)(3932,4)(3933,1)(3934,0)(3935,9)(3936,0)(3937,3)(3938,27)(3939,0)(3940,0)(3941,7)(3942,0)(3943,14)(3944,0)(3945,0)(3946,0)(3947,4)(3948,0)(3949,2)(3950,0)(3951,1)(3952,0)(3953,5)(3954,0)(3955,0)(3956,0)(3957,0)(3958,12)(3959,0)(3960,0)(3961,13)(3962,3)(3963,28)(3964,1)(3965,1)(3966,0)(3967,3)(3968,1)(3969,0)(3970,0)(3971,0)(3972,1)(3973,0)(3974,0)(3975,0)(3976,1)(3977,6)(3978,2)(3979,0)(3980,8)(3981,3)(3982,1)(3983,6)(3984,1)(3985,1)(3986,0)(3987,7)(3988,0)(3989,0)(3990,0)(3991,0)(3992,0)(3993,0)(3994,0)(3995,0)(3996,2)(3997,1)(3998,0)(3999,0)(4000,0)(4001,0)(4002,3)(4003,1)(4004,2)(4005,0)(4006,0)(4007,2)(4008,0)(4009,0)(4010,0)(4011,1)(4012,0)(4013,0)(4014,1)(4015,0)(4016,0)(4017,0)(4018,0)(4019,0)(4020,0)(4021,0)(4022,3)(4023,0)(4024,0)(4025,0)(4026,0)(4027,0)(4028,0)(4029,0)(4030,0)(4031,0)(4032,6)(4033,1)(4034,6)(4035,0)(4036,0)(4037,0)(4038,1)(4039,0)(4040,7)(4041,0)(4042,1)(4043,3)(4044,0)(4045,6)(4046,0)(4047,0)(4048,0)(4049,1)(4050,1)(4051,0)(4052,31)(4053,0)(4054,0)(4055,0)(4056,0)(4057,0)(4058,0)(4059,0)(4060,3)(4061,0)(4062,15)(4063,0)(4064,0)(4065,0)(4066,1)(4067,0)(4068,0)(4069,0)(4070,0)(4071,0)(4072,0)(4073,0)(4074,0)(4075,0)(4076,1)(4077,0)(4078,0)(4079,0)(4080,0)(4081,0)(4082,0)(4083,0)(4084,0)(4085,0)(4086,0)(4087,4)(4088,2)(4089,0)(4090,2)(4091,0)(4092,0)(4093,25)(4094,0)(4095,23)(4096,7)(4097,0)(4098,0)(4099,0)(4100,0)(4101,0)(4102,68)(4103,0)(4104,4)(4105,0)(4106,14)(4107,1)(4108,0)(4109,0)(4110,0)(4111,0)(4112,0)(4113,0)(4114,0)(4115,0)(4116,3)(4117,0)(4118,0)(4119,0)(4120,0)(4121,0)(4122,36)(4123,0)(4124,0)(4125,0)(4126,0)(4127,0)(4128,14)(4129,4)(4130,0)(4131,0)(4132,0)(4133,3)(4134,0)(4135,0)(4136,1)(4137,3)(4138,52)(4139,0)(4140,0)(4141,0)(4142,22)(4143,0)(4144,0)(4145,0)(4146,7)(4147,21)(4148,22)(4149,1)(4150,0)(4151,0)(4152,0)(4153,9)(4154,7)(4155,1)(4156,10)(4157,0)(4158,3)(4159,2)(4160,0)(4161,78)(4162,20)(4163,2)(4164,0)(4165,2)(4166,12)(4167,0)(4168,0)(4169,0)(4170,0)(4171,0)(4172,6)(4173,0)(4174,0)(4175,0)(4176,6)(4177,0)(4178,0)(4179,1)(4180,0)(4181,0)(4182,0)(4183,0)(4184,0)(4185,1)(4186,0)(4187,12)(4188,75)(4189,0)(4190,113)(4191,14)(4192,172)(4193,0)(4194,57)(4195,18)(4196,0)(4197,0)(4198,0)(4199,68)(4200,8)(4201,4)(4202,0)(4203,11)(4204,0)(4205,0)(4206,26)(4207,0)(4208,7)(4209,35)(4210,28)(4211,0)(4212,0)(4213,52)(4214,0)(4215,0)(4216,220)(4217,17)(4218,0)(4219,5)(4220,0)(4221,0)(4222,0)(4223,0)(4224,0)(4225,0)(4226,18)(4227,0)(4228,106)(4229,14)(4230,16)(4231,24)(4232,0)(4233,0)(4234,0)(4235,0)(4236,43)(4237,6)(4238,0)(4239,0)(4240,47)(4241,44)(4242,0)(4243,0)(4244,6)(4245,0)(4246,0)(4247,0)(4248,55)(4249,0)(4250,0)(4251,0)(4252,90)(4253,0)(4254,0)(4255,0)(4256,0)(4257,3)(4258,0)(4259,0)(4260,0)(4261,49)(4262,0)(4263,20)(4264,0)(4265,0)(4266,0)(4267,28)(4268,37)(4269,0)(4270,0)(4271,0)(4272,30)(4273,15)(4274,0)(4275,1)(4276,0)(4277,0)(4278,0)(4279,0)(4280,0)(4281,0)(4282,0)(4283,0)(4284,0)(4285,0)(4286,0)(4287,0)(4288,0)(4289,2)(4290,2)(4291,2)(4292,0)(4293,0)(4294,0)(4295,3)(4296,29)(4297,1)(4298,0)(4299,14)(4300,0)(4301,0)(4302,0)(4303,0)(4304,5)(4305,0)(4306,0)(4307,2)(4308,0)(4309,1)(4310,1)(4311,1)(4312,6)(4313,0)(4314,1)(4315,0)(4316,0)(4317,0)(4318,0)(4319,9)(4320,0)(4321,0)(4322,33)(4323,2)(4324,0)(4325,0)(4326,0)(4327,0)(4328,0)(4329,0)(4330,0)(4331,0)(4332,2)(4333,0)(4334,0)(4335,3)(4336,0)(4337,1)(4338,0)(4339,0)(4340,0)(4341,0)(4342,6)(4343,0)(4344,0)(4345,0)(4346,3)(4347,0)(4348,0)(4349,1)(4350,0)(4351,0)(4352,0)(4353,0)(4354,0)(4355,0)(4356,0)(4357,0)(4358,0)(4359,0)(4360,0)(4361,9)(4362,6)(4363,0)(4364,0)(4365,0)(4366,0)(4367,2)(4368,0)(4369,0)(4370,0)(4371,1)(4372,0)(4373,1)(4374,0)(4375,0)(4376,0)(4377,0)(4378,0)(4379,0)(4380,0)(4381,1)(4382,0)(4383,0)(4384,2)(4385,0)(4386,0)(4387,63)(4388,0)(4389,0)(4390,0)(4391,6)(4392,7)(4393,0)(4394,8)(4395,1)(4396,94)(4397,0)(4398,0)(4399,0)(4400,0)(4401,0)(4402,0)(4403,2)(4404,0)(4405,0)(4406,0)(4407,0)(4408,0)(4409,0)(4410,1)(4411,1)(4412,0)(4413,0)(4414,0)(4415,0)(4416,0)(4417,0)(4418,0)(4419,0)(4420,12)(4421,0)(4422,0)(4423,0)(4424,3)(4425,925)(4426,0)(4427,0)(4428,32)(4429,1150)(4430,0)(4431,0)(4432,31)(4433,6)(4434,0)(4435,0)(4436,0)(4437,72)(4438,0)(4439,24)(4440,0)(4441,0)(4442,0)(4443,0)(4444,0)(4445,0)(4446,0)(4447,0)(4448,0)(4449,0)(4450,0)(4451,0)(4452,0)(4453,31)(4454,0)(4455,9)(4456,0)(4457,0)(4458,0)(4459,0)(4460,0)(4461,4)(4462,2)(4463,0)(4464,0)(4465,1)(4466,0)(4467,6)(4468,0)(4469,0)(4470,0)(4471,0)(4472,11)(4473,23)(4474,8)(4475,23)(4476,3)(4477,1)(4478,0)(4479,209)(4480,567)(4481,326)(4482,3)(4483,26)(4484,12)(4485,1)(4486,0)(4487,32)(4488,0)(4489,0)(4490,0)(4491,0)(4492,0)(4493,0)(4494,0)(4495,0)(4496,0)(4497,29)(4498,0)(4499,5)(4500,0)(4501,2)(4502,0)(4503,0)(4504,1)(4505,0)(4506,0)(4507,6)(4508,6)(4509,6)(4510,5)(4511,0)(4512,0)(4513,0)(4514,0)(4515,0)(4516,0)(4517,0)(4518,0)(4519,432)(4520,216)(4521,0)(4522,0)(4523,0)(4524,401)(4525,57)(4526,0)(4527,58)(4528,0)(4529,1)(4530,4)(4531,0)(4532,0)(4533,0)(4534,0)(4535,0)(4536,0)(4537,0)(4538,0)(4539,0)(4540,2)(4541,5)(4542,8)(4543,24)(4544,4)(4545,157)(4546,11)(4547,0)(4548,11)(4549,0)(4550,0)(4551,34)(4552,23)(4553,19)(4554,214)(4555,7)(4556,6)(4557,0)(4558,30)(4559,0)(4560,0)(4561,0)(4562,4)(4563,0)(4564,0)(4565,0)(4566,0)(4567,0)(4568,0)(4569,20)(4570,0)(4571,62)(4572,62)(4573,27)(4574,125)(4575,27)(4576,1)(4577,13)(4578,0)(4579,0)(4580,0)(4581,39)(4582,42)(4583,85)(4584,0)(4585,28)(4586,28)(4587,14)(4588,208)(4589,0)(4590,0)(4591,0)(4592,208)(4593,0)(4594,0)(4595,28)(4596,0)(4597,0)(4598,0)(4599,0)(4600,776)(4601,2533)(4602,0)(4603,388)(4604,14)(4605,0)(4606,0)(4607,0)(4608,115)(4609,14)(4610,0)(4611,0)(4612,0)(4613,55)(4614,0)(4615,208)(4616,104)(4617,312)(4618,0)(4619,0)(4620,71)(4621,14)(4622,0)(4623,59)(4624,479)(4625,0)(4626,1737)(4627,0)(4628,14)(4629,119)(4630,119)(4631,211)(4632,11)(4633,0)(4634,0)(4635,0)(4636,0)(4637,12)(4638,0)(4639,1)(4640,0)(4641,0)(4642,0)(4643,0)(4644,0)(4645,6)(4646,6)(4647,6)(4648,0)(4649,0)(4650,6)(4651,0)(4652,0)(4653,0)(4654,216)(4655,401)(4656,0)(4657,30)(4658,87)(4659,0)(4660,0)(4661,0)(4662,58)(4663,0)(4664,0)(4665,4)(4666,0)(4667,0)(4668,0)(4669,0)(4670,0)(4671,0)(4672,12)(4673,12)(4674,5)(4675,0)(4676,0)(4677,0)(4678,0)(4679,0)(4680,3)(4681,27)(4682,0)(4683,1)(4684,0)(4685,0)(4686,0)(4687,6)(4688,0)(4689,14)(4690,0)(4691,13)(4692,0)(4693,0)(4694,0)(4695,30)(4696,1)(4697,0)(4698,3)(4699,27)(4700,0)(4701,0)(4702,0)(4703,30)(4704,30)(4705,246)(4706,90)(4707,586)(4708,30)(4709,1)(4710,1)(4711,0)(4712,1)(4713,0)(4714,11)(4715,0)(4716,0)(4717,0)(4718,8)(4719,2)(4720,0)(4721,1)(4722,0)(4723,7)(4724,47)(4725,0)(4726,1)(4727,5)(4728,0)(4729,0)(4730,0)(4731,0)(4732,0)(4733,0)(4734,0)(4735,0)(4736,255)(4737,0)(4738,1)(4739,0)(4740,2)(4741,0)(4742,0)(4743,0)(4744,7)(4745,8)(4746,5)(4747,0)(4748,5)(4749,0)(4750,2)(4751,4)(4752,0)(4753,5)(4754,0)(4755,0)(4756,0)(4757,0)(4758,0)(4759,3)(4760,2)(4761,1)(4762,4)(4763,11)(4764,60)(4765,11)(4766,0)(4767,12)(4768,5)(4769,25)(4770,0)(4771,0)(4772,0)(4773,10)(4774,0)(4775,1)(4776,0)(4777,0)(4778,0)(4779,0)(4780,5)(4781,0)(4782,0)(4783,1)(4784,9)(4785,5)(4786,0)(4787,0)(4788,0)(4789,0)(4790,0)(4791,7)(4792,0)(4793,9)(4794,0)(4795,0)(4796,26)(4797,0)(4798,0)(4799,0)(4800,0)(4801,0)(4802,1)(4803,0)(4804,1)(4805,0)(4806,0)(4807,0)(4808,0)(4809,0)(4810,0)(4811,0)(4812,0)(4813,0)(4814,0)(4815,0)(4816,156)(4817,0)(4818,0)(4819,0)(4820,0)(4821,0)(4822,52)(4823,1)(4824,0)(4825,4)(4826,0)(4827,0)(4828,3)(4829,0)(4830,1)(4831,0)(4832,2)(4833,0)(4834,2)(4835,0)(4836,2)(4837,0)(4838,0)(4839,0)(4840,1)(4841,24)(4842,0)(4843,2)(4844,7)(4845,0)(4846,0)(4847,0)(4848,0)(4849,0)(4850,0)(4851,0)(4852,0)(4853,5)(4854,34)(4855,0)(4856,1)(4857,129)(4858,222)(4859,526)(4860,2)(4861,0)(4862,0)(4863,0)(4864,0)(4865,0)(4866,99)(4867,0)(4868,0)(4869,283)(4870,0)(4871,0)(4872,57)(4873,65)(4874,58)(4875,98)(4876,29)(4877,0)(4878,0)(4879,66)(4880,192)(4881,0)(4882,0)(4883,14)(4884,1)(4885,190)(4886,95)(4887,0)(4888,0)(4889,270)(4890,43)(4891,0)(4892,1)(4893,56)(4894,15)(4895,0)(4896,71)(4897,99)(4898,0)(4899,158)(4900,300)(4901,28)(4902,42)(4903,28)(4904,0)(4905,28)(4906,14)(4907,14)(4908,0)(4909,14)(4910,15)(4911,0)(4912,1)(4913,3)(4914,0)(4915,327)(4916,0)(4917,0)(4918,16)(4919,0)(4920,0)(4921,0)(4922,14)(4923,0)(4924,0)(4925,0)(4926,0)(4927,0)(4928,7)(4929,0)(4930,0)(4931,5)(4932,0)(4933,0)(4934,2)(4935,0)(4936,2)(4937,2)(4938,0)(4939,0)(4940,0)(4941,0)(4942,0)(4943,0)(4944,2)(4945,0)(4946,25)(4947,2)(4948,0)(4949,5)(4950,0)(4951,5)(4952,0)(4953,5)(4954,0)(4955,0)(4956,0)(4957,0)(4958,0)(4959,0)(4960,4)(4961,0)(4962,0)(4963,0)(4964,0)(4965,0)(4966,3)(4967,3)(4968,0)(4969,0)(4970,1)(4971,0)(4972,1)(4973,0)(4974,0)(4975,0)(4976,2)(4977,0)(4978,3)(4979,489)(4980,16)(4981,16)(4982,0)(4983,48)(4984,16)(4985,64)(4986,105)(4987,0)(4988,21)(4989,0)(4990,655)(4991,133)(4992,0)(4993,292)(4994,161)(4995,0)(4996,0)(4997,161)(4998,38)(4999,134)(5000,0)(5001,328)(5002,0)(5003,0)(5004,0)(5005,13)(5006,0)(5007,4)(5008,8)(5009,0)(5010,0)(5011,0)(5012,1)(5013,0)(5014,0)(5015,0)(5016,2)(5017,2)(5018,6)(5019,70)(5020,0)(5021,192)(5022,130)(5023,391)(5024,0)(5025,262)(5026,65)(5027,65)(5028,0)(5029,33)(5030,0)(5031,33)(5032,316)(5033,133)(5034,60)(5035,0)(5036,0)(5037,0)(5038,257)(5039,21)(5040,0)(5041,65)(5042,146)(5043,1)(5044,5)(5045,8)(5046,0)(5047,0)(5048,2)(5049,4)(5050,7)(5051,2)(5052,5)(5053,0)(5054,0)(5055,0)(5056,6)(5057,0)(5058,0)(5059,1)(5060,0)(5061,1)(5062,14)(5063,0)(5064,0)(5065,3)(5066,0)(5067,0)(5068,0)(5069,0)(5070,0)(5071,0)(5072,0)(5073,0)(5074,0)(5075,0)(5076,0)(5077,0)(5078,0)(5079,0)(5080,0)(5081,0)(5082,0)(5083,1)(5084,0)(5085,6)(5086,0)(5087,1)(5088,0)(5089,0)(5090,0)(5091,0)(5092,2)(5093,0)(5094,0)(5095,0)(5096,0)(5097,1)(5098,0)(5099,0)(5100,3)(5101,6)(5102,0)(5103,0)(5104,0)(5105,0)(5106,0)(5107,1)(5108,0)(5109,0)(5110,1)(5111,0)(5112,0)(5113,0)(5114,5)(5115,0)(5116,0)(5117,0)(5118,0)(5119,0)(5120,0)(5121,0)(5122,0)(5123,0)(5124,9)(5125,0)(5126,0)(5127,0)(5128,0)(5129,1)(5130,0)(5131,0)(5132,0)(5133,0)(5134,3)(5135,4)(5136,0)(5137,0)(5138,0)(5139,0)(5140,0)(5141,0)(5142,4)(5143,0)(5144,0)(5145,0)(5146,0)(5147,0)(5148,0)(5149,0)(5150,0)(5151,0)(5152,0)(5153,1)(5154,7)(5155,0)(5156,0)(5157,0)(5158,14)(5159,0)(5160,0)(5161,0)(5162,0)(5163,0)
};
\addlegendentry{RSA}
 
\end{axis}
\end{tikzpicture}

%% file: ent-aes.tex
\begin{tikzpicture}[scale=0.5]

\begin{axis}[
    xtick={0,1000,2000,3000,4000,5000},
    legend pos=north east,
    ymajorgrids=true,
    grid style=dashed,
]

\addplot[color=blue]
coordinates {
(0,0)(1,0)(2,1)(3,9)(4,0)(5,0)(6,0)(7,0)(8,0)(9,0)(10,0)(11,2)(12,0)(13,0)(14,2)(15,2)(16,0)(17,0)(18,0)(19,0)(20,0)(21,0)(22,0)(23,0)(24,0)(25,0)(26,9)(27,0)(28,0)(29,0)(30,0)(31,0)(32,0)(33,0)(34,4)(35,0)(36,0)(37,20)(38,13)(39,4)(40,1)(41,0)(42,0)(43,0)(44,0)(45,1)(46,4)(47,0)(48,0)(49,3)(50,0)(51,0)(52,0)(53,0)(54,2)(55,3)(56,0)(57,1)(58,0)(59,0)(60,0)(61,2)(62,1)(63,0)(64,0)(65,16)(66,0)(67,0)(68,0)(69,0)(70,0)(71,0)(72,8)(73,0)(74,2)(75,4)(76,0)(77,0)(78,1)(79,0)(80,1)(81,3)(82,3)(83,15)(84,3)(85,0)(86,0)(87,7)(88,0)(89,3)(90,2)(91,0)(92,2)(93,0)(94,0)(95,0)(96,1)(97,0)(98,0)(99,1)(100,0)(101,0)(102,0)(103,0)(104,0)(105,0)(106,0)(107,0)(108,0)(109,0)(110,0)(111,0)(112,0)(113,3)(114,0)(115,0)(116,0)(117,0)(118,7)(119,0)(120,0)(121,0)(122,0)(123,0)(124,0)(125,0)(126,0)(127,10)(128,0)(129,1)(130,0)(131,0)(132,0)(133,0)(134,0)(135,0)(136,0)(137,0)(138,0)(139,0)(140,0)(141,3)(142,0)(143,0)(144,0)(145,0)(146,0)(147,298)(148,0)(149,0)(150,0)(151,0)(152,0)(153,0)(154,5)(155,0)(156,0)(157,0)(158,0)(159,2)(160,1)(161,0)(162,0)(163,0)(164,0)(165,0)(166,0)(167,0)(168,1)(169,0)(170,3)(171,0)(172,0)(173,2)(174,0)(175,0)(176,0)(177,0)(178,0)(179,0)(180,2)(181,0)(182,1)(183,0)(184,1)(185,0)(186,0)(187,2)(188,2)(189,0)(190,2)(191,2)(192,0)(193,0)(194,0)(195,4)(196,5)(197,0)(198,8)(199,0)(200,0)(201,5)(202,8)(203,0)(204,0)(205,0)(206,0)(207,0)(208,0)(209,0)(210,0)(211,0)(212,0)(213,0)(214,0)(215,0)(216,0)(217,0)(218,0)(219,14)(220,0)(221,0)(222,1)(223,0)(224,0)(225,0)(226,0)(227,0)(228,0)(229,0)(230,1)(231,1)(232,1)(233,0)(234,1)(235,6)(236,0)(237,0)(238,3)(239,0)(240,1)(241,1)(242,0)(243,0)(244,0)(245,0)(246,0)(247,0)(248,0)(249,0)(250,3)(251,2)(252,0)(253,0)(254,2)(255,0)(256,0)(257,0)(258,0)(259,0)(260,3)(261,23)(262,1)(263,2)(264,0)(265,10)(266,1)(267,0)(268,0)(269,1)(270,1)(271,3)(272,0)(273,0)(274,0)(275,2)(276,18)(277,1)(278,0)(279,0)(280,1)(281,0)(282,10)(283,1)(284,0)(285,5)(286,0)(287,0)(288,12)(289,15)(290,0)(291,0)(292,0)(293,4)(294,0)(295,0)(296,29)(297,0)(298,0)(299,1)(300,0)(301,7)(302,0)(303,0)(304,0)(305,0)(306,3)(307,0)(308,0)(309,0)(310,0)(311,0)(312,0)(313,0)(314,0)(315,3)(316,0)(317,4)(318,0)(319,0)(320,0)(321,1)(322,0)(323,0)(324,1)(325,0)(326,0)(327,0)(328,0)(329,0)(330,0)(331,0)(332,3)(333,0)(334,0)(335,0)(336,0)(337,0)(338,0)(339,2)(340,0)(341,5)(342,3)(343,0)(344,0)(345,53)(346,0)(347,0)(348,1)(349,1)(350,0)(351,0)(352,0)(353,0)(354,1)(355,2)(356,0)(357,0)(358,0)(359,0)(360,0)(361,0)(362,0)(363,0)(364,0)(365,3)(366,3)(367,0)(368,0)(369,0)(370,0)(371,0)(372,0)(373,0)(374,0)(375,0)(376,1)(377,0)(378,0)(379,0)(380,0)(381,0)(382,0)(383,0)(384,0)(385,0)(386,0)(387,0)(388,0)(389,0)(390,0)(391,0)(392,5)(393,9)(394,0)(395,0)(396,3)(397,2)(398,0)(399,46)(400,0)(401,0)(402,0)(403,0)(404,0)(405,8)(406,0)(407,3)(408,0)(409,0)(410,0)(411,0)(412,0)(413,0)(414,0)(415,0)(416,0)(417,0)(418,0)(419,0)(420,0)(421,0)(422,0)(423,0)(424,0)(425,0)(426,0)(427,0)(428,0)(429,0)(430,0)(431,0)(432,0)(433,0)(434,0)(435,0)(436,0)(437,14)(438,0)(439,0)(440,0)(441,0)(442,0)(443,0)(444,0)(445,0)(446,1)(447,5)(448,0)(449,0)(450,0)(451,0)(452,1)(453,3)(454,3)(455,0)(456,1)(457,0)(458,0)(459,0)(460,1)(461,0)(462,0)(463,0)(464,0)(465,0)(466,3)(467,0)(468,0)(469,0)(470,0)(471,1)(472,2)(473,2)(474,3)(475,0)(476,0)(477,0)(478,0)(479,4)(480,0)(481,0)(482,0)(483,1)(484,1)(485,0)(486,0)(487,0)(488,0)(489,0)(490,0)(491,2)(492,1)(493,0)(494,0)(495,0)(496,0)(497,0)(498,0)(499,0)(500,0)(501,2)(502,3)(503,133)(504,0)(505,0)(506,0)(507,0)(508,0)(509,9)(510,14)(511,0)(512,0)(513,0)(514,0)(515,0)(516,10)(517,9)(518,0)(519,0)(520,0)(521,0)(522,1)(523,1)(524,1)(525,1)(526,1)(527,1)(528,0)(529,0)(530,0)(531,0)(532,0)(533,0)(534,0)(535,0)(536,0)(537,0)(538,0)(539,6)(540,0)(541,0)(542,0)(543,0)(544,1)(545,8)(546,0)(547,5)(548,0)(549,0)(550,17)(551,6)(552,1)(553,0)(554,0)(555,2)(556,0)(557,1)(558,0)(559,0)(560,0)(561,0)(562,0)(563,3)(564,1)(565,0)(566,0)(567,3)(568,0)(569,0)(570,0)(571,0)(572,0)(573,0)(574,0)(575,0)(576,0)(577,2)(578,0)(579,0)(580,0)(581,0)(582,0)(583,2)(584,0)(585,0)(586,0)(587,0)(588,0)(589,0)(590,0)(591,0)(592,0)(593,0)(594,0)(595,0)(596,0)(597,0)(598,1)(599,0)(600,0)(601,0)(602,0)(603,0)(604,0)(605,0)(606,0)(607,0)(608,0)(609,0)(610,1)(611,0)(612,0)(613,0)(614,0)(615,0)(616,2)(617,2)(618,0)(619,0)(620,0)(621,0)(622,0)(623,0)(624,0)(625,0)(626,0)(627,0)(628,0)(629,0)(630,0)(631,0)(632,0)(633,0)(634,0)(635,0)(636,0)(637,0)(638,0)(639,0)(640,0)(641,1)(642,0)(643,0)(644,3)(645,0)(646,0)(647,0)(648,0)(649,0)(650,0)(651,0)(652,0)(653,0)(654,0)(655,0)(656,0)(657,0)(658,0)(659,0)(660,5)(661,1)(662,0)(663,0)(664,0)(665,0)(666,0)(667,0)(668,0)(669,0)(670,0)(671,0)(672,0)(673,0)(674,0)(675,0)(676,0)(677,2)(678,0)(679,0)(680,2)(681,1)(682,0)(683,4)(684,0)(685,0)(686,0)(687,0)(688,0)(689,0)(690,1)(691,2)(692,1)(693,0)(694,0)(695,0)(696,0)(697,0)(698,0)(699,0)(700,0)(701,0)(702,0)(703,0)(704,0)(705,3)(706,0)(707,1)(708,0)(709,17)(710,0)(711,0)(712,0)(713,1)(714,0)(715,0)(716,0)(717,0)(718,5)(719,0)(720,0)(721,0)(722,0)(723,0)(724,0)(725,0)(726,0)(727,0)(728,0)(729,0)(730,0)(731,0)(732,0)(733,0)(734,2)(735,0)(736,0)(737,2)(738,4)(739,0)(740,2)(741,0)(742,0)(743,1)(744,2)(745,0)(746,0)(747,0)(748,0)(749,0)(750,0)(751,0)(752,1)(753,0)(754,0)(755,0)(756,2)(757,0)(758,0)(759,0)(760,44)(761,0)(762,0)(763,0)(764,0)(765,0)(766,0)(767,1)(768,0)(769,0)(770,0)(771,0)(772,0)(773,0)(774,0)(775,0)(776,1)(777,0)(778,0)(779,0)(780,0)(781,0)(782,0)(783,2)(784,0)(785,0)(786,0)(787,0)(788,0)(789,1)(790,0)(791,0)(792,0)(793,0)(794,0)(795,0)(796,0)(797,0)(798,1)(799,0)(800,0)(801,0)(802,0)(803,1)(804,0)(805,2)(806,0)(807,1)(808,0)(809,0)(810,0)(811,0)(812,0)(813,4)(814,0)(815,0)(816,0)(817,0)(818,0)(819,0)(820,0)(821,0)(822,0)(823,0)(824,0)(825,0)(826,0)(827,7)(828,0)(829,2)(830,0)(831,2)(832,0)(833,0)(834,0)(835,0)(836,0)(837,0)(838,0)(839,0)(840,0)(841,0)(842,0)(843,0)(844,0)(845,0)(846,0)(847,0)(848,0)(849,0)(850,0)(851,0)(852,0)(853,0)(854,1)(855,0)(856,4)(857,0)(858,0)(859,0)(860,0)(861,0)(862,0)(863,0)(864,1)(865,0)(866,0)(867,0)(868,0)(869,0)(870,0)(871,0)(872,0)(873,0)(874,0)(875,0)(876,0)(877,0)(878,9)(879,10)(880,0)(881,0)(882,0)(883,1)(884,0)(885,0)(886,0)(887,0)(888,1)(889,0)(890,0)(891,0)(892,0)(893,0)(894,1)(895,0)(896,1)(897,1)(898,0)(899,0)(900,0)(901,0)(902,3)(903,2)(904,2)(905,3)(906,3)(907,7)(908,1)(909,0)(910,0)(911,0)(912,0)(913,0)(914,0)(915,0)(916,0)(917,0)(918,0)(919,0)(920,0)(921,12)(922,0)(923,0)(924,0)(925,0)(926,0)(927,0)(928,0)(929,0)(930,4)(931,0)(932,0)(933,0)(934,0)(935,0)(936,0)(937,2)(938,2)(939,0)(940,0)(941,2)(942,0)(943,0)(944,1)(945,2)(946,7)(947,2)(948,2)(949,2)(950,1)(951,2)(952,0)(953,2)(954,0)(955,0)(956,0)(957,0)(958,0)(959,0)(960,0)(961,0)(962,0)(963,0)(964,1)(965,0)(966,1)(967,0)(968,7)(969,0)(970,0)(971,0)(972,0)(973,2)(974,0)(975,1)(976,0)(977,0)(978,0)(979,0)(980,5)(981,0)(982,0)(983,0)(984,0)(985,4)(986,0)(987,0)(988,11)(989,1)(990,8)(991,0)(992,1)(993,0)(994,0)(995,0)(996,0)(997,1)(998,1)(999,0)(1000,0)(1001,1)(1002,1)(1003,0)(1004,2)(1005,4)(1006,0)(1007,0)(1008,0)(1009,0)(1010,0)(1011,0)(1012,0)(1013,0)(1014,0)(1015,14)(1016,0)(1017,0)(1018,9)(1019,1)(1020,0)(1021,0)(1022,0)(1023,0)(1024,0)(1025,1)(1026,7)(1027,0)(1028,0)(1029,0)(1030,0)(1031,0)(1032,0)(1033,0)(1034,0)(1035,0)(1036,0)(1037,2)(1038,0)(1039,0)(1040,0)(1041,0)(1042,0)(1043,0)(1044,27)(1045,0)(1046,25)(1047,0)(1048,0)(1049,0)(1050,0)(1051,0)(1052,0)(1053,0)(1054,1)(1055,0)(1056,0)(1057,0)(1058,1)(1059,18)(1060,0)(1061,0)(1062,4)(1063,0)(1064,25)(1065,0)(1066,0)(1067,1)(1068,0)(1069,5)(1070,0)(1071,0)(1072,0)(1073,0)(1074,1)(1075,0)(1076,0)(1077,0)(1078,0)(1079,2)(1080,10)(1081,6)(1082,0)(1083,0)(1084,0)(1085,0)(1086,0)(1087,0)(1088,0)(1089,0)(1090,10)(1091,0)(1092,0)(1093,0)(1094,0)(1095,0)(1096,0)(1097,0)(1098,0)(1099,0)(1100,0)(1101,0)(1102,0)(1103,0)(1104,0)(1105,0)(1106,0)(1107,0)(1108,0)(1109,0)(1110,0)(1111,0)(1112,0)(1113,58)(1114,0)(1115,0)(1116,0)(1117,0)(1118,0)(1119,0)(1120,0)(1121,0)(1122,0)(1123,0)(1124,0)(1125,0)(1126,0)(1127,0)(1128,3)(1129,0)(1130,0)(1131,0)(1132,0)(1133,0)(1134,12)(1135,6)(1136,6)(1137,0)(1138,0)(1139,0)(1140,108)(1141,8)(1142,0)(1143,2)(1144,0)(1145,0)(1146,0)(1147,0)(1148,0)(1149,0)(1150,0)(1151,0)(1152,0)(1153,0)(1154,0)(1155,0)(1156,3)(1157,0)(1158,0)(1159,0)(1160,0)(1161,5)(1162,0)(1163,0)(1164,0)(1165,0)(1166,0)(1167,5)(1168,0)(1169,0)(1170,0)(1171,3)(1172,0)(1173,0)(1174,0)(1175,0)(1176,7)(1177,0)(1178,0)(1179,0)(1180,0)(1181,0)(1182,0)(1183,0)(1184,0)(1185,7)(1186,1)(1187,0)(1188,0)(1189,0)(1190,12)(1191,0)(1192,0)(1193,0)(1194,0)(1195,0)(1196,0)(1197,1188)(1198,0)(1199,0)(1200,0)(1201,2)(1202,1)(1203,0)(1204,0)(1205,0)(1206,0)(1207,206)(1208,0)(1209,0)(1210,1108)(1211,84)(1212,3)(1213,0)(1214,0)(1215,4)(1216,21)(1217,59)(1218,7)(1219,11)(1220,3)(1221,0)(1222,0)(1223,0)(1224,2)(1225,0)(1226,4)(1227,10)(1228,71)(1229,49)(1230,0)(1231,0)(1232,0)(1233,0)(1234,0)(1235,0)(1236,0)(1237,0)(1238,0)(1239,203)(1240,0)(1241,0)(1242,5)(1243,0)(1244,0)(1245,0)(1246,0)(1247,0)(1248,2)(1249,0)(1250,0)(1251,381)(1252,0)(1253,0)(1254,0)(1255,0)(1256,0)(1257,13)(1258,0)(1259,2)(1260,12)(1261,0)(1262,14)(1263,237)(1264,1)(1265,48)(1266,1027)(1267,1194)(1268,255)(1269,224)(1270,3813)(1271,22)(1272,23)(1273,0)(1274,217)(1275,239)(1276,109)(1277,414)(1278,45)(1279,6)(1280,166)(1281,184)(1282,0)(1283,0)(1284,0)(1285,0)(1286,0)(1287,0)(1288,0)(1289,19)(1290,0)(1291,5)(1292,170)(1293,9)(1294,16)(1295,7)(1296,30)(1297,0)(1298,0)(1299,163)(1300,6)(1301,22)(1302,64)(1303,4)(1304,1286)(1305,4)(1306,34)(1307,2)(1308,109)(1309,17)(1310,13)(1311,0)(1312,17)(1313,0)(1314,1493)(1315,85)(1316,0)(1317,22)(1318,32)(1319,1)(1320,0)(1321,77)(1322,0)(1323,0)(1324,0)(1325,575)(1326,0)(1327,428)(1328,13)(1329,40)(1330,12)(1331,282)(1332,0)(1333,0)(1334,31)(1335,295)(1336,0)(1337,60)(1338,0)(1339,0)(1340,3)(1341,0)(1342,0)(1343,0)(1344,0)(1345,0)(1346,0)(1347,0)(1348,0)(1349,0)(1350,1)(1351,24)(1352,1)(1353,3)(1354,0)(1355,18)(1356,0)(1357,23)(1358,1)(1359,60)(1360,14)(1361,24)(1362,3)(1363,0)(1364,2)(1365,26)(1366,0)(1367,80)(1368,0)(1369,0)(1370,0)(1371,0)(1372,0)(1373,34)(1374,0)(1375,0)(1376,1)(1377,0)(1378,1)(1379,19)(1380,0)(1381,0)(1382,0)(1383,0)(1384,36)(1385,111)(1386,0)(1387,0)(1388,94)(1389,120)(1390,61)(1391,5)(1392,0)(1393,128)(1394,155)(1395,7)(1396,80)(1397,20)(1398,17)(1399,0)(1400,0)(1401,0)(1402,0)(1403,0)(1404,87)(1405,2)(1406,0)(1407,6)(1408,0)(1409,0)(1410,68)(1411,1)(1412,0)(1413,7)(1414,35)(1415,2)(1416,518)(1417,0)(1418,0)(1419,0)(1420,0)(1421,1)(1422,2)(1423,0)(1424,1)(1425,0)(1426,0)(1427,0)(1428,0)(1429,0)(1430,4)(1431,0)(1432,0)(1433,0)(1434,0)(1435,0)(1436,0)(1437,0)(1438,0)(1439,0)(1440,0)(1441,0)(1442,0)(1443,2)(1444,0)(1445,15)(1446,315)(1447,173)(1448,50)(1449,0)(1450,1)(1451,94)(1452,3)(1453,0)(1454,95)(1455,16)(1456,0)(1457,51)(1458,105)(1459,0)(1460,0)(1461,0)(1462,0)(1463,2)(1464,0)(1465,0)(1466,192)(1467,6)(1468,0)(1469,0)(1470,0)(1471,0)(1472,0)(1473,0)(1474,2)(1475,0)(1476,2)(1477,0)(1478,7)(1479,0)(1480,15)(1481,0)(1482,0)(1483,0)(1484,0)(1485,0)(1486,0)(1487,0)(1488,0)(1489,196)(1490,0)(1491,0)(1492,0)(1493,0)(1494,0)(1495,0)(1496,0)(1497,3)(1498,4)(1499,0)(1500,1)(1501,0)(1502,6)(1503,4)(1504,0)(1505,6)(1506,3)(1507,0)(1508,7657)(1509,3)(1510,0)(1511,0)(1512,7)(1513,17)(1514,102)(1515,79)(1516,34)(1517,1)(1518,5)(1519,5)(1520,0)(1521,5)(1522,46)(1523,412)(1524,22)(1525,0)(1526,0)(1527,0)(1528,0)(1529,3)(1530,0)(1531,0)(1532,0)(1533,0)(1534,0)(1535,0)(1536,0)(1537,0)(1538,0)(1539,0)(1540,0)(1541,76)(1542,0)(1543,12)(1544,12)(1545,20)(1546,28)(1547,0)(1548,2)(1549,2)(1550,0)(1551,0)(1552,14)(1553,7)(1554,0)(1555,0)(1556,0)(1557,4)(1558,9)(1559,0)(1560,0)(1561,0)(1562,0)(1563,0)(1564,2)(1565,1)(1566,0)(1567,0)(1568,19)(1569,0)(1570,2)(1571,0)(1572,0)(1573,0)(1574,7)(1575,11)(1576,0)(1577,5)(1578,0)(1579,0)(1580,137)(1581,0)(1582,214)(1583,388)(1584,200)(1585,212)(1586,32)(1587,1599)(1588,7)(1589,0)(1590,13)(1591,195)(1592,0)(1593,0)(1594,0)(1595,0)(1596,40)(1597,5)(1598,192)(1599,51)(1600,193)(1601,197)(1602,860)(1603,0)(1604,687)(1605,0)(1606,0)(1607,0)(1608,0)(1609,0)(1610,0)(1611,0)(1612,80)(1613,40)(1614,0)(1615,3)(1616,38)(1617,62)(1618,55)(1619,128)(1620,59)(1621,338)(1622,0)(1623,0)(1624,0)(1625,0)(1626,0)(1627,2)(1628,0)(1629,28)(1630,61)(1631,40)(1632,73)(1633,4)(1634,0)(1635,0)(1636,0)(1637,0)(1638,0)(1639,0)(1640,0)(1641,0)(1642,0)(1643,2)(1644,0)(1645,0)(1646,99)(1647,0)(1648,0)(1649,0)(1650,0)(1651,0)(1652,0)(1653,3)(1654,0)(1655,0)(1656,0)(1657,0)(1658,0)(1659,0)(1660,0)(1661,0)(1662,0)(1663,0)(1664,1)(1665,0)(1666,0)(1667,0)(1668,0)(1669,0)(1670,0)(1671,0)(1672,11)(1673,0)(1674,0)(1675,0)(1676,0)(1677,0)(1678,0)(1679,0)(1680,0)(1681,8)(1682,0)(1683,0)(1684,1)(1685,0)(1686,0)(1687,0)(1688,0)(1689,0)(1690,0)(1691,0)(1692,0)(1693,0)(1694,3)(1695,0)(1696,0)(1697,0)(1698,0)(1699,8)(1700,0)(1701,0)(1702,0)(1703,0)(1704,28)(1705,12)(1706,0)(1707,0)(1708,12)(1709,0)(1710,0)(1711,0)(1712,0)(1713,0)(1714,0)(1715,8)(1716,0)(1717,0)(1718,10)(1719,0)(1720,3)(1721,5)(1722,0)(1723,0)(1724,0)(1725,3)(1726,113)(1727,8)(1728,22)(1729,3)(1730,0)(1731,0)(1732,0)(1733,0)(1734,0)(1735,0)(1736,0)(1737,0)(1738,0)(1739,0)(1740,2)(1741,7)(1742,4)(1743,1)(1744,0)(1745,0)(1746,4)(1747,4)(1748,0)(1749,8)(1750,0)(1751,7)(1752,0)(1753,0)(1754,0)(1755,0)(1756,0)(1757,0)(1758,0)(1759,0)(1760,0)(1761,0)(1762,0)(1763,0)(1764,0)(1765,0)(1766,0)(1767,0)(1768,0)(1769,0)(1770,0)(1771,0)(1772,0)(1773,0)(1774,0)(1775,0)(1776,0)(1777,3)(1778,0)(1779,0)(1780,0)(1781,3)(1782,0)(1783,0)(1784,0)(1785,0)(1786,0)(1787,0)(1788,0)(1789,0)(1790,0)(1791,0)(1792,0)(1793,0)(1794,0)(1795,0)(1796,0)(1797,0)(1798,0)(1799,0)(1800,0)(1801,0)(1802,0)(1803,0)(1804,0)(1805,7)(1806,0)(1807,0)(1808,1)(1809,2)(1810,0)(1811,0)(1812,0)(1813,0)(1814,0)(1815,1)(1816,0)(1817,2)(1818,13)(1819,0)(1820,0)(1821,1)(1822,1)(1823,0)(1824,0)(1825,0)(1826,0)(1827,5)(1828,26)(1829,0)(1830,0)(1831,0)(1832,0)(1833,0)(1834,0)(1835,0)(1836,0)(1837,0)(1838,0)(1839,0)(1840,0)(1841,0)(1842,1)(1843,0)(1844,0)(1845,0)(1846,0)(1847,0)(1848,0)(1849,7)(1850,5)(1851,0)(1852,0)(1853,0)(1854,0)(1855,0)(1856,2)(1857,47)(1858,0)(1859,0)(1860,0)(1861,0)(1862,0)(1863,0)(1864,0)(1865,1)(1866,0)(1867,2)(1868,6)(1869,4)(1870,0)(1871,0)(1872,6)(1873,5)(1874,0)(1875,0)(1876,23)(1877,3)(1878,0)(1879,1)(1880,0)(1881,0)(1882,4)(1883,3)(1884,12)(1885,0)(1886,0)(1887,0)(1888,1)(1889,0)(1890,0)(1891,0)(1892,4)(1893,0)(1894,0)(1895,1)(1896,13)(1897,3)(1898,5235)(1899,0)(1900,39)(1901,11)(1902,18)(1903,0)(1904,0)(1905,0)(1906,0)(1907,1)(1908,0)(1909,2)(1910,0)(1911,0)(1912,0)(1913,0)(1914,0)(1915,0)(1916,0)(1917,1)(1918,0)(1919,0)(1920,0)(1921,0)(1922,0)(1923,0)(1924,0)(1925,0)(1926,0)(1927,0)(1928,0)(1929,0)(1930,0)(1931,0)(1932,0)(1933,0)(1934,0)(1935,0)(1936,0)(1937,0)(1938,0)(1939,0)(1940,1)(1941,0)(1942,0)(1943,0)(1944,0)(1945,0)(1946,0)(1947,0)(1948,0)(1949,0)(1950,0)(1951,0)(1952,0)(1953,0)(1954,5)(1955,19)(1956,0)(1957,0)(1958,0)(1959,253)(1960,16)(1961,0)(1962,0)(1963,0)(1964,0)(1965,0)(1966,0)(1967,0)(1968,0)(1969,1)(1970,5)(1971,3)(1972,0)(1973,0)(1974,0)(1975,0)(1976,1)(1977,3)(1978,0)(1979,0)(1980,0)(1981,0)(1982,0)(1983,0)(1984,0)(1985,1)(1986,0)(1987,0)(1988,0)(1989,0)(1990,0)(1991,0)(1992,0)(1993,0)(1994,2)(1995,0)(1996,3)(1997,0)(1998,0)(1999,0)(2000,0)(2001,0)(2002,26)(2003,0)(2004,1033)(2005,345)(2006,0)(2007,1)(2008,1)(2009,687)(2010,0)(2011,341)(2012,0)(2013,985)(2014,0)(2015,2)(2016,40)(2017,264)(2018,0)(2019,0)(2020,0)(2021,0)(2022,14349)(2023,0)(2024,14)(2025,0)(2026,0)(2027,35)(2028,0)(2029,25)(2030,323)(2031,326)(2032,233)(2033,2)(2034,247)(2035,111)(2036,13891)(2037,692)(2038,250)(2039,816)(2040,172)(2041,1491)(2042,703)(2043,0)(2044,28)(2045,154)(2046,0)(2047,1436)(2048,0)(2049,583)(2050,3020)(2051,0)(2052,121)(2053,1)(2054,0)(2055,0)(2056,0)(2057,2)(2058,119)(2059,0)(2060,94)(2061,4765)(2062,0)(2063,1)(2064,543)(2065,3)(2066,5558)(2067,0)(2068,25947)(2069,2290)(2070,982)(2071,0)(2072,3663)(2073,2083)(2074,1506)(2075,39)(2076,1366)(2077,20755)(2078,0)(2079,39)(2080,23)(2081,58565)(2082,5369)(2083,5)(2084,31400)(2085,6919)(2086,466)(2087,2)(2088,28)(2089,218)(2090,181)(2091,1797)(2092,0)(2093,0)(2094,323)(2095,612)(2096,18550)(2097,2328)(2098,244)(2099,1660)(2100,112)(2101,205)(2102,1973)(2103,4314)(2104,35)(2105,41)(2106,1719)(2107,688)(2108,2)(2109,0)(2110,0)(2111,0)(2112,0)(2113,364)(2114,553)(2115,1152)(2116,0)(2117,296)(2118,0)(2119,903)(2120,68)(2121,12)(2122,0)(2123,0)(2124,84)(2125,1)(2126,80)(2127,1848)(2128,0)(2129,24087)(2130,462)(2131,5)(2132,38)(2133,3)(2134,0)(2135,0)(2136,12888)(2137,0)(2138,0)(2139,0)(2140,0)(2141,63)(2142,8)(2143,79)(2144,0)(2145,18)(2146,236)(2147,290)(2148,5)(2149,0)(2150,5)(2151,6)(2152,0)(2153,0)(2154,0)(2155,0)(2156,0)(2157,0)(2158,0)(2159,0)(2160,5)(2161,0)(2162,0)(2163,0)(2164,0)(2165,0)(2166,0)(2167,0)(2168,0)(2169,20)(2170,0)(2171,14)(2172,0)(2173,0)(2174,0)(2175,0)(2176,5)(2177,0)(2178,0)(2179,37)(2180,11)(2181,4)(2182,36)(2183,5)(2184,271)(2185,34)(2186,38)(2187,23)(2188,0)(2189,0)(2190,0)(2191,0)(2192,0)(2193,0)(2194,0)(2195,0)(2196,0)(2197,0)(2198,0)(2199,0)(2200,0)(2201,0)(2202,8)(2203,9)(2204,5)(2205,0)(2206,0)(2207,0)(2208,0)(2209,0)(2210,0)(2211,62)(2212,0)(2213,47)(2214,3)(2215,915)(2216,2521)(2217,0)(2218,43)(2219,3)(2220,0)(2221,695)(2222,67)(2223,0)(2224,0)(2225,1)(2226,0)(2227,3)(2228,4)(2229,0)(2230,1)(2231,0)(2232,1)(2233,2)(2234,0)(2235,1)(2236,0)(2237,20)(2238,2)(2239,0)(2240,5)(2241,1)(2242,8473)(2243,228)(2244,0)(2245,0)(2246,0)(2247,8847)(2248,54)(2249,11)(2250,0)(2251,1)(2252,5)(2253,9)(2254,1177)(2255,285)(2256,96)(2257,0)(2258,0)(2259,1734)(2260,13)(2261,66)(2262,1897)(2263,100)(2264,32)(2265,0)(2266,10)(2267,1)(2268,0)(2269,5)(2270,50)(2271,26)(2272,0)(2273,200)(2274,0)(2275,0)(2276,483)(2277,0)(2278,142)(2279,8)(2280,22)(2281,3984)(2282,10305)(2283,408)(2284,12)(2285,0)(2286,1884)(2287,0)(2288,1481)(2289,0)(2290,0)(2291,1)(2292,0)(2293,21)(2294,2)(2295,0)(2296,0)(2297,0)(2298,0)(2299,0)(2300,0)(2301,0)(2302,12)(2303,23)(2304,4)(2305,0)(2306,1)(2307,0)(2308,0)(2309,245)(2310,76)(2311,44)(2312,13)(2313,0)(2314,1)(2315,110)(2316,15)(2317,6)(2318,2)(2319,0)(2320,3)(2321,0)(2322,0)(2323,0)(2324,0)(2325,0)(2326,11)(2327,1016)(2328,1)(2329,3)(2330,0)(2331,0)(2332,0)(2333,0)(2334,0)(2335,0)(2336,0)(2337,0)(2338,2)(2339,0)(2340,5)(2341,0)(2342,3)(2343,94)(2344,0)(2345,26)(2346,40)(2347,88)(2348,1)(2349,1122)(2350,10)(2351,9)(2352,2)(2353,29)(2354,4)(2355,56)(2356,157)(2357,2792)(2358,0)(2359,1)(2360,5)(2361,0)(2362,0)(2363,0)(2364,19)(2365,13)(2366,4)(2367,19)(2368,10)(2369,2)(2370,0)(2371,4)(2372,0)(2373,4)(2374,0)(2375,3)(2376,3)(2377,7)(2378,32)(2379,8)(2380,0)(2381,0)(2382,1)(2383,0)(2384,2)(2385,8)(2386,3)(2387,10)(2388,98)(2389,0)(2390,5)(2391,1)(2392,124)(2393,3)(2394,1)(2395,0)(2396,0)(2397,0)(2398,0)(2399,0)(2400,0)(2401,4)(2402,0)(2403,9)(2404,0)(2405,0)(2406,420)(2407,0)(2408,3)(2409,5)(2410,7)(2411,0)(2412,0)(2413,152)(2414,0)(2415,0)(2416,7)(2417,1)(2418,0)(2419,0)(2420,4)(2421,3)(2422,0)(2423,0)(2424,1)(2425,8)(2426,0)(2427,0)(2428,0)(2429,0)(2430,0)(2431,0)(2432,0)(2433,0)(2434,0)(2435,0)(2436,0)(2437,1)(2438,0)(2439,0)(2440,2)(2441,0)(2442,177)(2443,27)(2444,7)(2445,3)(2446,1)(2447,0)(2448,0)(2449,7)(2450,0)(2451,0)(2452,0)(2453,13)(2454,30)(2455,1)(2456,0)(2457,0)(2458,0)(2459,0)(2460,0)(2461,4)(2462,0)(2463,27)(2464,0)(2465,1)(2466,4)(2467,28)(2468,3)(2469,2)(2470,0)(2471,0)(2472,0)(2473,0)(2474,0)(2475,0)(2476,0)(2477,0)(2478,25)(2479,0)(2480,144)(2481,142)(2482,26)(2483,67)(2484,9)(2485,0)(2486,0)(2487,0)(2488,0)(2489,4)(2490,0)(2491,1)(2492,6)(2493,50)(2494,0)(2495,0)(2496,15)(2497,89)(2498,0)(2499,3)(2500,18)(2501,68)(2502,64)(2503,6)(2504,188)(2505,120)(2506,39)(2507,1)(2508,0)(2509,11)(2510,13)(2511,4)(2512,27)(2513,0)(2514,0)(2515,0)(2516,0)(2517,0)(2518,0)(2519,0)(2520,0)(2521,0)(2522,0)(2523,0)(2524,4)(2525,1)(2526,20)(2527,23)(2528,3)(2529,0)(2530,2)(2531,0)(2532,2)(2533,37)(2534,3)(2535,0)(2536,16)(2537,6)(2538,0)(2539,3)(2540,187)(2541,0)(2542,33)(2543,4)(2544,0)(2545,0)(2546,24)(2547,101)(2548,31)(2549,15)(2550,0)(2551,1)(2552,14)(2553,2)(2554,5)(2555,23)(2556,0)(2557,0)(2558,1)(2559,0)(2560,0)(2561,0)(2562,3)(2563,0)(2564,0)(2565,0)(2566,0)(2567,0)(2568,0)(2569,0)(2570,0)(2571,0)(2572,3)(2573,0)(2574,2)(2575,0)(2576,3)(2577,67)(2578,13)(2579,0)(2580,0)(2581,0)(2582,0)(2583,10)(2584,2)(2585,0)(2586,1)(2587,403)(2588,17)(2589,1)(2590,0)(2591,362)(2592,1)(2593,0)(2594,1)(2595,1)(2596,22)(2597,1)(2598,1)(2599,6)(2600,1)(2601,7)(2602,1)(2603,0)(2604,4)(2605,0)(2606,1)(2607,0)(2608,0)(2609,0)(2610,0)(2611,8)(2612,0)(2613,67)(2614,8)(2615,7)(2616,0)(2617,9)(2618,0)(2619,0)(2620,0)(2621,0)(2622,1)(2623,0)(2624,1)(2625,0)(2626,4)(2627,0)(2628,0)(2629,0)(2630,0)(2631,0)(2632,3)(2633,22)(2634,49)(2635,12)(2636,12)(2637,135)(2638,0)(2639,103)(2640,2)(2641,0)(2642,45)(2643,0)(2644,0)(2645,0)(2646,0)(2647,3)(2648,0)(2649,0)(2650,247)(2651,129)(2652,9)(2653,280)(2654,65)(2655,0)(2656,0)(2657,3)(2658,0)(2659,0)(2660,0)(2661,0)(2662,0)(2663,7)(2664,1)(2665,8)(2666,0)(2667,0)(2668,13)(2669,2)(2670,11)(2671,319)(2672,0)(2673,34)(2674,185)(2675,0)(2676,0)(2677,83)(2678,8)(2679,0)(2680,0)(2681,52)(2682,4)(2683,7)(2684,0)(2685,0)(2686,5)(2687,181)(2688,6)(2689,5)(2690,2)(2691,0)(2692,2)(2693,0)(2694,25)(2695,0)(2696,2)(2697,0)(2698,0)(2699,0)(2700,0)(2701,0)(2702,7)(2703,0)(2704,0)(2705,13)(2706,70)(2707,0)(2708,0)(2709,1)(2710,0)(2711,0)(2712,1)(2713,0)(2714,0)(2715,4)(2716,0)(2717,0)(2718,0)(2719,7)(2720,0)(2721,0)(2722,1)(2723,0)(2724,885)(2725,136)(2726,7)(2727,7707)(2728,1512)(2729,318)(2730,235)(2731,399)(2732,7)(2733,386)(2734,4)(2735,0)(2736,341)(2737,221)(2738,169)(2739,0)(2740,0)(2741,0)(2742,0)(2743,0)(2744,0)(2745,0)(2746,0)(2747,1)(2748,0)(2749,0)(2750,0)(2751,1)(2752,63)(2753,19)(2754,134)(2755,0)(2756,0)(2757,0)(2758,15)(2759,0)(2760,27)(2761,0)(2762,2)(2763,0)(2764,0)(2765,74)(2766,12)(2767,4)(2768,7)(2769,1)(2770,0)(2771,0)(2772,0)(2773,0)(2774,0)(2775,0)(2776,6)(2777,0)(2778,0)(2779,0)(2780,0)(2781,121)(2782,140)(2783,134)(2784,96)(2785,24)(2786,2)(2787,29)(2788,6)(2789,0)(2790,1)(2791,4)(2792,0)(2793,0)(2794,0)(2795,49)(2796,0)(2797,0)(2798,0)(2799,1)(2800,0)(2801,2)(2802,0)(2803,2)(2804,0)(2805,0)(2806,0)(2807,0)(2808,0)(2809,0)(2810,1)(2811,1)(2812,0)(2813,0)(2814,0)(2815,0)(2816,0)(2817,0)(2818,0)(2819,0)(2820,0)(2821,0)(2822,4)(2823,2)(2824,25)(2825,2)(2826,1)(2827,2)(2828,0)(2829,8)(2830,3)(2831,0)(2832,2)(2833,0)(2834,4)(2835,0)(2836,0)(2837,0)(2838,0)(2839,2)(2840,4)(2841,0)(2842,2)(2843,0)(2844,0)(2845,0)(2846,16)(2847,1565)(2848,49)(2849,0)(2850,0)(2851,5)(2852,1827)(2853,1047)(2854,130)(2855,0)(2856,200)(2857,274)(2858,169)(2859,412)(2860,13)(2861,9)(2862,0)(2863,0)(2864,0)(2865,3)(2866,0)(2867,0)(2868,0)(2869,0)(2870,10)(2871,0)(2872,5)(2873,3)(2874,0)(2875,63)(2876,13)(2877,0)(2878,1)(2879,0)(2880,2)(2881,0)(2882,0)(2883,132)(2884,68)(2885,8)(2886,6)(2887,6)(2888,0)(2889,0)(2890,8)(2891,0)(2892,5)(2893,2)(2894,0)(2895,0)(2896,3)(2897,1)(2898,0)(2899,1657)(2900,0)(2901,814)(2902,1)(2903,2)(2904,0)(2905,412)(2906,944)(2907,193)(2908,111)(2909,0)(2910,2)(2911,0)(2912,93)(2913,0)(2914,0)(2915,417)(2916,34)(2917,2)(2918,162)(2919,0)(2920,0)(2921,7)(2922,59)(2923,1018)(2924,159)(2925,92)(2926,237)(2927,117)(2928,4941)(2929,1156)(2930,0)(2931,1)(2932,0)(2933,77)(2934,125)(2935,118)(2936,54)(2937,0)(2938,0)(2939,0)(2940,0)(2941,1)(2942,11)(2943,1)(2944,0)(2945,0)(2946,0)(2947,79)(2948,1)(2949,0)(2950,1)(2951,0)(2952,7)(2953,0)(2954,28)(2955,329)(2956,1)(2957,6)(2958,0)(2959,0)(2960,0)(2961,11)(2962,0)(2963,0)(2964,3)(2965,1)(2966,0)(2967,0)(2968,266)(2969,0)(2970,316)(2971,317)(2972,212)(2973,0)(2974,0)(2975,0)(2976,4)(2977,0)(2978,8)(2979,0)(2980,537)(2981,305)(2982,48)(2983,7770)(2984,0)(2985,910)(2986,251)(2987,15)(2988,19)(2989,2)(2990,5)(2991,0)(2992,448)(2993,0)(2994,11)(2995,128)(2996,0)(2997,257)(2998,1486)(2999,3)(3000,4776)(3001,0)(3002,0)(3003,333)(3004,0)(3005,0)(3006,0)(3007,1)(3008,1241)(3009,0)(3010,2)(3011,0)(3012,0)(3013,0)(3014,1)(3015,636)(3016,0)(3017,0)(3018,0)(3019,259)(3020,6)(3021,0)(3022,0)(3023,0)(3024,1856)(3025,1)(3026,0)(3027,0)(3028,4)(3029,0)(3030,4)(3031,0)(3032,115)(3033,0)(3034,3)(3035,2)(3036,0)(3037,0)(3038,229)(3039,25)(3040,284)(3041,210)(3042,71)(3043,0)(3044,8)(3045,2)(3046,0)(3047,5)(3048,0)(3049,0)(3050,15)(3051,0)(3052,86)(3053,0)(3054,0)(3055,0)(3056,24)(3057,5164)(3058,54)(3059,0)(3060,0)(3061,39)(3062,0)(3063,38)(3064,0)(3065,0)(3066,0)(3067,2)(3068,0)(3069,0)(3070,0)(3071,0)(3072,8)(3073,0)(3074,0)(3075,0)(3076,0)(3077,0)(3078,0)(3079,0)(3080,0)(3081,0)(3082,0)(3083,416)(3084,0)(3085,0)(3086,38)(3087,0)(3088,19)(3089,0)(3090,0)(3091,0)(3092,1948)(3093,0)(3094,2)(3095,0)(3096,0)(3097,0)(3098,0)(3099,0)(3100,0)(3101,2)(3102,1660)(3103,0)(3104,31)(3105,0)(3106,0)(3107,3)(3108,0)(3109,30)(3110,0)(3111,6)(3112,1)(3113,1)(3114,542)(3115,0)(3116,0)(3117,0)(3118,0)(3119,21)(3120,5)(3121,0)(3122,48)(3123,148)(3124,0)(3125,2)(3126,0)(3127,6)(3128,0)(3129,0)(3130,0)(3131,6)(3132,0)(3133,0)(3134,0)(3135,0)(3136,0)(3137,26)(3138,0)(3139,0)(3140,174)(3141,91)(3142,0)(3143,0)(3144,0)(3145,0)(3146,5)(3147,0)(3148,0)(3149,0)(3150,0)(3151,0)(3152,10)(3153,108)(3154,0)(3155,4)(3156,2)(3157,0)(3158,0)(3159,0)(3160,4)(3161,0)(3162,0)(3163,0)(3164,0)(3165,0)(3166,3539)(3167,0)(3168,0)(3169,0)(3170,0)(3171,0)(3172,0)(3173,0)(3174,0)(3175,0)(3176,0)(3177,0)(3178,0)(3179,0)(3180,0)(3181,0)(3182,0)(3183,0)(3184,0)(3185,0)(3186,0)(3187,0)(3188,0)(3189,0)(3190,0)(3191,0)(3192,0)(3193,0)(3194,0)(3195,0)(3196,0)(3197,2)(3198,0)(3199,0)(3200,0)(3201,0)(3202,7)(3203,1)(3204,1)(3205,11)(3206,5)(3207,10)(3208,0)(3209,2)(3210,37)(3211,0)(3212,0)(3213,0)(3214,0)(3215,0)(3216,0)(3217,0)(3218,0)(3219,0)(3220,0)(3221,2)(3222,0)(3223,0)(3224,1)(3225,0)(3226,0)(3227,6)(3228,0)(3229,0)(3230,2)(3231,60)(3232,35)(3233,6)(3234,0)(3235,0)(3236,86)(3237,0)(3238,0)(3239,0)(3240,0)(3241,2627)(3242,0)(3243,234)(3244,4)(3245,3)(3246,9)(3247,2)(3248,0)(3249,0)(3250,78)(3251,82)(3252,307)(3253,27)(3254,1255)(3255,248)(3256,62)(3257,0)(3258,0)(3259,0)(3260,0)(3261,26)(3262,6)(3263,4)(3264,0)(3265,0)(3266,25)(3267,0)(3268,51)(3269,1)(3270,166)(3271,65)(3272,14)(3273,19)(3274,14)(3275,1)(3276,0)(3277,7)(3278,14)(3279,0)(3280,0)(3281,4)(3282,0)(3283,267)(3284,0)(3285,0)(3286,0)(3287,149)(3288,0)(3289,10)(3290,0)(3291,72)(3292,0)(3293,11)(3294,0)(3295,490)(3296,0)(3297,0)(3298,2)(3299,0)(3300,0)(3301,0)(3302,2)(3303,0)(3304,0)(3305,111)(3306,0)(3307,0)(3308,1615)(3309,0)(3310,0)(3311,0)(3312,0)(3313,0)(3314,0)(3315,0)(3316,0)(3317,0)(3318,3)(3319,1)(3320,1)(3321,1)(3322,0)(3323,0)(3324,0)(3325,0)(3326,0)(3327,0)(3328,8)(3329,42)(3330,14)(3331,53)(3332,0)(3333,0)(3334,26)(3335,0)(3336,0)(3337,0)(3338,0)(3339,0)(3340,11)(3341,348)(3342,0)(3343,0)(3344,0)(3345,0)(3346,0)(3347,10)(3348,0)(3349,0)(3350,3)(3351,0)(3352,0)(3353,1)(3354,4)(3355,0)(3356,0)(3357,2070)(3358,106)(3359,22)(3360,0)(3361,0)(3362,0)(3363,3)(3364,22)(3365,0)(3366,5)(3367,45)(3368,0)(3369,0)(3370,0)(3371,0)(3372,0)(3373,0)(3374,34)(3375,27)(3376,0)(3377,538)(3378,0)(3379,0)(3380,0)(3381,27)(3382,39)(3383,300)(3384,248)(3385,58)(3386,43)(3387,0)(3388,0)(3389,5)(3390,0)(3391,88)(3392,90)(3393,120)(3394,1)(3395,0)(3396,0)(3397,0)(3398,0)(3399,0)(3400,20)(3401,0)(3402,1)(3403,0)(3404,0)(3405,1)(3406,3)(3407,5)(3408,58)(3409,0)(3410,2)(3411,0)(3412,0)(3413,0)(3414,0)(3415,0)(3416,9)(3417,3)(3418,29)(3419,0)(3420,0)(3421,0)(3422,0)(3423,1)(3424,0)(3425,0)(3426,0)(3427,0)(3428,28)(3429,0)(3430,0)(3431,5)(3432,0)(3433,0)(3434,44)(3435,0)(3436,1)(3437,3)(3438,0)(3439,34)(3440,1)(3441,12)(3442,0)(3443,1)(3444,1)(3445,0)(3446,1)(3447,0)(3448,0)(3449,0)(3450,33)(3451,23)(3452,0)(3453,0)(3454,0)(3455,0)(3456,0)(3457,0)(3458,0)(3459,0)(3460,0)(3461,0)(3462,0)(3463,7)(3464,0)(3465,0)(3466,0)(3467,0)(3468,0)(3469,277)(3470,0)(3471,2)(3472,0)(3473,0)(3474,0)(3475,0)(3476,0)(3477,31)(3478,1)(3479,0)(3480,0)(3481,0)(3482,0)(3483,0)(3484,7)(3485,0)(3486,0)(3487,3)(3488,8)(3489,0)(3490,0)(3491,0)(3492,1)(3493,0)(3494,0)(3495,15)(3496,223)(3497,0)(3498,0)(3499,1)(3500,3)(3501,0)(3502,0)(3503,0)(3504,0)(3505,4)(3506,0)(3507,0)(3508,0)(3509,0)(3510,0)(3511,0)(3512,2)(3513,61)(3514,66)(3515,54)(3516,0)(3517,0)(3518,16)(3519,0)(3520,0)(3521,167)(3522,0)(3523,193)(3524,48)(3525,5)(3526,0)(3527,6)(3528,0)(3529,5)(3530,291)(3531,538)(3532,0)(3533,2)(3534,0)(3535,0)(3536,0)(3537,0)(3538,0)(3539,0)(3540,1)(3541,238)(3542,211)(3543,0)(3544,0)(3545,1)(3546,6)(3547,1)(3548,0)(3549,1)(3550,1)(3551,1)(3552,1)(3553,1)(3554,248)(3555,0)(3556,314)(3557,0)(3558,1)(3559,8)(3560,0)(3561,9)(3562,1)(3563,13)(3564,5)(3565,13)(3566,0)(3567,0)(3568,0)(3569,2)(3570,0)(3571,0)(3572,0)(3573,0)(3574,7)(3575,0)(3576,0)(3577,0)(3578,5)(3579,0)(3580,0)(3581,1)(3582,1)(3583,0)(3584,1)(3585,0)(3586,0)(3587,0)(3588,2)(3589,8)(3590,0)(3591,0)(3592,0)(3593,0)(3594,0)(3595,0)(3596,0)(3597,1)(3598,0)(3599,0)(3600,129)(3601,1)(3602,0)(3603,0)(3604,19)(3605,0)(3606,0)(3607,0)(3608,0)(3609,0)(3610,7)(3611,10)(3612,2)(3613,52)(3614,1)(3615,0)(3616,0)(3617,4)(3618,3)(3619,1)(3620,0)(3621,0)(3622,0)(3623,0)(3624,0)(3625,0)(3626,0)(3627,1)(3628,0)(3629,6)(3630,0)(3631,0)(3632,0)(3633,0)(3634,0)(3635,0)(3636,1)(3637,8)(3638,233)(3639,0)(3640,0)(3641,5)(3642,0)(3643,0)(3644,0)(3645,0)(3646,0)(3647,1)(3648,0)(3649,0)(3650,0)(3651,0)(3652,0)(3653,0)(3654,2)(3655,2)(3656,4)(3657,23)(3658,0)(3659,9)(3660,0)(3661,0)(3662,0)(3663,0)(3664,0)(3665,0)(3666,0)(3667,0)(3668,1)(3669,0)(3670,4)(3671,0)(3672,0)(3673,0)(3674,0)(3675,0)(3676,0)(3677,3)(3678,0)(3679,2)(3680,0)(3681,0)(3682,8)(3683,0)(3684,150)(3685,3)(3686,0)(3687,3)(3688,2)(3689,8)(3690,1)(3691,6)(3692,5)(3693,0)(3694,0)(3695,0)(3696,1)(3697,0)(3698,4)(3699,0)(3700,0)(3701,9)(3702,0)(3703,1)(3704,0)(3705,0)(3706,4)(3707,2)(3708,0)(3709,1)(3710,0)(3711,0)(3712,0)(3713,4)(3714,0)(3715,0)(3716,1)(3717,35)(3718,0)(3719,1)(3720,3)(3721,0)(3722,0)(3723,0)(3724,0)(3725,0)(3726,0)(3727,1)(3728,6)(3729,0)(3730,0)(3731,2)(3732,0)(3733,0)(3734,0)(3735,321)(3736,0)(3737,0)(3738,0)(3739,0)(3740,0)(3741,0)(3742,0)(3743,2)(3744,0)(3745,1)(3746,47)(3747,29)(3748,6)(3749,0)(3750,24)(3751,1)(3752,1)(3753,0)(3754,1)(3755,13)(3756,3)(3757,0)(3758,0)(3759,0)(3760,6)(3761,0)(3762,3)(3763,0)(3764,0)(3765,0)(3766,1)(3767,0)(3768,0)(3769,0)(3770,1)(3771,0)(3772,0)(3773,0)(3774,0)(3775,0)(3776,0)(3777,0)(3778,0)(3779,0)(3780,0)(3781,0)(3782,0)(3783,0)(3784,1)(3785,0)(3786,141)(3787,1)(3788,1)(3789,0)(3790,0)(3791,9)(3792,0)(3793,0)(3794,1)(3795,0)(3796,0)(3797,0)(3798,0)(3799,2)(3800,0)(3801,0)(3802,0)(3803,0)(3804,0)(3805,0)(3806,8)(3807,0)(3808,3)(3809,86)(3810,7)(3811,0)(3812,0)(3813,0)(3814,0)(3815,4)(3816,0)(3817,0)(3818,0)(3819,33)(3820,0)(3821,0)(3822,16)(3823,0)(3824,0)(3825,272)(3826,0)(3827,126)(3828,0)(3829,0)(3830,0)(3831,0)(3832,0)(3833,0)(3834,167)(3835,0)(3836,9)(3837,0)(3838,3)(3839,0)(3840,46)(3841,2)(3842,0)(3843,5)(3844,9)(3845,0)(3846,0)(3847,0)(3848,0)(3849,0)(3850,0)(3851,0)(3852,0)(3853,0)(3854,0)(3855,1)(3856,0)(3857,0)(3858,1)(3859,5)(3860,0)(3861,0)(3862,0)(3863,4)(3864,1)(3865,0)(3866,1)(3867,1)(3868,0)(3869,3)(3870,3)(3871,20)(3872,0)(3873,0)(3874,3)(3875,4)(3876,0)(3877,2)(3878,0)(3879,0)(3880,0)(3881,0)(3882,0)(3883,0)(3884,0)(3885,15)(3886,0)(3887,0)(3888,1)(3889,0)(3890,0)(3891,0)(3892,7)(3893,0)(3894,3)(3895,0)(3896,5)(3897,2)(3898,10)(3899,31)(3900,4)(3901,0)(3902,17)(3903,0)(3904,1)(3905,0)(3906,0)(3907,0)(3908,0)(3909,0)(3910,0)(3911,0)(3912,0)(3913,0)(3914,20)(3915,0)(3916,0)(3917,0)(3918,2)(3919,2)(3920,0)(3921,0)(3922,73)(3923,419)(3924,10)(3925,0)(3926,0)(3927,0)(3928,0)(3929,0)(3930,0)(3931,0)(3932,0)(3933,0)(3934,1)(3935,2)(3936,2)(3937,0)(3938,0)(3939,0)(3940,0)(3941,0)(3942,0)(3943,2)(3944,10)(3945,0)(3946,11)(3947,0)(3948,2)(3949,0)(3950,0)(3951,9)(3952,0)(3953,0)(3954,0)(3955,0)(3956,0)(3957,0)(3958,0)(3959,0)(3960,0)(3961,88)(3962,56)(3963,346)(3964,0)(3965,11)(3966,0)(3967,0)(3968,0)(3969,0)(3970,0)(3971,0)(3972,0)(3973,0)(3974,0)(3975,0)(3976,0)(3977,0)(3978,0)(3979,0)(3980,11)(3981,81)(3982,0)(3983,0)(3984,0)(3985,8)(3986,7)(3987,0)(3988,1)(3989,9)(3990,0)(3991,1)(3992,33)(3993,0)(3994,0)(3995,0)(3996,0)(3997,0)(3998,0)(3999,0)(4000,0)(4001,1)(4002,9)(4003,6)(4004,0)(4005,97)(4006,0)(4007,0)(4008,0)(4009,0)(4010,0)(4011,2)(4012,0)(4013,0)(4014,0)(4015,0)(4016,0)(4017,214)(4018,0)(4019,0)(4020,0)(4021,0)(4022,0)(4023,0)(4024,0)(4025,0)(4026,0)(4027,0)(4028,0)(4029,0)(4030,0)(4031,0)(4032,11)(4033,0)(4034,7)(4035,65)(4036,0)(4037,12)(4038,5)(4039,0)(4040,0)(4041,44)(4042,0)(4043,48)(4044,3)(4045,0)(4046,0)(4047,46)(4048,0)(4049,62)(4050,22)(4051,0)(4052,0)(4053,1)(4054,12)(4055,69)(4056,0)(4057,1)(4058,46)(4059,10)(4060,0)(4061,0)(4062,0)(4063,0)(4064,0)(4065,0)(4066,0)(4067,17)(4068,5)(4069,0)(4070,42)(4071,2)(4072,3)(4073,0)(4074,0)(4075,0)(4076,0)(4077,0)(4078,0)(4079,0)(4080,0)(4081,0)(4082,0)(4083,0)(4084,0)(4085,11)(4086,69)(4087,3)(4088,2)(4089,0)(4090,0)(4091,0)(4092,0)(4093,36)(4094,0)(4095,27)(4096,8)(4097,0)(4098,25)(4099,13)(4100,8)(4101,0)(4102,0)(4103,1635)(4104,0)(4105,192)(4106,0)(4107,30)(4108,21)(4109,0)(4110,1994)(4111,0)(4112,0)(4113,0)(4114,0)(4115,0)(4116,0)(4117,0)(4118,0)(4119,0)(4120,0)(4121,0)(4122,3)(4123,0)(4124,5)(4125,0)(4126,0)(4127,0)(4128,0)(4129,0)(4130,3)(4131,97)(4132,182)(4133,30)(4134,0)(4135,0)(4136,220)(4137,0)(4138,2)(4139,0)(4140,45)(4141,0)(4142,22)(4143,0)(4144,0)(4145,60)(4146,0)(4147,17)(4148,5)(4149,0)(4150,0)(4151,159)(4152,14)(4153,19)(4154,250)(4155,0)(4156,0)(4157,113)(4158,0)(4159,5)(4160,0)(4161,0)(4162,0)(4163,0)(4164,238)(4165,0)(4166,0)(4167,32)(4168,0)(4169,22)(4170,196)(4171,160)(4172,1)(4173,0)(4174,0)(4175,0)(4176,0)(4177,42)(4178,8)(4179,1)(4180,2)(4181,0)(4182,0)(4183,0)(4184,20)(4185,4)(4186,0)(4187,0)(4188,4)(4189,144)(4190,0)(4191,56)(4192,0)(4193,138)(4194,23)(4195,0)(4196,96)(4197,23)(4198,1)(4199,0)(4200,3)(4201,0)(4202,0)(4203,0)(4204,0)(4205,0)(4206,0)(4207,0)(4208,6)(4209,0)(4210,3)(4211,2)(4212,19)(4213,0)(4214,0)(4215,0)(4216,0)(4217,0)(4218,0)(4219,0)(4220,0)(4221,0)(4222,0)(4223,0)(4224,0)(4225,0)(4226,0)(4227,0)(4228,0)(4229,0)(4230,0)(4231,0)(4232,0)(4233,0)(4234,19)(4235,0)(4236,0)(4237,8)(4238,0)(4239,372)(4240,0)(4241,0)(4242,2)(4243,11)(4244,1)(4245,6)(4246,0)(4247,0)(4248,4)(4249,0)(4250,0)(4251,0)(4252,1)(4253,40)(4254,0)(4255,52)(4256,0)(4257,0)(4258,3)(4259,12)(4260,0)(4261,4)(4262,154)(4263,110)(4264,0)(4265,14)(4266,130)(4267,20)(4268,4)(4269,0)(4270,18)(4271,0)(4272,29)(4273,218)(4274,0)(4275,0)(4276,11)(4277,497)(4278,0)(4279,26)(4280,5)(4281,16)(4282,0)(4283,0)(4284,0)(4285,0)(4286,0)(4287,0)(4288,0)(4289,0)(4290,0)(4291,0)(4292,0)(4293,0)(4294,0)(4295,19)(4296,118)(4297,139)(4298,575)(4299,0)(4300,0)(4301,24)(4302,0)(4303,0)(4304,0)(4305,0)(4306,0)(4307,0)(4308,0)(4309,3)(4310,0)(4311,9)(4312,23)(4313,0)(4314,7)(4315,0)(4316,204)(4317,0)(4318,8)(4319,4)(4320,7)(4321,78)(4322,23)(4323,1)(4324,0)(4325,0)(4326,20)(4327,0)(4328,22)(4329,18)(4330,0)(4331,0)(4332,1)(4333,0)(4334,13)(4335,0)(4336,0)(4337,0)(4338,13)(4339,0)(4340,146)(4341,141)(4342,4)(4343,84)(4344,0)(4345,2)(4346,30)(4347,2)(4348,0)(4349,1)(4350,278)(4351,127)(4352,347)(4353,5)(4354,0)(4355,0)(4356,129)(4357,6)(4358,0)(4359,0)(4360,8)(4361,137)(4362,87)(4363,0)(4364,26)(4365,0)(4366,39)(4367,13)(4368,0)(4369,58)(4370,35)(4371,16)(4372,17)(4373,26)(4374,2)(4375,0)(4376,0)(4377,0)(4378,0)(4379,0)(4380,0)(4381,1)(4382,0)(4383,0)(4384,0)(4385,0)(4386,0)(4387,0)(4388,0)(4389,0)(4390,0)(4391,0)(4392,2)(4393,0)(4394,0)(4395,0)(4396,4)(4397,0)(4398,7)(4399,1)(4400,2)(4401,0)(4402,5)(4403,1)(4404,0)(4405,1)(4406,0)(4407,0)(4408,1)(4409,4)(4410,0)(4411,0)(4412,4)(4413,5)(4414,151)(4415,0)(4416,0)(4417,11)(4418,3)(4419,2)(4420,63)(4421,639)(4422,8)(4423,6)(4424,0)(4425,33)(4426,0)(4427,29)(4428,0)(4429,443)(4430,10)(4431,0)(4432,0)(4433,0)(4434,35)(4435,0)(4436,0)(4437,0)(4438,11)(4439,0)(4440,36)(4441,0)(4442,0)(4443,0)(4444,0)(4445,0)(4446,0)(4447,0)(4448,0)(4449,0)(4450,0)(4451,0)(4452,0)(4453,0)(4454,130)(4455,23)(4456,2)(4457,0)(4458,0)(4459,6)(4460,0)(4461,1)(4462,0)(4463,0)(4464,0)(4465,0)(4466,0)(4467,0)(4468,0)(4469,12)(4470,0)(4471,0)(4472,58)(4473,49)(4474,19)(4475,27)(4476,4)(4477,0)(4478,1)(4479,4)(4480,0)(4481,141)(4482,0)(4483,16)(4484,8)(4485,12)(4486,4)(4487,1)(4488,13)(4489,33)(4490,1)(4491,2)(4492,3)(4493,0)(4494,0)(4495,0)(4496,0)(4497,4)(4498,3)(4499,0)(4500,3)(4501,9)(4502,0)(4503,0)(4504,1)(4505,0)(4506,0)(4507,0)(4508,0)(4509,14)(4510,0)(4511,0)(4512,0)(4513,0)(4514,0)(4515,0)(4516,1)(4517,13)(4518,83)(4519,53)(4520,6)(4521,0)(4522,15)(4523,0)(4524,34)(4525,18)(4526,0)(4527,0)(4528,98)(4529,261)(4530,0)(4531,29)(4532,0)(4533,17)(4534,0)(4535,0)(4536,0)(4537,0)(4538,0)(4539,0)(4540,71)(4541,23)(4542,0)(4543,0)(4544,6)(4545,0)(4546,1)(4547,59)(4548,8)(4549,504)(4550,39)(4551,15)(4552,51)(4553,21)(4554,41)(4555,0)(4556,0)(4557,2)(4558,23)(4559,9)(4560,0)(4561,3)(4562,0)(4563,0)(4564,0)(4565,0)(4566,3)(4567,9)(4568,3)(4569,6)(4570,0)(4571,0)(4572,0)(4573,3)(4574,0)(4575,0)(4576,29)(4577,10)(4578,0)(4579,0)(4580,0)(4581,46)(4582,0)(4583,0)(4584,28)(4585,14)(4586,0)(4587,0)(4588,17)(4589,2)(4590,4)(4591,3)(4592,0)(4593,0)(4594,151)(4595,0)(4596,0)(4597,0)(4598,0)(4599,0)(4600,0)(4601,3)(4602,2)(4603,0)(4604,9)(4605,54)(4606,0)(4607,0)(4608,0)(4609,2)(4610,6)(4611,2)(4612,7)(4613,0)(4614,0)(4615,0)(4616,0)(4617,195)(4618,9)(4619,0)(4620,20)(4621,0)(4622,41)(4623,0)(4624,1)(4625,0)(4626,6)(4627,0)(4628,4)(4629,0)(4630,0)(4631,4)(4632,0)(4633,0)(4634,8)(4635,0)(4636,0)(4637,0)(4638,0)(4639,0)(4640,0)(4641,0)(4642,35)(4643,4)(4644,0)(4645,0)(4646,30)(4647,10)(4648,16)(4649,15)(4650,0)(4651,0)(4652,0)(4653,12)(4654,0)(4655,6)(4656,18)(4657,0)(4658,0)(4659,0)(4660,0)(4661,0)(4662,0)(4663,0)(4664,3)(4665,6)(4666,1)(4667,0)(4668,0)(4669,6)(4670,2)(4671,0)(4672,0)(4673,2)(4674,0)(4675,6)(4676,0)(4677,0)(4678,1)(4679,0)(4680,1)(4681,10)(4682,0)(4683,289)(4684,0)(4685,93)(4686,113)(4687,31)(4688,12)(4689,0)(4690,119)(4691,606)(4692,63)(4693,0)(4694,0)(4695,0)(4696,13)(4697,6)(4698,0)(4699,5)(4700,0)(4701,0)(4702,0)(4703,0)(4704,6)(4705,12)(4706,0)(4707,0)(4708,15)(4709,0)(4710,0)(4711,0)(4712,0)(4713,17)(4714,0)(4715,32)(4716,3)(4717,4)(4718,0)(4719,0)(4720,0)(4721,0)(4722,0)(4723,0)(4724,0)(4725,81)(4726,0)(4727,0)(4728,0)(4729,0)(4730,0)(4731,0)(4732,1)(4733,0)(4734,0)(4735,0)(4736,0)(4737,1)(4738,6)(4739,0)(4740,0)(4741,0)(4742,207)(4743,0)(4744,0)(4745,68)(4746,0)(4747,0)(4748,0)(4749,0)(4750,77)(4751,0)(4752,0)(4753,0)(4754,0)(4755,0)(4756,46)(4757,0)(4758,310)(4759,0)(4760,0)(4761,0)(4762,0)(4763,0)(4764,0)(4765,10)(4766,0)(4767,6)(4768,15)(4769,0)(4770,0)(4771,12)(4772,0)(4773,4)(4774,0)(4775,0)(4776,13)(4777,9)(4778,10)(4779,6)(4780,0)(4781,0)(4782,7)(4783,0)(4784,5)(4785,0)(4786,0)(4787,0)(4788,7)(4789,0)(4790,0)(4791,6)(4792,0)(4793,0)(4794,0)(4795,0)(4796,0)(4797,0)(4798,13)(4799,0)(4800,28)(4801,0)(4802,0)(4803,0)(4804,0)(4805,34)(4806,0)(4807,70)(4808,20)(4809,12)(4810,578)(4811,0)(4812,0)(4813,158)(4814,0)(4815,0)(4816,0)(4817,0)(4818,4)(4819,0)(4820,44)(4821,0)(4822,0)(4823,11)(4824,64)(4825,37)(4826,0)(4827,408)(4828,0)(4829,0)(4830,8)(4831,0)(4832,0)(4833,3)(4834,6)(4835,2)(4836,5)(4837,0)(4838,0)(4839,0)(4840,10)(4841,5)(4842,20)(4843,5)(4844,10)(4845,10)(4846,0)(4847,0)(4848,8)(4849,0)(4850,0)(4851,0)(4852,0)(4853,0)(4854,0)(4855,1)(4856,0)(4857,0)(4858,6)(4859,0)(4860,0)(4861,0)(4862,0)(4863,1)(4864,0)(4865,0)(4866,0)(4867,0)(4868,0)(4869,0)(4870,0)(4871,0)(4872,0)(4873,0)(4874,0)(4875,0)(4876,0)(4877,27)(4878,0)(4879,0)(4880,0)(4881,3)(4882,2)(4883,0)(4884,0)(4885,0)(4886,0)(4887,0)(4888,0)(4889,0)(4890,6)(4891,3)(4892,0)(4893,3)(4894,3)(4895,0)(4896,36)(4897,17)(4898,418)(4899,65)(4900,0)(4901,0)(4902,0)(4903,2)(4904,0)(4905,0)(4906,0)(4907,0)(4908,0)(4909,0)(4910,0)(4911,0)(4912,0)(4913,0)(4914,116)(4915,4)(4916,0)(4917,0)(4918,0)(4919,3)(4920,0)(4921,0)(4922,1)(4923,0)(4924,0)(4925,0)(4926,0)(4927,0)(4928,4)(4929,3)(4930,0)(4931,0)(4932,0)(4933,2)(4934,2)(4935,0)(4936,0)(4937,0)(4938,3)(4939,3)(4940,0)(4941,0)(4942,2)(4943,0)(4944,0)(4945,0)(4946,4)(4947,0)(4948,0)(4949,1)(4950,0)(4951,8)(4952,0)(4953,0)(4954,1)(4955,0)(4956,0)(4957,0)(4958,0)(4959,5)(4960,3)(4961,1)(4962,0)(4963,6)(4964,0)(4965,2)(4966,0)(4967,2)(4968,0)(4969,0)(4970,3)(4971,18)(4972,0)(4973,0)(4974,5)(4975,0)(4976,0)(4977,0)(4978,0)(4979,96)(4980,0)(4981,82)(4982,2)(4983,1)(4984,3)(4985,0)(4986,2)(4987,2)(4988,0)(4989,0)(4990,0)(4991,0)(4992,0)(4993,0)(4994,0)(4995,0)(4996,1)(4997,3)(4998,39)(4999,0)(5000,0)(5001,104)(5002,0)(5003,0)(5004,5)(5005,0)(5006,0)(5007,4)(5008,0)(5009,0)(5010,0)(5011,0)(5012,0)(5013,0)(5014,0)(5015,0)(5016,353)(5017,32)(5018,91)(5019,0)(5020,0)(5021,0)(5022,500)(5023,0)(5024,7)(5025,0)(5026,0)(5027,82)(5028,164)(5029,0)(5030,0)(5031,0)(5032,0)(5033,0)(5034,0)(5035,0)(5036,0)(5037,0)(5038,0)(5039,4)(5040,0)(5041,0)(5042,3)(5043,0)(5044,0)(5045,1)(5046,0)(5047,1)(5048,0)(5049,1)(5050,68)(5051,24)(5052,0)(5053,0)(5054,0)(5055,0)(5056,14)(5057,71)(5058,71)(5059,28)(5060,0)(5061,0)(5062,0)(5063,0)(5064,2)(5065,0)(5066,1)(5067,1)(5068,4)(5069,0)(5070,0)(5071,0)(5072,5)(5073,0)(5074,0)(5075,10)(5076,0)(5077,0)(5078,8)(5079,0)(5080,0)(5081,0)(5082,0)(5083,0)(5084,0)(5085,0)(5086,0)(5087,0)(5088,0)(5089,2)(5090,1)(5091,0)(5092,0)(5093,1)(5094,0)(5095,0)(5096,0)(5097,0)(5098,0)(5099,0)(5100,0)(5101,0)(5102,0)(5103,0)(5104,4)(5105,0)(5106,0)(5107,2)(5108,2)(5109,3)(5110,0)(5111,0)(5112,9)(5113,1)(5114,0)(5115,0)(5116,1)(5117,0)(5118,0)(5119,0)(5120,0)(5121,9)(5122,0)(5123,0)(5124,0)(5125,0)(5126,0)(5127,0)(5128,0)(5129,0)(5130,0)(5131,3)(5132,0)(5133,0)(5134,7)(5135,1)(5136,0)(5137,0)(5138,0)(5139,0)(5140,0)(5141,15)(5142,0)(5143,0)(5144,0)(5145,0)(5146,0)(5147,7)(5148,3)(5149,0)(5150,2)(5151,4)(5152,0)(5153,0)(5154,0)(5155,1)(5156,0)(5157,0)(5158,2)(5159,0)(5160,0)(5161,0)(5162,0)(5163,0)(5164,0)(5165,0)(5166,0)(5167,0)(5168,7)(5169,0)(5170,9)(5171,0)(5172,0)(5173,0)(5174,0)(5175,14)(5176,0)(5177,2)(5178,0)(5179,0)(5180,0)(5181,0)(5182,0)(5183,0)(5184,0)(5185,0)(5186,0)(5187,0)(5188,0)(5189,0)(5190,0)(5191,0)(5192,0)(5193,0)(5194,0)(5195,0)(5196,51)(5197,129)(5198,66)(5199,3)(5200,0)(5201,0)(5202,3)(5203,0)(5204,3)(5205,0)(5206,2)(5207,0)(5208,2)(5209,0)(5210,2)(5211,0)(5212,0)(5213,0)(5214,19)(5215,0)(5216,0)(5217,7)(5218,1)(5219,0)(5220,2)(5221,0)(5222,104)(5223,0)(5224,0)(5225,0)(5226,0)(5227,0)(5228,1)(5229,0)(5230,0)(5231,0)(5232,0)(5233,2379)(5234,2)(5235,32)(5236,0)(5237,0)(5238,100)(5239,0)(5240,0)(5241,0)(5242,0)(5243,0)(5244,0)(5245,0)(5246,50)(5247,0)(5248,14)(5249,50)(5250,0)(5251,0)(5252,3)(5253,0)(5254,0)(5255,0)(5256,4)(5257,0)(5258,118)(5259,61)(5260,0)(5261,110)(5262,2)(5263,128)(5264,0)(5265,160)(5266,0)(5267,15)(5268,0)(5269,157)(5270,158)(5271,0)(5272,0)(5273,0)(5274,14)(5275,222)(5276,180)(5277,71)(5278,14)(5279,0)(5280,71)(5281,71)(5282,72)(5283,57)(5284,0)(5285,28)(5286,0)(5287,0)(5288,118)(5289,0)(5290,1)(5291,4)(5292,0)(5293,0)(5294,0)(5295,0)(5296,0)(5297,0)(5298,151)(5299,70)(5300,0)(5301,14)(5302,150)(5303,0)(5304,3)(5305,0)(5306,4)(5307,0)(5308,0)(5309,9)(5310,0)(5311,0)(5312,5)(5313,0)(5314,0)(5315,2)(5316,0)(5317,0)(5318,1)(5319,0)(5320,0)(5321,0)(5322,0)(5323,0)(5324,0)(5325,0)(5326,1)(5327,0)(5328,0)(5329,0)(5330,0)(5331,4)(5332,0)(5333,323)(5334,3)(5335,3)(5336,26)(5337,2163)(5338,0)(5339,151)(5340,0)(5341,0)(5342,0)(5343,0)(5344,1)(5345,0)(5346,166)(5347,24)(5348,0)(5349,0)(5350,0)(5351,95)(5352,0)(5353,0)(5354,13)(5355,7)(5356,22)(5357,185)(5358,0)(5359,16)(5360,101)(5361,36)(5362,0)(5363,0)(5364,3)(5365,378)(5366,1)(5367,6)(5368,3)(5369,0)(5370,26)(5371,13)(5372,1)(5373,179)(5374,0)(5375,1)(5376,0)(5377,0)(5378,0)(5379,0)(5380,0)(5381,0)(5382,0)(5383,0)(5384,2)(5385,0)(5386,5)(5387,0)(5388,0)(5389,0)(5390,0)(5391,0)(5392,0)(5393,4)(5394,0)(5395,0)(5396,0)(5397,0)(5398,3)(5399,3)(5400,0)(5401,0)(5402,0)(5403,0)(5404,2)(5405,2)(5406,262)(5407,787)(5408,128)(5409,0)(5410,0)(5411,0)(5412,16)(5413,96)(5414,92)(5415,75)(5416,7)(5417,206)(5418,194)(5419,17)(5420,0)(5421,161)(5422,108)(5423,4)(5424,46)(5425,157)(5426,0)(5427,0)(5428,1)(5429,0)(5430,0)(5431,0)(5432,43)(5433,0)(5434,0)(5435,0)(5436,0)(5437,4)(5438,0)(5439,2)(5440,0)(5441,0)(5442,0)(5443,2)(5444,0)(5445,0)(5446,3)(5447,3)(5448,11)(5449,106)(5450,98)(5451,0)(5452,0)(5453,302)(5454,0)(5455,94)(5456,47)(5457,100)(5458,10)(5459,1556)(5460,98)(5461,0)(5462,0)(5463,0)(5464,0)(5465,146)(5466,196)(5467,0)(5468,94)(5469,0)(5470,142)(5471,0)(5472,4)(5473,7)(5474,13)(5475,0)(5476,0)(5477,0)(5478,0)(5479,4)(5480,6)(5481,18)(5482,5)(5483,0)(5484,0)(5485,0)(5486,6)(5487,0)(5488,0)(5489,1)(5490,0)(5491,1)(5492,14)(5493,0)(5494,0)(5495,3)(5496,0)(5497,1)(5498,0)(5499,0)(5500,0)(5501,0)(5502,0)(5503,0)(5504,0)(5505,0)(5506,0)(5507,0)(5508,0)(5509,0)(5510,0)(5511,0)(5512,0)(5513,0)(5514,0)(5515,3)(5516,0)(5517,0)(5518,0)(5519,0)(5520,0)(5521,2)(5522,0)(5523,0)(5524,0)(5525,0)(5526,1)(5527,0)(5528,0)(5529,3)(5530,6)(5531,0)(5532,0)(5533,0)(5534,0)(5535,0)(5536,1)(5537,0)(5538,0)(5539,1)(5540,0)(5541,0)(5542,0)(5543,5)(5544,0)(5545,0)(5546,0)(5547,0)(5548,0)(5549,0)(5550,0)(5551,0)(5552,0)(5553,9)(5554,0)(5555,0)(5556,0)(5557,0)(5558,1)(5559,0)(5560,0)(5561,0)(5562,0)(5563,3)(5564,4)(5565,0)(5566,0)(5567,0)(5568,0)(5569,0)(5570,0)(5571,4)(5572,0)(5573,0)(5574,0)(5575,0)(5576,0)(5577,0)(5578,0)(5579,0)(5580,0)(5581,0)(5582,1)(5583,7)(5584,0)(5585,0)(5586,0)(5587,14)(5588,0)(5589,0)(5590,0)(5591,0)(5592,0)
};
\addlegendentry{AES}

\end{axis}
\end{tikzpicture}%

%% file: cnn_arch.tex
\begin{figure}[!h]
	\centering
	\begin{tikzpicture}[scale=.35,every node/.style={minimum size=1cm}]

	\begin{scope}[xshift=0cm,yshift=0cm]
	\draw[draw=base03,fill=blue,thick]
		(0,0) grid (7,4) rectangle (0,0);
	\draw[draw=base03,fill=red,thick]
		(0,0) grid (3,1) rectangle (0,0);
	\end{scope}

	\foreach \x in {-8,6} {%
            	\begin{scope}[xshift=\x cm,yshift=6cm]
		\draw[draw=base03,fill=violet,thick]
                        (0,0) grid (9,4) rectangle (0,0);
		\draw[draw=base03,fill=red,thick]
			(0,0) grid (0,0) rectangle (1,1);
		\draw[draw=base03,fill=green, thick]
			(2,1) grid (7,2) rectangle (2,1);
		\end{scope}
	}
	\foreach \x in {-6,8} {%
		\begin{scope}[xshift=\x cm,yshift=12cm]
		\draw[draw=base03,fill=cyan,thick]
			(0,0) grid (5,4) rectangle (0,0);
		\draw[draw=base03,fill=green,thick]
			(0,1) grid (5,2) rectangle (0,1);
		\draw[draw=base03,fill=red,thick]
			(2,3) grid (4,4) rectangle (2,3);
		\end{scope}
	}
	\foreach \x in {-6,7} {%
		\begin{scope}[xshift=\x cm,yshift=18cm]
		\draw[draw=base03,fill=violet,thick]
			(0,0) grid (6,4) rectangle (0,0);
		\draw[draw=base03,fill=red,thick]
			(0,1) grid (1,2) rectangle (0,1);
		\draw[draw=base03,fill=yellow,thick]
			(5,0) grid (6,2) rectangle (5,0);
		\end{scope}
		\begin{scope}[xshift=\x cm,yshift=24cm]
		\draw[draw=base03,fill=blue,thick]
			(0,0) grid (6,2) rectangle (0,0);
		\draw[draw=base03,fill=green,thick]
			(0,1) grid (3,2) rectangle (0,1);
		\draw[draw=base03,fill=yellow,thick]
			(5,0) grid (6,1) rectangle (5,0);
		\end{scope}
	}
	\foreach \x in {-4,8} {%
		\begin{scope}[xshift=\x cm,yshift=28cm]
		\draw[draw=base03,fill=cyan,thick]
			(0,0) grid (3,2) rectangle (0,0);
		\draw[draw=base03,fill=green,thick]
			(0,0) grid (1,1) rectangle (0,0);
            	\draw[draw=base03,fill=white,thick]
                	(2,0) grid (3,2) rectangle (2,0);
		\end{scope}
        }
  
        \begin{scope}[xshift=2cm,yshift=31cm]
            \draw[draw=base03,fill=white,thick]
                (0,0) grid (4,1) rectangle (0,0);
        \end{scope}

	\begin{scope}[xshift=0cm,yshift=-5cm]
	\node[fill=red,
		draw,
		minimum height=.1cm,
		minimum width=.1cm,
		label=west:Convolution
	] at (0,2) {};
	\node[fill=green,
		draw,
		minimum height=.1cm,
		minimum width=.1cm,
		label=west:k-Max Pooling
	] at (12,2) {};
	\node[fill=yellow,
		draw,
		minimum height=.1cm,
		minimum width=.1cm,
		label=west:Folding
	] at (0,0) {};
	\node[fill=white,
		draw,
		minimum height=.1cm,
		minimum width=.1cm,
		label=west:Fully Connected
	] at (12,0) {};
	\end{scope}


    \end{tikzpicture}
    \caption{\label{fig:model} Dynamic Convolutional Neural Network (DCNN)}
    \medskip
    \small
    The architecture of a DCNN as illustrated by \cite{DBLP:journals/corr/KalchbrennerGB14}. In this case, the model is intended for a seven word input sentence of embedding $d=4$ with two convolutional layers ($m = 3$ and $m = 2$), two (dynamic) k-max pooling layers ($k = 5$ and $k = 3$) within two feature maps.    
\end{figure}

%% file: accuracy.tex
\begin{figure}[t]
	\centering
	\begin{tikzpicture}

\begin{axis}[
    xlabel={Epoch},
    ylabel={Accuracy},
    legend pos=north west,
    ymajorgrids=true,
    grid style=dashed,
]

\addplot[color=blue]
coordinates {
(1,16.6666666667)(2,33.3333333333)(3,33.3333333333)(4,42.0)(5,58.0)(6,58.0)(7,64.6666666667)(8,74.0)(9,77.3333333333)(10,78.0)(11,78.6666666667)(12,78.6666666667)(13,78.6666666667)(14,78.6666666667)(15,78.6666666667)(16,78.6666666667)(17,78.6666666667)(18,78.6666666667)(19,79.3333333333)(20,79.3333333333)(21,79.3333333333)(22,79.3333333333)(23,79.3333333333)(24,79.3333333333)(25,79.3333333333)(26,79.3333333333)(27,79.3333333333)(28,79.3333333333)(29,79.3333333333)(30,79.3333333333)(31,79.3333333333)(32,79.3333333333)(33,84.6666666667)(34,87.3333333333)(35,87.3333333333)(36,87.3333333333)(37,87.3333333333)(38,87.3333333333)(39,92.0)(40,92.0)(41,92.0)(42,92.0)(43,92.0)(44,92.0)(45,92.0)(46,92.0)(47,92.0)(48,92.0)(49,92.0)(50,92.0)(51,91.3333333333)(52,91.3333333333)(53,91.3333333333)(54,91.3333333333)(55,91.3333333333)(56,91.3333333333)(57,91.3333333333)(58,91.3333333333)(59,91.3333333333)(60,91.3333333333)(61,91.3333333333)(62,91.3333333333)(63,91.3333333333)(64,91.3333333333)(65,91.3333333333)(66,91.3333333333)(67,91.3333333333)(68,91.3333333333)(69,91.3333333333)(70,91.3333333333)(71,91.3333333333)(72,91.3333333333)(73,91.3333333333)(74,91.3333333333)(75,91.3333333333)(76,91.3333333333)(77,91.3333333333)(78,91.3333333333)(79,91.3333333333)(80,91.3333333333)(81,91.3333333333)(82,91.3333333333)(83,91.3333333333)(84,91.3333333333)(85,91.3333333333)(86,91.3333333333)(87,91.3333333333)(88,91.3333333333)(89,91.3333333333)(90,91.3333333333)(91,91.3333333333)(92,91.3333333333)(93,91.3333333333)(94,91.3333333333)(95,91.3333333333)(96,91.3333333333)(97,91.3333333333)(98,91.3333333333)(99,91.3333333333)(100,91.3333333333)(101,91.3333333333)(102,91.3333333333)(103,91.3333333333)(104,91.3333333333)(105,91.3333333333)(106,91.3333333333)(107,91.3333333333)(108,91.3333333333)(109,91.3333333333)(110,91.3333333333)(111,91.3333333333)(112,91.3333333333)(113,91.3333333333)(114,91.3333333333)(115,91.3333333333)(116,91.3333333333)(117,91.3333333333)(118,91.3333333333)(119,91.3333333333)(120,91.3333333333)(121,91.3333333333)(122,91.3333333333)(123,91.3333333333)(124,91.3333333333)(125,91.3333333333)(126,91.3333333333)(127,91.3333333333)(128,91.3333333333)(129,91.3333333333)(130,91.3333333333)(131,91.3333333333)(132,91.3333333333)(133,91.3333333333)(134,91.3333333333)(135,91.3333333333)(136,91.3333333333)(137,91.3333333333)(138,91.3333333333)(139,91.3333333333)(140,91.3333333333)(141,91.3333333333)(142,91.3333333333)(143,91.3333333333)(144,91.3333333333)(145,91.3333333333)(146,91.3333333333)(147,91.3333333333)(148,91.3333333333)(149,91.3333333333)(150,91.3333333333)(151,91.3333333333)(152,91.3333333333)(153,91.3333333333)(154,91.3333333333)(155,91.3333333333)(156,91.3333333333)(157,91.3333333333)(158,91.3333333333)(159,91.3333333333)(160,91.3333333333)(161,91.3333333333)(162,91.3333333333)(163,91.3333333333)(164,91.3333333333)(165,91.3333333333)(166,91.3333333333)(167,91.3333333333)(168,91.3333333333)(169,91.3333333333)(170,91.3333333333)(171,91.3333333333)(172,91.3333333333)(173,91.3333333333)(174,91.3333333333)(175,91.3333333333)(176,91.3333333333)(177,91.3333333333)(178,91.3333333333)(179,91.3333333333)(180,91.3333333333)(181,91.3333333333)(182,91.3333333333)(183,91.3333333333)(184,91.3333333333)(185,91.3333333333)(186,91.3333333333)(187,91.3333333333)(188,91.3333333333)(189,91.3333333333)(190,91.3333333333)(191,91.3333333333)(192,91.3333333333)(193,91.3333333333)(194,91.3333333333)(195,91.3333333333)(196,91.3333333333)(197,91.3333333333)(198,91.3333333333)(199,91.3333333333)(200,91.3333333333)
};

 
\end{axis}
	\end{tikzpicture}
	\caption{Accuracy}
	\label{fig:accuracy}
	\medskip
	\small
\end{figure}

%% file: loss.tex
\begin{figure}[!h]
	\centering
	\begin{tikzpicture}
	
\begin{axis}[
    xlabel={Epoch},
    ylabel={Loss},
    legend pos=north west,
    ymajorgrids=true,
    grid style=dashed,
]

\addplot[color=blue]
coordinates {
(1,10.4208867105)(2,3.54709836245)(3,2.49701992989)(4,1.82397160769)(5,1.43311875979)(6,1.19170774698)(7,1.04767643134)(8,0.928929870129)(9,0.849987079302)(10,0.783912360668)(11,0.737599990368)(12,0.694495043755)(13,0.663360946178)(14,0.634204741319)(15,0.607301534017)(16,0.581317017873)(17,0.562984978358)(18,0.543091324965)(19,0.526758409341)(20,0.511348976294)(21,0.49542066733)(22,0.48448233525)(23,0.471139218012)(24,0.46061521848)(25,0.449605449041)(26,0.441346580187)(27,0.432134724458)(28,0.424323595365)(29,0.416519887447)(30,0.409754830996)(31,0.403712936242)(32,0.398227467537)(33,0.393072562218)(34,0.387631996473)(35,0.382697094282)(36,0.379256293774)(37,0.374848823547)(38,0.371861948967)(39,0.367107718786)(40,0.36467404445)(41,0.361048485438)(42,0.358396440347)(43,0.356146982511)(44,0.353112214406)(45,0.350838128726)(46,0.348096529643)(47,0.346560227871)(48,0.344988881747)(49,0.342361463706)(50,0.340565017064)(51,0.339166916211)(52,0.337667791843)(53,0.337078268528)(54,0.334806743463)(55,0.333399202824)(56,0.331588227749)(57,0.33059271733)(58,0.329080308278)(59,0.328821618557)(60,0.327389864127)(61,0.325973951022)(62,0.325373023351)(63,0.324868667126)(64,0.323185049693)(65,0.323312126795)(66,0.322028078238)(67,0.321320043405)(68,0.320612010956)(69,0.320513776938)(70,0.319735908508)(71,0.319512310028)(72,0.318820685546)(73,0.317756039302)(74,0.316770845254)(75,0.315983622074)(76,0.315458637873)(77,0.314942479928)(78,0.314327776432)(79,0.314089023272)(80,0.314094823996)(81,0.314049825668)(82,0.313600503604)(83,0.312807788054)(84,0.312146238486)(85,0.311612041791)(86,0.310949191252)(87,0.310612769127)(88,0.310046453476)(89,0.309523847103)(90,0.308986432552)(91,0.308665304979)(92,0.308878088792)(93,0.308586746852)(94,0.308492782911)(95,0.308048229218)(96,0.30765969038)(97,0.307377621333)(98,0.306815571785)(99,0.306474717458)(100,0.306192189058)(101,0.305982268651)(102,0.305350179672)(103,0.305388417244)(104,0.305023016135)(105,0.305322937171)(106,0.304928580125)(107,0.304613521894)(108,0.304450887839)(109,0.304386614164)(110,0.303951394558)(111,0.303636771838)(112,0.303348594507)(113,0.303165516059)(114,0.302883589268)(115,0.302566858133)(116,0.302258486748)(117,0.301905886332)(118,0.301690202554)(119,0.301829241117)(120,0.301660993099)(121,0.301389846007)(122,0.301276929379)(123,0.300533151627)(124,0.300493408839)(125,0.300438973904)(126,0.30063801686)(127,0.300356249809)(128,0.300268266996)(129,0.299990942478)(130,0.299675424894)(131,0.299506925742)(132,0.299580977758)(133,0.299486906528)(134,0.299353706042)(135,0.299030293624)(136,0.299131000837)(137,0.299040827751)(138,0.298642200629)(139,0.298620378971)(140,0.298458778858)(141,0.298525598049)(142,0.298570979436)(143,0.298200043043)(144,0.298149270217)(145,0.297947038015)(146,0.297996121248)(147,0.297908996741)(148,0.297834521135)(149,0.297700354258)(150,0.297609206041)(151,0.297325871785)(152,0.297285206318)(153,0.297139006456)(154,0.297052094936)(155,0.297054817677)(156,0.296980622609)(157,0.296587167581)(158,0.29650414149)(159,0.296559971968)(160,0.296308396657)(161,0.296045589447)(162,0.296044406891)(163,0.295965127945)(164,0.296094043255)(165,0.296161585649)(166,0.29586378177)(167,0.295864402453)(168,0.295618231297)(169,0.295448635419)(170,0.295336367289)(171,0.295325770378)(172,0.295229818821)(173,0.295061469078)(174,0.294995590051)(175,0.294918276469)(176,0.294669408003)(177,0.294492212931)(178,0.294641476472)(179,0.294363030593)(180,0.2943019104)(181,0.294367969831)(182,0.294297254086)(183,0.294205362002)(184,0.294039355119)(185,0.294215842883)(186,0.294310154915)(187,0.294238858223)(188,0.29395762682)(189,0.293773727417)(190,0.293799756368)(191,0.293674414158)(192,0.293686238925)(193,0.293584829171)(194,0.293474515279)(195,0.293450963497)(196,0.293507437706)(197,0.293371779124)(198,0.293456209501)(199,0.293230075041)(200,0.293039073944)
};

 
\end{axis}

	\end{tikzpicture}
	\caption{Loss}
	\label{fig:loss}
	\medskip
	\small
\end{figure}

%% file: accuracy-noentropy.tex
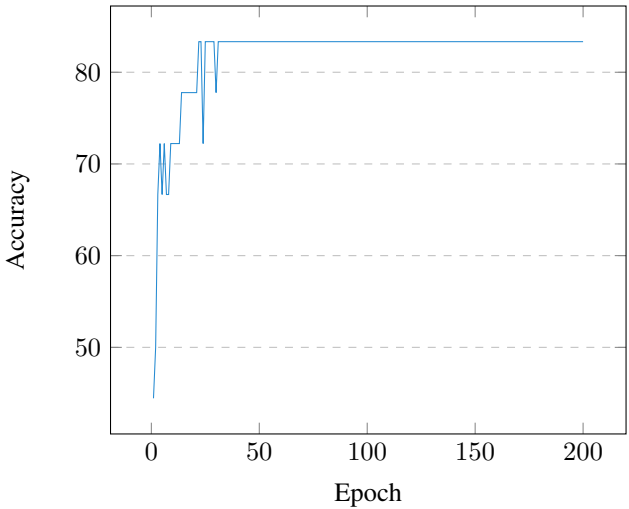
\begin{figure}[!h]
	\centering
	\begin{tikzpicture}

\begin{axis}[
    xlabel={Epoch},
    ylabel={Accuracy},
    legend pos=north west,
    ymajorgrids=true,
    grid style=dashed,
]

\addplot[color=blue]
coordinates {
(1,44.4444444444)(2,50.0)(3,66.6666666667)(4,72.2222222222)(5,66.6666666667)(6,72.2222222222)(7,66.6666666667)(8,66.6666666667)(9,72.2222222222)(10,72.2222222222)(11,72.2222222222)(12,72.2222222222)(13,72.2222222222)(14,77.7777777778)(15,77.7777777778)(16,77.7777777778)(17,77.7777777778)(18,77.7777777778)(19,77.7777777778)(20,77.7777777778)(21,77.7777777778)(22,83.3333333333)(23,83.3333333333)(24,72.2222222222)(25,83.3333333333)(26,83.3333333333)(27,83.3333333333)(28,83.3333333333)(29,83.3333333333)(30,77.7777777778)(31,83.3333333333)(32,83.3333333333)(33,83.3333333333)(34,83.3333333333)(35,83.3333333333)(36,83.3333333333)(37,83.3333333333)(38,83.3333333333)(39,83.3333333333)(40,83.3333333333)(41,83.3333333333)(42,83.3333333333)(43,83.3333333333)(44,83.3333333333)(45,83.3333333333)(46,83.3333333333)(47,83.3333333333)(48,83.3333333333)(49,83.3333333333)(50,83.3333333333)(51,83.3333333333)(52,83.3333333333)(53,83.3333333333)(54,83.3333333333)(55,83.3333333333)(56,83.3333333333)(57,83.3333333333)(58,83.3333333333)(59,83.3333333333)(60,83.3333333333)(61,83.3333333333)(62,83.3333333333)(63,83.3333333333)(64,83.3333333333)(65,83.3333333333)(66,83.3333333333)(67,83.3333333333)(68,83.3333333333)(69,83.3333333333)(70,83.3333333333)(71,83.3333333333)(72,83.3333333333)(73,83.3333333333)(74,83.3333333333)(75,83.3333333333)(76,83.3333333333)(77,83.3333333333)(78,83.3333333333)(79,83.3333333333)(80,83.3333333333)(81,83.3333333333)(82,83.3333333333)(83,83.3333333333)(84,83.3333333333)(85,83.3333333333)(86,83.3333333333)(87,83.3333333333)(88,83.3333333333)(89,83.3333333333)(90,83.3333333333)(91,83.3333333333)(92,83.3333333333)(93,83.3333333333)(94,83.3333333333)(95,83.3333333333)(96,83.3333333333)(97,83.3333333333)(98,83.3333333333)(99,83.3333333333)(100,83.3333333333)(101,83.3333333333)(102,83.3333333333)(103,83.3333333333)(104,83.3333333333)(105,83.3333333333)(106,83.3333333333)(107,83.3333333333)(108,83.3333333333)(109,83.3333333333)(110,83.3333333333)(111,83.3333333333)(112,83.3333333333)(113,83.3333333333)(114,83.3333333333)(115,83.3333333333)(116,83.3333333333)(117,83.3333333333)(118,83.3333333333)(119,83.3333333333)(120,83.3333333333)(121,83.3333333333)(122,83.3333333333)(123,83.3333333333)(124,83.3333333333)(125,83.3333333333)(126,83.3333333333)(127,83.3333333333)(128,83.3333333333)(129,83.3333333333)(130,83.3333333333)(131,83.3333333333)(132,83.3333333333)(133,83.3333333333)(134,83.3333333333)(135,83.3333333333)(136,83.3333333333)(137,83.3333333333)(138,83.3333333333)(139,83.3333333333)(140,83.3333333333)(141,83.3333333333)(142,83.3333333333)(143,83.3333333333)(144,83.3333333333)(145,83.3333333333)(146,83.3333333333)(147,83.3333333333)(148,83.3333333333)(149,83.3333333333)(150,83.3333333333)(151,83.3333333333)(152,83.3333333333)(153,83.3333333333)(154,83.3333333333)(155,83.3333333333)(156,83.3333333333)(157,83.3333333333)(158,83.3333333333)(159,83.3333333333)(160,83.3333333333)(161,83.3333333333)(162,83.3333333333)(163,83.3333333333)(164,83.3333333333)(165,83.3333333333)(166,83.3333333333)(167,83.3333333333)(168,83.3333333333)(169,83.3333333333)(170,83.3333333333)(171,83.3333333333)(172,83.3333333333)(173,83.3333333333)(174,83.3333333333)(175,83.3333333333)(176,83.3333333333)(177,83.3333333333)(178,83.3333333333)(179,83.3333333333)(180,83.3333333333)(181,83.3333333333)(182,83.3333333333)(183,83.3333333333)(184,83.3333333333)(185,83.3333333333)(186,83.3333333333)(187,83.3333333333)(188,83.3333333333)(189,83.3333333333)(190,83.3333333333)(191,83.3333333333)(192,83.3333333333)(193,83.3333333333)(194,83.3333333333)(195,83.3333333333)(196,83.3333333333)(197,83.3333333333)(198,83.3333333333)(199,83.3333333333)(200,83.3333333333)
};
 
\end{axis}
	\end{tikzpicture}
	\caption{No Entropy}
	\label{fig:accuracy-ent}
	\medskip
	\small
\end{figure}

%% file: confusion-validation.tex
\begin{center}
	\begin{tabular}{c|cccccc|}  
				& AES & RC4 & BLF & MD5 & RSA & R/A \\
		\hhline{-------}
		AES		& 37\cellcolor[gray]{.8}		& 0	& 0	& 0	& 13 & 0	\\
		RC4		& 0	 & 50\cellcolor[gray]{.8}		& 0	& 0	& 0	 & 0	\\
		BLF 	& 0	 & 8 & 42\cellcolor[gray]{.8}		& 0	& 0	 & 0	\\
		MD5		& 0	 & 2 & 7 & 41\cellcolor[gray]{.8}		& 0	 & 0	\\
		RSA		& 19 & 0 & 0 & 0	& 30\cellcolor[gray]{.8}	 & 1	\\
		R/A		& 0	 & 0 & 0 & 0	& 0	& 50\cellcolor[gray]{.8}		\\
		\hhline{~------}
	\end{tabular}
	\captionof{table}{Validation Results}
	\label{tbl:confusion}
	\medskip
	\small
	$x = predicted, y = actual$
\end{center}